\documentclass[twocolumn]{aastex62}

\hypersetup{linkcolor=red}
\newcommand\st{\bgroup\markoverwith
{\textcolor{magenta}{\rule[0.5ex]{2pt}{1pt}}}\ULon}

\usepackage{amsmath, bm, hyperref, ulem}
\usepackage{xcolor,colortbl, array, makecell} 
\newcolumntype{?}[1]{!{\vrule width #1}}
\definecolor{Gray}{gray}{0.85}
\definecolor{LightCyan}{rgb}{0.88,1,1}
\usepackage{savesym} \savesymbol{tablenum}
\usepackage[detect-weight=true, detect-family=true]{siunitx} \restoresymbol{SIX}{tablenum}
\usepackage[T1]{fontenc}
\usepackage[overload]{textcase}
\graphicspath{{./}{Figures/}}

\def\vsini{\hbox{$v$\,sin\,$i$}}      

\shorttitle{WASP-96\MakeLowercase{b}} \shortauthors{McGruder et al.}

\begin{document} 
\title{ACCESS: Confirmation of a Clear Atmosphere for WASP-96b and a Comparison of Light Curve Detrending Techniques}

\correspondingauthor{Chima D. McGruder} \email{chima.mcgruder@cfa.harvard.edu}

\author[0000-0002-6167-3159]{Chima D. McGruder}\altaffiliation{NSF Graduate Research Fellow}\affiliation{Center for Astrophysics ${\rm \mid}$ Harvard {\rm \&} Smithsonian, 60 Garden St, Cambridge, MA 02138, USA}

\author[0000-0003-3204-8183]{Mercedes L\'opez-Morales} \affiliation{Center for Astrophysics ${\rm \mid}$ Harvard {\rm \&} Smithsonian, 60 Garden St, Cambridge, MA 02138, USA}

\author[0000-0002-4207-6615]{James Kirk}\affiliation{Center for Astrophysics ${\rm \mid}$ Harvard {\rm \&} Smithsonian, 60 Garden St, Cambridge, MA 02138, USA}

\author[0000-0001-9513-1449]{N\'estor Espinoza} \affiliation{Space Telescope Science Institute (STScI), 3700 San Martin Dr, Baltimore, MD 21218, USA}

\author[0000-0002-3627-1676]{Benjamin V.\ Rackham}
\altaffiliation{51 Pegasi b Fellow}
\affiliation{Department of Earth, Atmospheric and Planetary Sciences, and Kavli Institute for Astrophysics and Space Research, Massachusetts Institute of Technology, Cambridge, MA 02139, USA}

\author[0000-0003-4157-832X]{Munazza K. Alam}
\affiliation{Carnegie Earth \& Planets Laboratory, 5241 Broad Branch Road NW, Washington, DC 20015}

\author[0000-0002-0832-710X]{Natalie Allen} \affiliation{William H. Miller III Department of Physics and Astronomy, Johns Hopkins University, 3400 N Charles St, Baltimore, MD 21218, USA}

\author[0000-0002-6500-3574]{Nikolay Nikolov}\affiliation{Space Telescope Science Institute (STScI), 3700 San Martin Dr, Baltimore, MD 21218, USA}

\author[0000-0001-6205-6315]{Ian C. Weaver}
\affiliation{Center for Astrophysics ${\rm \mid}$ Harvard {\rm \&} Smithsonian, 60 Garden St, Cambridge, MA 02138, USA}

\author[0000-0003-3455-8814]{Kevin Ortiz Ceballos}\altaffiliation{NSF Graduate Research Fellow}\altaffiliation{Ford Foundation Pre-doctoral Fellow}\affiliation{Center for Astrophysics ${\rm \mid}$ Harvard {\rm \&} Smithsonian, 60 Garden St, Cambridge, MA 02138, USA}

\author[0000-0003-0412-9664]{David J. Osip}\affiliation{Las Campanas Observatory, Carnegie Institution of Washington, Colina el Pino, Casilla 601 La Serena, Chile}

\author[0000-0003-3714-5855]{D\'aniel Apai}\affiliation{ Steward Observatory, The University of Arizona, 933 N. Cherry Avenue, Tucson, AZ 85721, USA}\affiliation{Lunar and Planetary Laboratory, The University of Arizona, 1629 E. University Boulevard, Tucson, AZ 85721}

\author[0000-0002-5389-3944]{Andr\'es Jord\'an}
\affiliation{Facultad de Ingeniera y Ciencias, Universidad Adolfo Ib\'{a}\~{n}ez, Av. Diagonal las Torres 2640, Pe\~{n}alol\'{e}n, Santiago, Chile}
\altaffiliation{Millennium Institute for Astrophysics, Chile} \altaffiliation{Data Observatory Foundation, Chile}

\author[0000-0002-9843-4354]{Jonathan J. Fortney} \affiliation{Department of Astronomy \& Astrophysics, University of California, Santa Cruz, CA 95064}

\begin{abstract} 
One of the strongest ${\rm Na~I}$ features was observed in WASP-96b. To confirm this novel detection, we provide a new 475--825\,nm transmission spectrum obtained with Magellan/IMACS, which indeed confirms the presence of a broad sodium absorption feature. We find the same result when reanalyzing the 400--825\,nm VLT/FORS2 data. We also utilize synthetic data to test the effectiveness of two common detrending techniques: (1) a Gaussian processes (GP) routine, and (2) common-mode correction followed by polynomial correction (CMC+Poly). We find that both methods poorly reproduce the absolute transit depths but maintain their true spectral shape. This emphasizes the importance of fitting for offsets when combining spectra from different sources or epochs. Additionally, we find that for our datasets both methods give consistent results, but CMC+Poly is more accurate and precise. We combine the Magellan/IMACS and VLT/FORS2 spectra with literature 800--1644\,nm HST/WFC3 spectra, yielding a global spectrum from 400--1644\,nm. We used the \texttt{PLATON} and \texttt{Exoretrievals} retrieval codes to interpret this spectrum, and find that both yield relatively deeper pressures where the atmosphere is optically thick at log-pressures between $1.3^{+1.0}_{-1.1}$ and 0.29$^{+1.86}_{-2.02}$\,bars, respectively. \texttt{Exoretrievals} finds a solar to super-solar ${\rm Na~I}$ and ${\rm H_2O}$ log-mixing ratios of $-5.4^{+2.0}_{-1.9}$ and $-4.5^{+2.0}_{-2.0}$, respectively, while \texttt{PLATON} finds an overall metallicity of $log_{10}(Z/Z_{\odot}) = -0.49^{+1.0}_{-0.37}$\,dex. Therefore, our findings are in agreement with literature and support the inference that the terminator of WASP-96b has few aerosols obscuring prominent features in the optical to near-infrared (near-IR) spectrum.
\end{abstract}

\keywords{planets and satellites: atmospheres --- 
planets and satellites: --- 
stars: activity --- stars: starspots --- techniques: spectroscopic}


\section{Introduction} \label{sec:intro} 
In-depth studies of exoplanetary atmospheres (exo-atmospheres) is a key pathway to obtaining more detailed insights about the formation and evolution of planetary systems. Many of these planets are in extreme environments not found in the solar system and understanding how their atmospheres are sculpted by such unique environments gives us detailed insights on the complex chemistry and physics at play. Examples include the cause of observed temperature inversions in hot Jupiters \citep[e.g.][]{Baxter:2020, Gandhi:2019}, the cause and composition of high altitude aerosols (clouds and/or hazes) \citep[e.g.][]{Moses:2011, wakeford:2019b, Gao:2020}, and observed super sonic wind speeds \citep[e.g.][]{Fromang:2016}. Observing exo-atmospheres also can improve our understanding of the formation and evolutionary processes that exoplanets undergo, e.g., host disk dissipation rate \citep[e.g.,][]{Powell:2021} and hot Jupiter migration timescales \citep[e.g.,][]{Moses:2013, Powell:2021}. 

To improve our understanding of exoplanets, observations have had to push the performance of instruments, which were not designed with exo-atmosphere studies in mind, to their limits. This has also been the case for data analysis techniques aimed at removing both instrumental and astrophysical systematics both from space-based observatories \citep[e.g.][]{Pont:2013,Sing:2019} and ground-based ones \citep[e.g.][]{Gibson:2012,Yan2020}.

As expected in any developing field, there have been a number of cases with disagreeing results (e.g., \citealt{Southworth2017} and \citealt{Diamond-Lowe2018}; \citealt{Sedaghati2017} and \citealt{Espinoza2019}; and \citealt{Sing:2015} and \citealt{Gibson2017WASP31}, \citealt{Gibson2019WASP31}, and \citealt{McGruder:2020}). These discrepancies are attributed to imperfect understanding of systematics, whether it be instrumental, observational, or astrophysical in nature. This highlights the importance of confirming features of interest via independent analyses in order to isolate spurious signals and instill more confidence in agreeing detections. This is particularly important in attempts to find correlations between atmospheric features and other system parameters, such as the cause of high-altitude aerosol formation in gas giants. There have been several studies attempting to find such a correlation \citep[i.e.][]{Sing:2016,Heng:2016,Stevenson:2016,Fu:2017,Tsiaras:2018, Dymont:2021}, with no clear answer yet \citep[e.g.][]{Alam2020}.

WASP-96b \citep[M=0.48\,M\textsubscript{J}, $R=1.2\,R\textsubscript{J}$, $P=3.425$\,days, G8 host star,][]{Hellier:2014}, is one of few exoplanets observed to-date to have little evidence supporting the presence of high-altitude aerosols in both optical \citep{Nikolov:2018} and near-infrared \citep[IR,][]{Yip:2021} transmission spectra. An exoplanet with a transmission spectrum that can be modeled excluding high-altitude aerosols is also called a ``clear'' atmosphere, which does not mean the planet has no aerosols but little to no aerosols obscure the summed terminator's spectrum. Other planetary atmosphere that can be explained without including high-altitude aerosols are 
WASP-17b \citep{Sing:2016}, WASP-39b \citep[]{Fischer:2016,Nikolov2016,Wakeford:2018,Kirk:2019}, WASP-62b \citep{Alam:2021}, and WASP-94Ab \citep{Ahrer:2022}. Finding planets like these is essential for understanding the evolutionary, chemical, and physical processes underway in this class of planets because aerosols mute the features needed to probe exo-atmospheres. This is particularly challenging given that it is estimated only ${\sim} 7\%$ of hot Jupiters have clear atmospheres \citep{wakeford:2019b}. As such, it is vital to find and thoroughly study clear planets like WASP-96b. The optical transmission spectrum of WASP-96b was first observed by \cite{Nikolov:2018} with VLT/FORS2 and recently \cite{Yip:2021} combined the VLT/FORS2 spectrum with a near-IR spectrum derived using HST/WFC3 observations from GO program 15469 (PI: Nikolay Nikolov) to obtain a better picture of the planet's clear nature. These studies report strong ${\rm Na~I}$ and ${\rm H_2O}$ features with abundances of $-3.88^{+1.05}_{-0.82}$ and $-3.65^{+0.90}_{-0.94}$, respectively. 

In this paper, we present a new optical transmission spectrum of WASP-96b derived from new observations obtained as part of ACCESS\footnote{The Atmospheric Characterization Collaboration for Exoplanet Spectroscopic Studies (ACCESS) survey on the Magellan Telescopes \citep{Jordan:2013, Rackham:2017,Bixel:2019,Espinoza2019,Weaver:2020,McGruder:2020,Weaver:2021,Kirk:2021}}. 

These observations are described in Section~\ref{sec:Observs}. We then combined the ACCESS observations with an independent re-analysis of the original VLT/FORS observations and the available near-IR observations to derive a new 400--1644\,nm optical to near-infrared transmission spectrum for this planet and re-inspect its atmospheric properties via retrieval models. 

To understand individual and relative performances of commonly used detrending methods, we also conduct
a detailed comparison of two often-used techniques in ground-based transmission spectroscopy: (1) a Gaussian processes routine \citep{Gibson:2012,Weaver:2020,Yan2020,McGruder:2020,Weaver:2021}, and (2) common-mode correction followed by polynomial correction \citep{Gibson:2013,Nikolov2016,Gibson2017WASP31,Nikolov:2018,Carter2020}. We discuss these detrending techniques and their performance comparison in Section~\ref{sec:LC_analysis}.

The remainder of this paper is structured as follows.
Section~\ref{sec:Tran_Spec} describes our methods for constructing the transmission spectra, and Section~\ref{sec:Activity} outlines how we consider stellar activity affecting the transmission spectra. The spectra are then interpreted using retrievals in Section~\ref{sec:Retrievals}. In Section~\ref{sec:Retrieval_Interp}, we present and discuss the retrieval analysis results, and contextualize WASP-96b within the exoplanet population in Section~\ref{sec:Contxt}. We conclude in Section~\ref{sec:Conclusion}.

\section{Observations} \label{sec:Observs}

\subsection{VLT/FORS2 Transits}
We used two transit observations of WASP-96b on the nights of UT170729 and UT170822 (UTYYMMDD) with the FOcal Reducer and Spectrograph (FORS2) \footnote{\href{https://www.eso.org/sci/facilities/paranal/instruments/fors.html}{\texttt{FORS instrument website}}} mounted on the 8.2\,m Very Large Telescope (VLT) in the European Southern Observatory on Cerro Paranal, Chile. These observations were originally collected, reduced, and published by \cite{Nikolov:2018}. For our re-analysis of this data, described in Section~\ref{sec:LC_analysis}, we used the same extracted spectra produced in \cite{Nikolov:2018}, where additional detail of the data collection and extraction process can be found. Both observations were taken with the multi-object spectroscopy (MOS) mode and the same two-slit mask. Each slit in the mask was $\SI{22}{\arcsecond}$ by $\SI{120}{\arcsecond}$ and centered on the target and comparison star. Given the wide slits, the spectral resolution of the observations were determined by the seeing, with an average resolving power of R $\sim$ 600. The first transit was observed using the bluer dispersive element, GRIS600B (600B), which had approximate spectral coverage of 360--620\,nm. Contrary to \cite{Nikolov:2018}, we excluded the spectrum from 360--400\,nm to avoid systematics caused by the low counts, where that region had over 85\% fewer counts than the rest of the spectrum. The second night used the redder GRIS600RI (600RI) grism and the GG435 filter, producing an approximate spectral coverage of 520--835\,nm. 

\subsection{Magellan/IMACS Transits} \label{subsec:Setup}
We observed two transits of WASP-96b as part of the ACCESS Survey \footnote{ACCESS generally acquires $\geq$ 3 transits of a target to reduce systematics and increase precision, but determined two sufficient when considering the already observed VLT/FORS2 transits.} on the nights of UT170804 and UT171108 with the Inamori-Magellan Areal Camera \& Spectrograph \cite[IMACS;][]{2011Dressler} on the 6.5-m Baade Magellan Telescope at Las Campanas Observatory in Chile. Both transits were observed using the 8K\,$\times$\,8K CCD mosaic camera at the f/2 focus, which provides a \SI{27.4}{\arcmin} field of view (FoV). We used a 300\,line/mm grating at blaze angle of 17.5$^{\circ}$, yielding a spectral coverage of 0.44--0.97\,$\mu$m (without a filter). To reduce readout time and improve the duty cycle of the observations, we used 2$\times$2 binning. Both observations used the FAST readout mode, which had a 29\,s readout time and a 3\,s overhead. The first observation had no filter, but we applied the GG495 filter (coverage 0.49--0.97\,$\mu$m) for the second night to prevent second-order light contamination. The effect of second-order contamination on the first observation is discussed in Appendix~\ref{Appx:2ndOrderCorr}.

For each observation we used the MOS mode and designed a custom science mask with $\SI{10}{\arcsecond}$ by $\SI{90}{\arcsecond}$ slits for the target and a number of comparison stars in the field. Identical masks, but with slit widths of $\SI{0.5}{\arcsecond}$ instead of $\SI{10}{\arcsecond}$, were also designed for wavelength calibrations. The average seeing-limited resolving power of the observations were R $\sim$ 900 

The comparison stars were selected using the same prescription described in \cite{Rackham:2017}. We consider a nearby star as a suitable comparison if it has a color difference with WASP-96 of $D < 1$, where 
\begin{equation*}
D=\sqrt{[(B-V)_c - (B-V)_t]^2 + [(J-K)_c - (J-K)_t]^2}
\end{equation*}
The uppercase letters in the equation correspond to the Johnson-Cousin apparent magnitudes of the stars, and the subscripts $t$ and $c$ indicate the target and comparison, respectively. The comparisons used in both the ACCESS and VLT/FORS2 observations are summarized in Table~\ref{tab:comp_stars}.

\begin{deluxetable*}{cccccccc}[htb]
    \caption{Target and comparison star magnitudes and coordinates from the UCAC4 catalog \citep{2013UCAC4}. We used comparisons 14 and 15 for the analysis of both IMACS transits and comparison 1 for both FORS2 transits. Comparison 3 was in the IMACS slit but was over saturated and not used.} 
    \label{tab:comp_stars}
    \tablehead{\colhead{Star} & \colhead{RA} & \colhead{Dec} & \colhead{B} & \colhead{V} & \colhead{J} & \colhead{K} & \colhead{D}}
    \startdata 
  WASP-96 & 00:04:11.12 & -47:21:38.25 & 13.25 & 12.51 & 11.27 & 10.91 & 0 \\
  COMP1   & 00:04:18.87 & -47:16:31.05 & 13.52 & 12.88 & 11.76 & 11.41 & 0.1 \\
  COMP3   & 00:04:02.03 & -47:14:49.87 & 12.19 & 11.28 & 9.1   & 9.67 & 0.94 \\
  COMP14  & 00:05:12.82 & -47:03:37.02 & 13.87 & 12.98 & 11.40 & 10.85 & 0.24 \\
  COMP15  & 00:04:25.01 & -47:00:42.96 & 13.86 & 13.17 & 11.93 & 11.44 & 0.14
    \enddata 
\end{deluxetable*}

We limited observations of each transit to airmass below 2 during the full transit window to minimize differential atmospheric refraction effects, so we ended up with two full transits with additional out-of-transit baselines of 1.29 and 1.35 hours for the UT170804 and UT171108 transits, respectively. To maintain count levels of 25,000--35,000\,
ADU\footnote{This is well within the linearity limits of the IMACS CCDs \citep{Bixel:2019}.}, given the average seeing conditions at each night, we held exposure times constant to 60 seconds for transit UT170804 (average seeing of $\SI{1.55}{\arcsecond}$) and 45 seconds for transit UT171108 (average seeing of $\SI{0.71}{\arcsecond}$). A summary of each transit's observing conditions are given in Table~\ref{tab:ObsLog}.

Finally, we collected bias frames, quartz lamp flats, HeNeAr calibrations (using the $\SI{0.5}{\arcsecond}$ by $\SI{90}{\arcsecond}$ masks), each with the same binning as the science observations.

\begin{deluxetable}{ccccccccc}[htb]
    \caption{Observing log for WASP-96b data from \textit{Magellan/IMACS}.} 
    \label{tab:ObsLog}
    \tablehead{{Night Obs.} & {Obs. Start/} & {Airmass} & {Frames} &{min/max}\\ {Start (UTC)} & {End (UTC)} & & & {seeing ["]}}
    \startdata 
    2017 Aug. 04 & 02:55/06:38 & 1.93--1.09 & 148  & 1.4/1.71 \\
    2017 Nov. 08 & 00:01/03:49 & 1.11--1.18 & 180 & 0.68/0.8 \\
    \enddata 
\end{deluxetable}

\subsubsection{ACCESS Reduction Pipeline} \label{subsec:Pipline}
We reduced the data using the ACCESS custom pipeline introduced in \cite{Jordan:2013} and described in detail in \cite{Espinoza:2017}. 
We first subtracted the bias level from each frame using the median of the overscan region. 
The pipeline also has the option of flat-fielding each frame using a master flat, but as with previous ACCESS datasets \citep[e.g.][]{Rackham:2017,Bixel:2019,McGruder:2020,Weaver:2021,Kirk:2021}, we found that flat-fielding worsens the photometric precision of the transit light curves, so we opted to not flat-field the data. The ineffectiveness of flat-fielding is likely due to the lack of sufficient flats needed for the high precision analysis. We collected 10 to 25 flats per night with 1 to 5 second exposures.

We then performed a bad-pixel correction using a bad-pixel map obtained from the flats. Any pixel with 10 times higher or lower photon noise relative to neighboring pixels was added to the map. The correction was done to the science images by replacing the value of each flagged pixel with a value interpolated from the neighboring pixels in the dispersion direction of the detector. 

We traced each spectrum in the images by first using a second-order polynomial to identify each resolution element's centroid (relative to both spatial and dispersion directions). Then a fourth-order polynomial was fit on the centroids in order to ensure the smoothness of the trace.

The sky background was subtracted using the median of the spectral counts outside of the central apertures for a given wavelength element. The appropriate aperture sizes were empirically determined to be $\SI{3.6}{\arcsecond}$ (9\,pixels) in radius for UT170804 and $\SI{3.2}{\arcsecond}$ (8\,pixels) for UT171108. The extracted spectrum was produced by summing the spectral profile within the central aperture in the spatial direction and subtracting the sky background.

A Lorentzian profile was used to fit each spectral line in the HeNeAr calibration spectra. The wavelengths of each line as a function of pixel position was then fit using sixth-order polynomials. The pipeline iteratively excluded deviant datapoints until the root mean square error value of the fit was less than 2\,km\,s$^{-1}$ ($\sim$ 0.05\,{\AA}). This wavelength map was applied to each extracted spectrum. 

We identified and removed cosmic rays by first creating a global median spectrum using the median of all the normalized spectra for each exposure. This normalized, median spectrum was used to compare against each individual normalized spectrum, and any counts in a spectrum more than $10\sigma$ greater than the global median spectrum (on a corresponding wavelength) were replaced by interpolating the counts at the appropriate wavelength from the preceding and following spectra in the time series. 

It should also be noted that the pipeline is capable of doing optimal extraction \citep{Marsh1989OptExtract}. This could potentially be used in place of bad-pixel and cosmic-ray removal, as it effectively highlights and removes deviant counts (Allen et. al., in prep.). However, we tested this for both WASP-96 observations, and found very little difference between resulting spectra. 
 
\begin{figure*}[htb]
    \centering
    \includegraphics[width=1\textwidth]{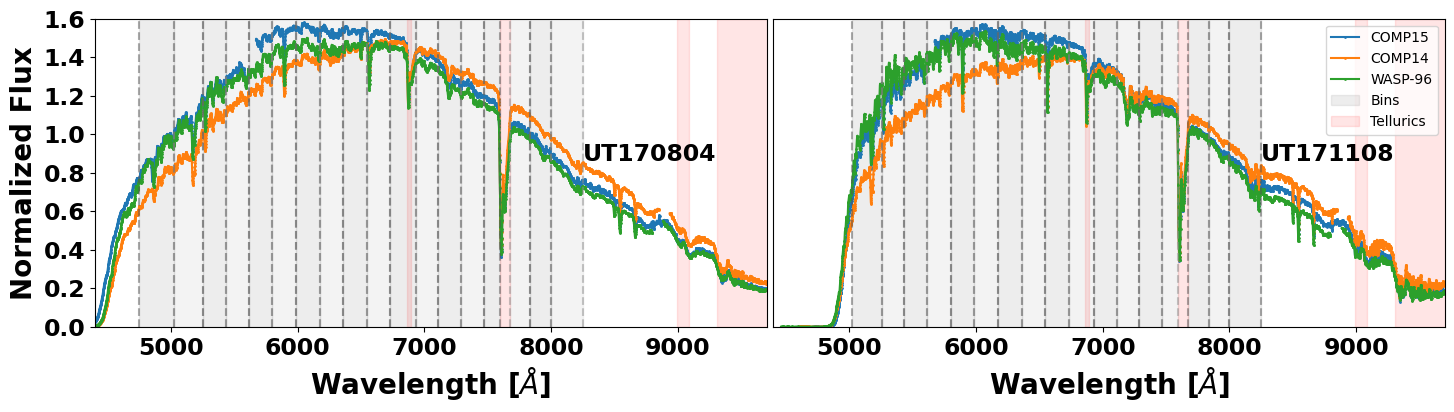}
    \includegraphics[width=1\textwidth]{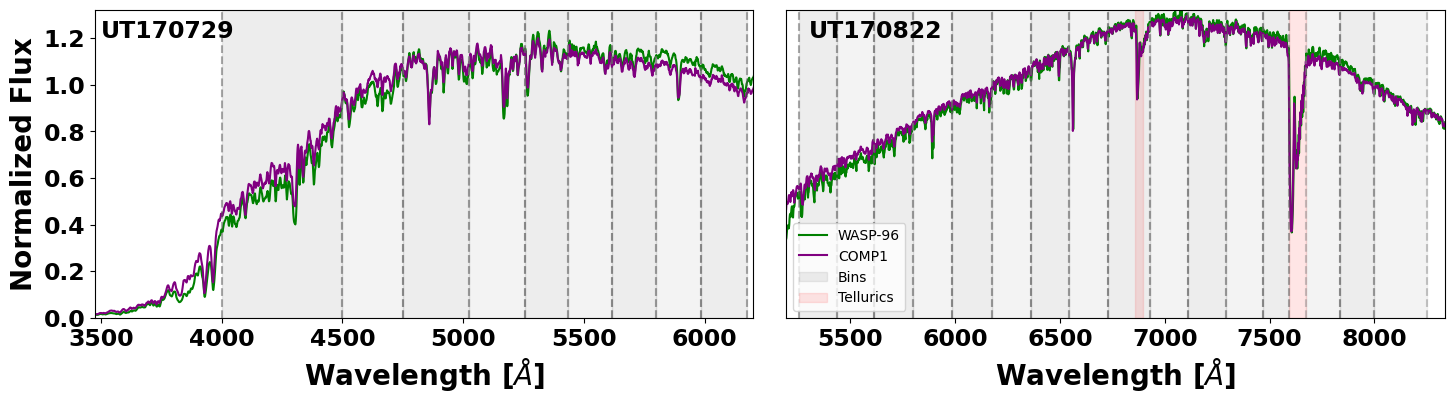}
    \caption{Median extracted spectra of WASP-96 and comparisons, excluding the oversaturated COMP 3. The top row shows IMACS data (UT170804 and UT171108) and the bottom shows FORS2 data from \citealt{Nikolov:2018} (UT170729 and UT170822). Each spectroscopic bin used is shaded in light gray and demarked by dotted lines. Telluric regions with less than 0.1 transmission are shaded in light red. The only gap in the binning scheme is the strong telluric region from 7594.0--7672.0{\AA}.} 
    \label{fig:ExtraSpec} 
\end{figure*}

Figure~\ref{fig:ExtraSpec} shows the median extracted spectrum for both nights. The final wavelength range used was 475--825\,nm for UT170804 and 525--825\,nm for UT171108 (see Fig.~\ref{fig:ExtraSpec}). Given the diminishing throughput on the edge of the spectra, the bluest wavelengths had too few counts to be useful. We also excluded data from 759.4--767.2\,nm, a region of strong tellurics. Though the sky subtraction and dividing by the comparison(s) should in principle negate the telluric features, that region still had a drastic decrease in counts (and likely residual telluric-induced systematics). This made it difficult to properly fit a light-curve at that wavelength range. Unfortunately, this coincides with the ${\rm K~I}$ doublet (centered at 768.15\,nm). We also excluded wavelengths longer than 825\,nm due to the diminishing throughput and increase in tellurics (see Fig.~\ref{fig:ExtraSpec}). We could have attempted to obtain useful information from that range, but given that the available HST/WFC3/IR/G102 data \citep{Yip:2021} is not afflicted by any of these effects, we instead omit that wavelength range from the ACCESS analysis. 

When using the f/2 mode, each stellar spectrum is dispersed on two CCD chips, causing a gap in the spectra (see Figure~\ref{fig:ExtraSpec}). Fortunately, all utilized spectra of the target and COMP14 fell on one chip. For COMP15, the gap was approximately between 5580-5675{\AA}; however, that region still had COMP14 for systematic corrections. 

\subsection{HST WFC3 Transits} \label{subsec:HSTSpectrum}

The HST data were obtained directly from \cite{Yip:2021} (GO program 15469, P.I. N. Nikolov). It consisted of two transits, one on UT181218 with the G102 grism (0.8--1.1\,$\mu$m) and another on UT181228 G141 (1.1--1.7\,$\mu$m) grism. The raw spatially scanned spectra were reduced with \texttt{Iraclis} \citep{Tsiaras:2016}. The quadratic limb-darkening parameters were obtained using the models of \cite{Claret:2013} and the inclination (i\,=\,1.486\,radians), semimajor axis relative to stellar radius ($a/R_s$\,=\,8.84), and orbital period (P\,=\,3.4252602\,days) were all held fixed to parameters determined from \cite{Hellier:2014} and \cite{Nikolov:2018}. 


\subsection{Photometric Monitoring} \label{subsec:Photo_Monitor}

We compiled and analyzed available time-series photometry of the host star WASP-96 from the Transiting Exoplanet Survey Satellite \citep[TESS,][]{TESS2014} and the All-Sky Automated Survey for Supernovae \citep[ASAS-SN,][]{Kochanek:2017} to constrain the presence of star spots that could contaminate the transmission spectrum of WASP-96b.

TESS observed WASP-96 in 2-minute-cadence mode over 27 days during Sector 2 (between 23 August and 20 September 2018) and over 24 days in Sector 29 (between 26 August and 20 September 2020). To model its photometric modulation, we used the Pre-search Data Conditioning Simple Aperture Photometry \citep[][PDCSAP]{Jenkins2016} light curve obtained from the
\href{https://archive.stsci.edu/missions-and-data/tess}{\texttt{Barbara A. Mikulski Archive for Space Telescopes (MAST) portal}}. We masked the transit data using a period of 3.42526 days, initial mid-transit time (t0) of 2458354.319946 BJD, and a transit window of 3.04 hours (25\% buffer in duration) \footnote{where the transit duration of 2.445 hours was calculated using the system parameters of \cite{Yip:2021}}. 
The resulting out-of-transit light curves have a median absolute deviation (MAD) of 2.76 parts-per-thousand (ppt) and a peak-to-peak difference of 25.5\,ppt for Sector 2 (17676 data points), and a MAD of 2.9\,ppt and peak-to-peak difference of 37.3\,ppt for Sector 29 (14252 data points). For our analysis of the TESS data, we binned every 100 observations together.

ASAS-SN observed WASP-96 in two filters. $V$ filter observations were collected between April 2014 and September 2018. $g$ filter observations were collected between September 2017 and February 2020. To remove deviating observations, we sigma-clipped points that deviated by more than $3\sigma$ from the mean. This led to excluding 11 of 921 and 27 of 1626 observations for the V and g filters, respectively. Because of the lower cadence of the ASAS-SN observations ($\sim$3 per day) compared to the TESS observations, their larger variance, and since we do not expect significant stellar modulation in a day, we weighted averaged observations taken on the same night. We then removed observations with uncertainties larger than 3 times the mean uncertainty, which led to six binned observations removed in V band, and seven in g band. The final light curves included 351 \& 520 datapoints with MADs of 10.4 \& 9.5\,ppt and peak-to-peak differences of 83.5 \& 96.5\,ppt, respectively.


\section{Light-curve Analysis and Comparison of Detrending Techniques}\label{sec:LC_analysis}
All the wavelengths used in each extracted spectrum obtained in Section \ref{subsec:Pipline} were summed for every exposure to construct a photometric white-light curve. Furthermore, binned-light curves were constructed by partitioning the spectra into distinct spectro-photometric bins that were used to determine the change in transit depth over wavelength (i.e., the transmission spectrum). The binning scheme was designed by considering spectro-photometric precision, the overlap of spectral bands from different observations, high telluric absorption regions, and the desire to properly probe for atmospheric features, such as a scattering slope, sodium, and potassium lines. The bins used to construct the transmission spectrum are shaded grey in Figure \ref{fig:ExtraSpec} and their wavelength coverages are listed in the first column of Table~\ref{tab:Optical_Spec}. An overview of the white-light and binned light curve detrending technique implemented for our final light curves is provided in sections \ref{sec:LC} \& \ref{sec:Binned}. However a more detailed description of our detrending routines are given in Appendix \ref{Appx:Detrending_Routines}. This section ends (\ref{sec:Synth}) with an explanation of how we determined the best detrending methods for our data. 

\subsection{White Light Curve Fitting} \label{sec:LC} 
Our first step in detrending the VLT/FORS2 and Magellan/IMACS white-light curves was a sigma clipping of the raw light curves (target divided by mean of comparisons). This was done by individually evaluating each data point in the light curve with a moving average of 11 points centered around the point of interest. If the value of the specific point deviated by more than $3\sigma$ from the mean it was removed. We opted for this method because it does not depend on an initial transit model fit, which is often used as a sigma-clipping criteria. It resulted in zero, two, three, and two points being removed from each transit in chronological order. 

We then used the PCA+GP routine (see Appendix~\ref{appx:PCA}), to fit the sigma-clipped white-light curves. The transit parameters used in the fits were quadratic limb darkening (LD) coefficients, $q_1, q_2$, planet orbital period, $P$, semi-major axis relative to the stellar radius, $a/R_s$, the planet-to-star radius ratio, $R_p/R_s$, impact parameter, $b$, mid-transit time, $t_0$, eccentricity, $e$, and the argument of periastron, $\omega$. $q_1$ and $q_2$ are parameterizations of the more commonly used $u_1$ and $u_2$ quadratic LD parameters, where $q_1=(u_1 + u_2)^2$ and $q_2=u_1/2(u_1 + u_2)$ \citep{Kipping:2013}. 
For our analysis we had to express $b$ in terms of inclination, $i$, as
\begin{align}\label{equ:inclination}
    i=\cos^{-1}\left(\frac{b}{a}*\frac{(1 + e*\sin(\omega))}{1 - e^2}\right).
\end{align}

\begin{figure*}[htb]
    \centering
    \includegraphics[width=1\textwidth]{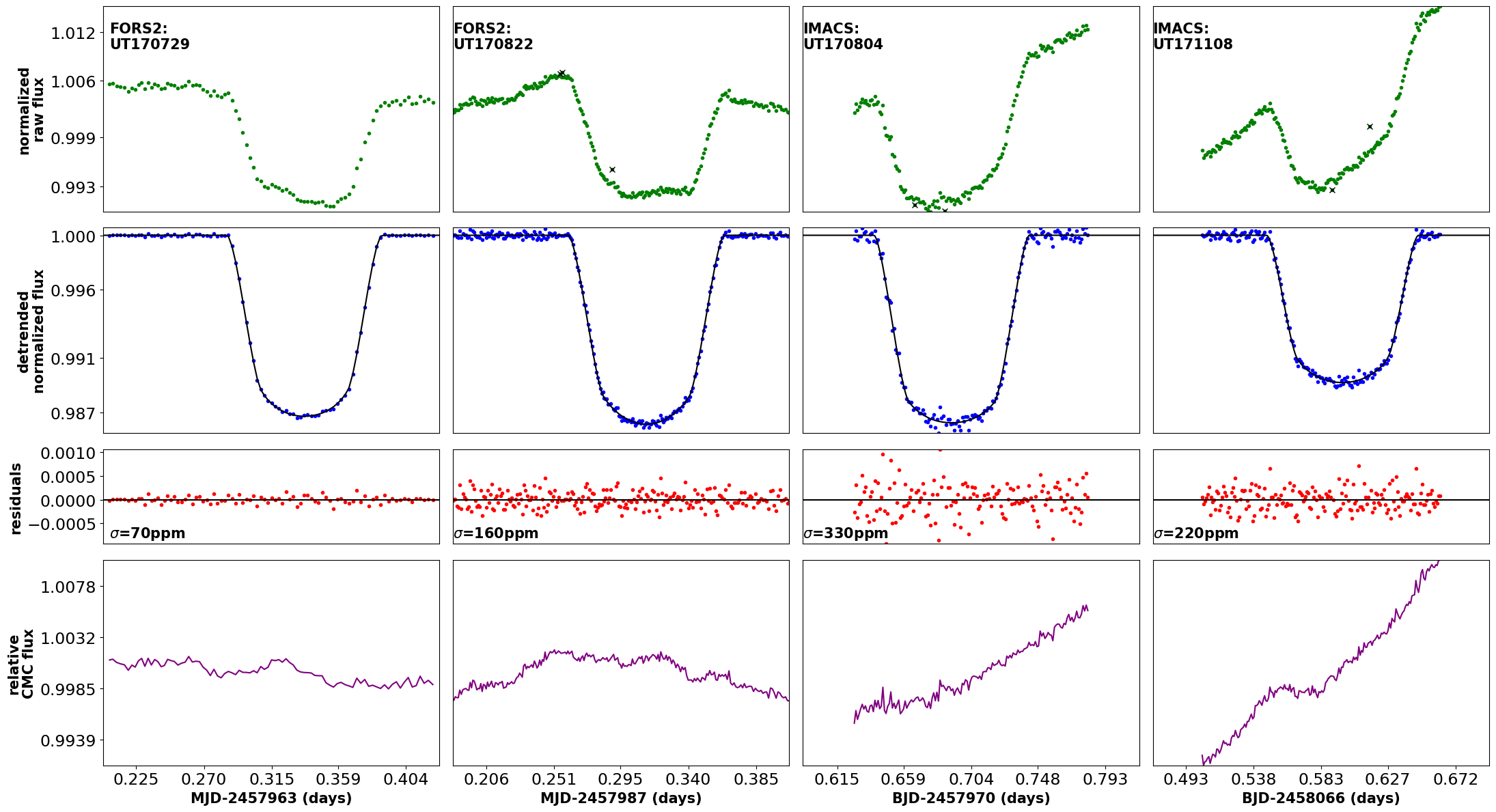}
    \caption{\textbf{Top:} White light curves (LC) composed from dividing the target by the sum of comparison stars, i.e., the raw LC. The first two columns are for the VLT/FORS2 transits, showing the exact data of \cite{Nikolov:2018}. The last two columns are new Magellan/IMACS observations directly from ACCESS’s custom pipeline. The data points marked with an x were identified as outliers and are not included in the rest of the analysis. \textbf{Middle:} The detrended white LC (blue points) produced utilizing our PCA+GP routine, the corresponding best-fit transit LC (solid black line) with orbital parameters shown in Table \ref{tab:wlc_GP}, and the residuals (red points) produce by subtracting the detrended data from the model. The value of $\sigma$ given on each residual panel is the standard deviation of the residuals in ppm. \textbf{Bottom:} The common mode correction term, which was produced by dividing the top LC by the best fit transit model (the black LC in the middle panel, see Appendix~\ref{Appx:CMC} for detailed discussions). This term was used to correct for common systematics in the binned light curves.}
    \label{fig:WLCs} 
\end{figure*}

We adopted a circular orbit for WASP-96b, based on \cite{Hellier:2014}, and used a quadratic LD law as in \cite{Nikolov:2018}. We also held $P$ (3.4252602\,days), $a/R_s$ (8.84), and $i$ (1.486\,radians) fixed to the values used by \cite{Yip:2021} in order to later combine the optical and HST spectra. The quadratic LD parameters were set to be uniform from 0 to 1, and $R_p/R_s$ had a normal prior with mean value of 0.115 and a standard deviation of 0.02. Here priors were based off of the $R_p/R_s$ values obtained by \cite{Nikolov:2018}. For $t_0$ we used normal priors with uncertainties of 0.005 days (7.2 minutes) and mean values of $t_0\,-\,2,450,000$ =7963.33672[MJD], 7970.69261[BJD], 7987.31195[MJD], and 8066.59828[BJD] days for the transits in chronological order \footnote{small priors were used for the FORS2 data because they were based off of the $t_0$ values found by \cite{Nikolov:2018}. In order to use consistent priors for all transits, we fit the IMACS data first with wider priors (0.5 days, 12 hours), then with the 0.005 day priors after using the initial fit to obtain estimates on $t_0$.}. The best-fit white light curves and their corresponding orbital parameters are shown in Figure~\ref{fig:WLCs} and Table~\ref{tab:wlc_GP}. We note that the difference in depth for the last transit (UT171108) is because the orbital parameters were held fixed. When allowed to be free, more consistent depths are obtained, and the other system parameter still agree within their uncertainty ranges. However, the absolute depth is not as relevant, as will be discussed in Section~\ref{sec:Synth}. 


\begin{deluxetable*}{CcRRRR}[htb]
    \caption{Fitted white light curve values. The period (P=3.4252602), semi-major axis (relative to stellar radius, $a/R_s$=8.84), and inclination (i=85.14) were all held fixed to parameters used by \cite{Yip:2021}. The mid-transit times are in terms of MJD for the FORS2 observations (UT170729 and UT170822) and BJD for the IMACS observations (UT170804 and UT171108). Note: the difference in depth for the last transit is because the orbital parameters were held fixed, when allowed to be free more consistent depths are obtained.}
    \label{tab:wlc_GP}
    \tablehead{\colhead{parameter} & \colhead{definition} & \colhead{UT170729} & \colhead{UT170804} &
    \colhead{UT170822} & \colhead{UT171108}}
    \startdata 
    R_p/R_s                & planet radius/star radius & 0.1148^{+0.0029}_{-0.0033}    & 0.1168^{+0.0046}_{-0.0048}       & 0.1188^{+0.0015}_{-0.0018}    & 0.1018^{+0.0049}_{-0.0051}   \\
    t_0-2.45e6      & mid-transit (JD) & 7963.33662^{+0.00050}_{-0.00048}       & 7970.69093^{+0.00049}_{-0.00046}       & 7987.31216$\pm$0.00024 & 8066.59703^{+0.00069}_{-0.00071} \\
    q_1              & LD coeff 1       & 0.37^{+0.10}_{-0.12}                & 0.23^{+0.11}_{-0.09}          & 0.31^{+0.06}_{-0.05}    & 0.32^{+0.18}_{-0.15}    \\
    q_2              & LD coeff 2       & 0.29^{+0.24}_{-0.15}                & 0.33^{+0.34}_{-0.23}          & 0.63^{+0.18}_{-0.22}    & 0.31^{+0.32}_{-0.21}    
    \enddata 
\end{deluxetable*}

\subsection{Binned light curve fitting} \label{sec:Binned}
The next step on the path to produce the transmission spectrum is fitting the binned light curves to obtain values of $R_p/R_s$ as a function of wavelength. 
With the uncorrected binned-light curves in hand, we first used the same 3$\sigma$ clipping process discussed in Section~\ref{sec:LC} for each bin. Then the bins were detrended with a combination of common mode correction and polynomial fitting (CMC+Poly), where the CMC term (shown in Figure \ref{fig:WLCs}) was obtained from a PCA+GP fit of the white-light curve (detail in Appendix~\ref{Appx:CMC}). We also fixed all parameters, aside from $q_1$, $q_2$, and $R_p/R_s$, to values obtained by the PCA+GP white-light fit obtained in Section~\ref{sec:LC} \footnote{$P$, $a/R_s$, and $i$ were already fixed to the values used by \cite{Yip:2021}}. The bins used and their corresponding radii, along with plots of the raw (target/comparisons) and detrended light curves can all be found in Appendix~\ref{Appx:Lgt_curves}. We determined the CMC+Poly routine was best to detrend the binned light curves for this dataset, by testing its effectiveness against synthetic data, which is discussed in the following section (Sec. \ref{sec:Synth}).

\subsection{Comparing the performance of GP and CMC+Poly with Synthetic data} \label{sec:Synth}
A number of techniques have been used to detrend the light curves of exoplanet transits (i.e. PCA e.g. \citealt{Jordan:2013}, GP e.g. \citealt{Gibson:2012}, CMC e.g. \citealt{Gibson:2013}), but there are few instances in the literature that compare the effectiveness of one method over another \citep[][but they still do not compare against synthetic data]{Gibson:2013, Panwar2022}. Furthermore, using two separate analysis techniques could produce widely different results (e.g. \citealt{Sing:2015} and \citealt{Gibson2017WASP31}). Thus, it is important to ensure the method utilized for a particular dataset yields as accurate results as possible and provides consistent uncertainty measurements. To instill confidence in our analysis procedures of the binned light curves, we synthesize transmission spectra to compare the precisions and accuracies of obtained transit depths to synthesized values. We test the synthetic data against two common transit reduction techniques: 1) a Gaussian Processes (GP) routine, and 2) Common-Mode-Correction followed by polynomial correction (CMC+Poly). 

To test both methods, we created 500 synthetic light curves 
similar to the VLT/FORS2 observations. A detailed description of how the synthetic data were created can be found in Appendix~\ref{Appx:Synethic_Spectra}, but in general we simulated 50 flat (constant $R_p/R_s$) transmission spectra out of 10 bins each. All of the 10 bins had approximately the same shot noise levels ($\sim$\,400\,ppm), which was assigned to produce a white-light curve noise level of about 125\,ppm, consistent to the VLT/FORS2 white-light photon noise limits ($\sim$63 and 162 for transit UT170729 and UT170822, respectively). All bins in a given spectrum had the same overarching systematic generated by a random draw in a GP distribution, where the GP was constructed to be correlated to a few of the UT170729 observation's auxiliary parameters. Then each individual bin had additional systematics generated from up to a second-order polynomial fit using other auxiliary parameters with random polynomial coefficients. Each bin also had their own quadratic limb darkening coefficients. Because the VLT/FORS2 observations only had one comparison star, we also only created one comparison star in our synthetic data. As outlined in Appendix~\ref{appx:PCA}, when there is only one comparison star, PCA cannot be done and the PCA+GP routine becomes a GP routine with the comparison star used as a linear regressor. This is why in the synthetic analysis we refer to this method as a GP routine, but when using the same routine with the Magellan/IMACS data, which has two comparison stars, we refer to it as PCA+GP. Images of each step in the synthetic data production process are shown in Appendix~\ref{Appx:Synethic_Spectra}.

After the synthetic spectra were produced, we fit all 50 white light curves using the GP method in the same way described in Appendix~\ref{appx:PCA}. We used the results of each white light curve as priors for the analysis of the binned light curves (see Appendix~\ref{Appx:Synethic_Spectra}). With those white-light parameters we produced the transmission spectra following first the GP process to detrend the bins, and then with the CMC+Poly process, which are both discussed in Appendix~\ref{Appx:Detrending_Routines}. Both methods used to detrend the binned-light curves utilized the parameters determined from the GP detrended white-light data, and the CMC+Poly method used that white-light curve model to produce the CMC term. This allowed us to compare the effectiveness and accuracy of both methods given that each true depth is known. Additionally, because all 50 spectra were flat with the same inputted depth, we could collectively compare the results from every reduced bin. 

\begin{figure*}[htb]
    \centering
    \includegraphics[width=.99\textwidth]{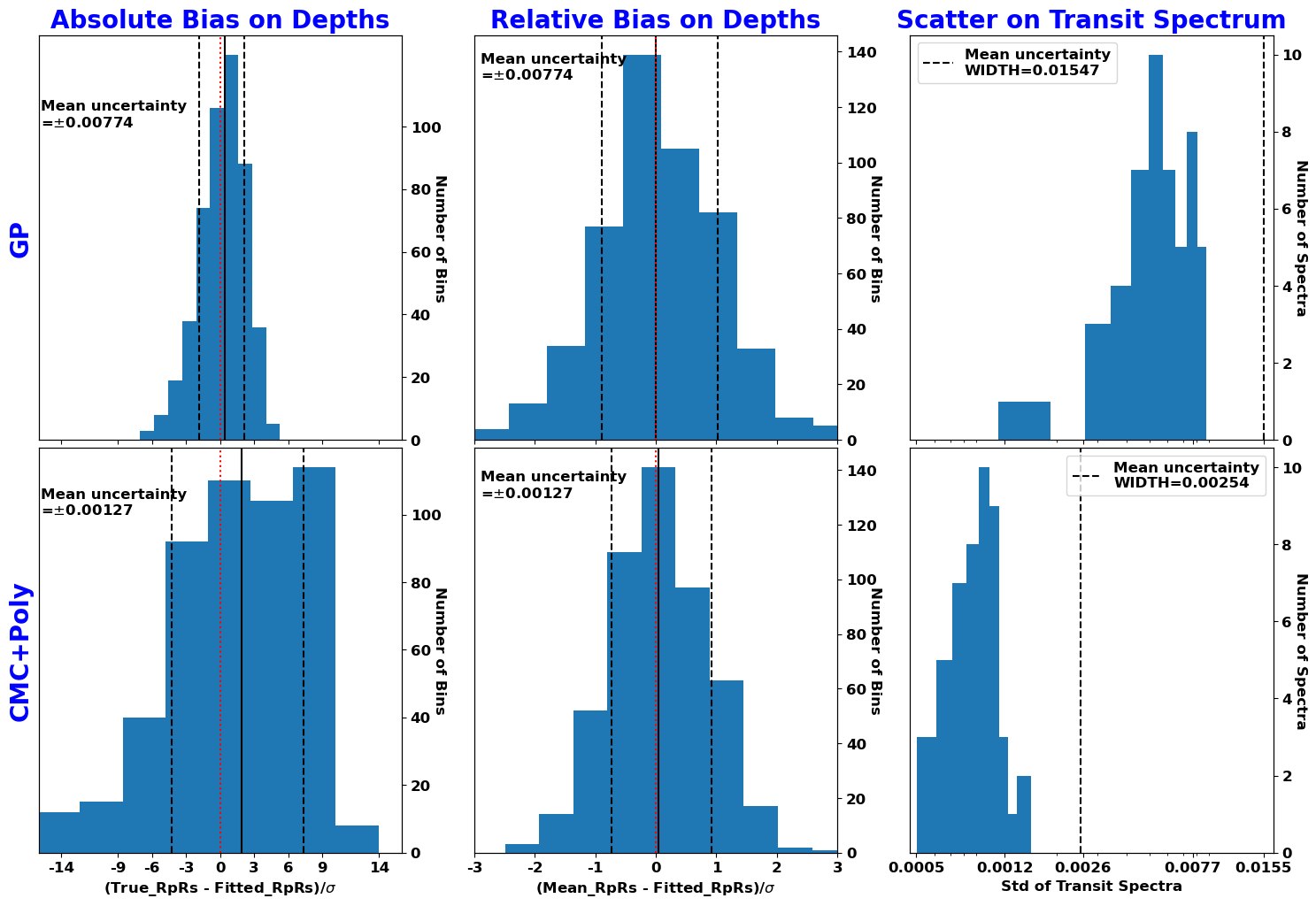}
    \caption{Histograms used to outline the biases and precisions of both the GP (top row) and the CMC+GP (bottom row) routines. \textbf{Left column:} We fit each of the 500 total synthetic binned light curves using the GP and CMC+Poly routines; then take the difference of each fitted binned depth from the true depth ($R_p/R_s$=0.1157) used to produce all synthetic data. That difference was divided by the uncertainty of the fitted depth, in order to obtain the bias from the input depth, relative to the uncertainties. The distribution of the relative differences are plotted. \textbf{Middle column:} This is similar to the left column, but instead is the difference of each fitted binned depth from the mean depth of all 10 bins in its corresponding transmission spectra. This provides a relative bias on obtained transit depths. In this case `relative' means relative to the transmission spectrum's mean depth. For both the left and middle columns, the black dashed line represents the average 16 and 84 percentiles of the histograms, the middle solid black line is the 50 percentile, and the red dotted line is 0. This corresponds to the true value of the depth, thus when the dashed black line and dotted red line do not overlap the routine has an inherent bias on that obtained depth. \textbf{Right column:} This is the standard deviation of each of the 50 transmission spectra. It shows the measured scatter of the transmission spectra, where the true scatter is 0, because the spectra were made to be flat. Here the black dashed line is the mean width of all bins uncertainties. Note that though both routines produce scatter well under their respective mean uncertainty widths, because the mean uncertainty width is over 6 times larger for the GP routine, the CMC+Poly routine is much more precise.}
    \label{fig:SynthHist} 
\end{figure*}

To first understand the accuracies of both methods, we plot a histogram of the difference of each bin's obtained depth relative to the true depth ($R_p/R_s=0.1157$), shown in the first column of Figure~\ref{fig:SynthHist}. From this we see that though the true depth is on average obtained using the GP method (first row), neither method consistently reproduced the true depths, where only 46.32$\pm$2.82\%\footnote{uncertainties obtained through bootstrapping.} of the bins detrended using the GP method obtained a depth 1$\sigma$ from its average uncertainties levels. The mean uncertainty of obtained $R_p/R_s$ with the GP method was 0.00774. This is significantly worse using the CMC+Poly method, which only has 12.54$\pm$2.00\% of the bins within its 1$\sigma$ level, where the mean uncertainty with the CMC+Poly method was 0.00127. This suggests that if one were to fit any given transmission spectrum with the GP routine there is only a 46.32$\pm$2.82\% likelihood of those obtained depths being consistent with the physical depths. This is only worse using the CMC+Poly method.  
One likely strong contribution for this is that the bins are already biased by the white-light fit, given that the bins reduction is dependent on the white-light's. This is especially the case for the CMC method, because the common mode term, which drives the binned light curves correction model, can only be determined from the white-light fit. To support this we compared the obtained binned depths relative to their corresponding white-light fits' depths and find much more consistency of the depths with 41.79$\pm$2.82\% and 57.28$\pm$2.88\% within the 1$\sigma$ average uncertainty for the CMC+Poly and GP methods, respectively. 

Still, neither method can consistently re-obtain the true white-light depths. However, when comparing each bin to a corresponding mean depth determined by averaging all 10 bins for a given transmission spectrum (column two of Figure~\ref{fig:SynthHist}), we find 78.21$\pm$2.42\% and 70.72$\pm$2.67\% are within 1$\sigma$ for the CMC+Poly and GP methods, respectively. This implies that though the absolute depth is not consistently obtained, a relative depth is for both detrending methods. If that is the case, it provides justification for the required offsets often needed when combining transmission spectra from different nights and/or different instruments \citep[e.g.][]{McGruder:2020,Weaver:2021,Yip:2021}. These results would also explain why the inconsistency in white-light depth found for transit UT171108 (Table \ref{tab:wlc_GP} and Fig. \ref{fig:WLCs}) does not imply that the transmission spectrum of that night is incorrect. The likely scenario for that dataset is the GP routine misestimated the white-light depth (likely because of the fixed parameters), but this is common for the majority of the synthetic data as well. However, the relative binned depths can still be preserved using this white-light depth for the CMC correction, because the difference in depths from one bin to another is still maintained. 

To further highlight that the structure of the spectrum is preserved we plot the standard deviation of all bins in a given spectrum. Since, each synthesized spectrum is flat, each bin (of a given spectrum) should not significantly vary from one another. This is exactly what we find, where every spectrum has a bin standard deviation significantly lower than that spectrum's average $R_p/R_s$ uncertainty width (see column three of Figure~\ref{fig:SynthHist}). Furthermore, the third column shows that the standard deviation is higher for the GP method, implying that the CMC+Poly method is more consistent. Additionally, because the CMC+Poly method inherently produces lower uncertainties (average $R_p/R_s$ uncertainty of
0.00127 for CMC+Poly compared to 0.00774 for GP), for this set of data, the CMC+Poly method is consistently more accurate and precise than the GP method. As such we elect to use the CMC+Poly method to detrend all binned light curves.

It is important to understand that these finding are only for this specific dataset, which is constructed to mimic the particular VLT/FORS2 observations of WASP-96b; i.e. the synthetic data has a relatively low shot noise level ($\sim$400 for bins), the bulk of the systematics are dominated by the white-light systematics (see Appendix~\ref{Appx:Synethic_Spectra}), and there was only one comparison star used. Therefore, this should not be extrapolated to every dataset. For example, we reduced the VLT/FORS2 data with just a CMC correction and found a similar fit compared to using CMC+Poly, outlining how little chromatic systematics persists in the given VLT/FORS2 data. If chromatic systematics are more dominant in a dataset or there are multiple comparison stars, then it could be possible that the data would be better reduced with a method like PCA+GP, which is less heavily dependent on the initial white-light fit. For this reason, we still ran the ACCESS data through both the CMC+Poly and PCA+GP routines as described below to confirm CMC+Poly still performed better for those data. In summary, one should explore the best detrending method for specific data before assigning one.  

\section{Optical Transmission Spectrum from the VLT/FOR2 and ACCESS Data} \label{sec:Tran_Spec}
\begin{figure*}[htb]
    \centering
    \includegraphics[width=.8\textwidth]{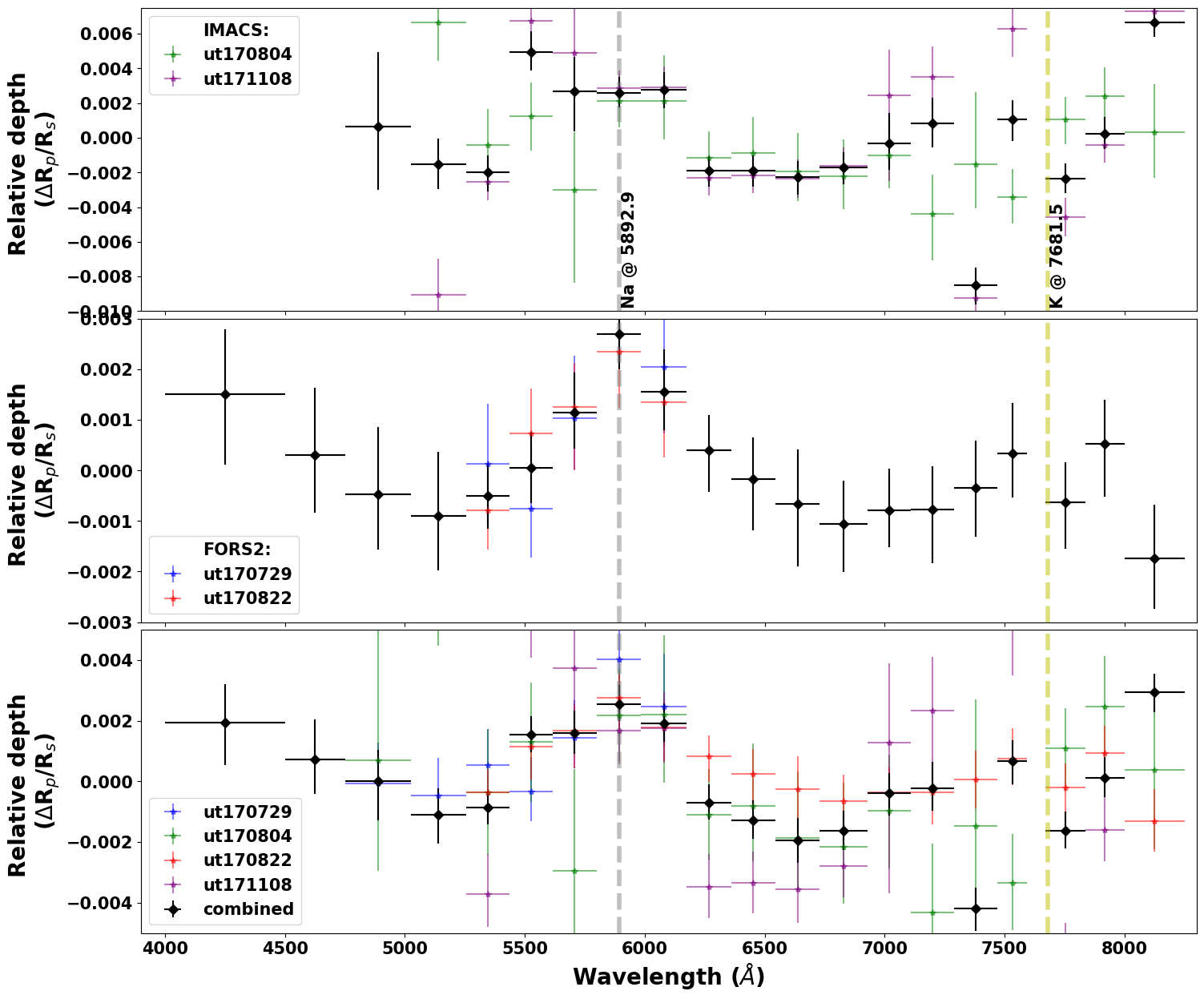}
    \caption{The transmission spectra using only the two Magellan/IMACS transits (top), only the two VLT/FORS2 transits (middle), and combining all four transits (bottom). In all three plots the black points correspond to the final spectrum produced by taking the weighted average of all overlapping bins. When combining each spectrum an offset is applied, so the means of each individual spectra are consistent, before the bins are averaged together. The center of the ${\rm Na~I}$ and ${\rm K~I}$ absorption are plotted as dotted gray and yellow lines, respectively.}
    \label{fig:OpticalSpec} 
\end{figure*}

We produced the transmission spectrum by plotting the $R_p/R_s$ found from each detrended binned-light curve against that wavelength interval (bin). Three optical transmission spectra were created. One from combining the two FORS2 transits where we weighted averaged all overlapping bins, another from combining the two IMACS transits, and the third from combining all four transits (global optical spectrum). These are plotted in Figure~\ref{fig:OpticalSpec}. As justified in Section~\ref{sec:Synth}, we fit an offset when combining each of the spectra. For a given combined spectrum a white-light depth was determined for each individual spectrum using a weighted mean of only overlapping bins. The weighted mean of these white-light depths was then used as the central depth for which each individual spectra was offset to. The average precision of each bin for the combined IMACS, FORS2, and global optical spectra are 0.00129, 0.00094, and 0.00076 $R_p/R_s$, respectively.

The IMACS data has more scatter and lower precision than the FORS2 data (see Fig.~\ref{fig:OpticalSpec}), even though both use two transits. One explanations to this is that the size of Magellan Baade (6.5\,m) is smaller than VLT (8.2\,m), meaning less collecting area and more shot noise. Additionally, the difference in comparisons used could cause the IMACS data to have more systematics. The comparison \cite{Nikolov:2018} used (D=0.1) was more similar to the target than either of ours (D=0.24 and 0.14; see Table~\ref{tab:comp_stars}). The importance of the comparisons is emphasized by \cite{Ahrer:2022}, where they were able to detect a strong sodium signal and a super-Rayleigh scattering slope in the atmosphere of WASP 94Ab with just one comparison and one transit, partially attributed to the comparison being nearly identical in spectral type and location in the sky. Furthermore, for both IMACS transits the baselines, which is needed to properly correct for systematics, were relatively short (seen in Fig~\ref{fig:WLCs}), where ideally the baseline should be the same length as the transit duration.

Likely the largest cause of deviation between the two datasets is that the IMACS data has more chromatic systematics. This can be seen when looking at the binned light curves in Appendix~\ref{Appx:Lgt_curves}. Why the chromatic systematics are stronger for the IMACS observations is unclear. It might be due to chromatic differences in comparisons, or some other issue. 
Regardless, the CMC method relies on the assumption that the bulk of the systematics found in the white-light curve can be applied to each of the binned light curves. If each bin's systematics are more unique the initial CMC is not as effective, and might even introduce more spurious signals that the polynomial fit has to correct for. Concern over this led us to analyze the IMACS data with both CMC+Poly and PCA+GP routines. In doing so we found that the two transmission spectra were consistent to one another, with each bin deviating from one another by 0.562$\sigma$ on average. However, the uncertainties of the PCA+GP method were nearly 5 times larger than the CMC+Poly's. Given both methods produced similar features, the accuracy of the CMC+Poly method is supported by Section~\ref{sec:Synth}, and the desire to use the same routine for all datasets we choose to continue using the CMC+Poly routine for the IMACS data.

\section{Stellar Activity} \label{sec:Activity}
Activity in the host star can introduce signals into the planetary transmission spectrum \citep{Pont:2008,Berta:2011,Oshagh:2014, Rackham:2017,Alam:2018,Wakeford:2019a,Rackham2022}. Therefore, in order to prevent misinterpretation of WASP-96b’s transmission spectrum, we parameterize the host star’s level of activity. The three proxies for stellar activity we explore are 1)\,rotational period \citep{Pallavicini:1981}, 2)\,Ca\,\textsc{ii} lines
\citep{Vaughan:1980, Middelkoop:1981, Noyes:1984}, and 3)\,photometric modulation \citep[e.g.,][]{Kipping:2012, Weaver:2020}.

\subsection{Rotational Period} \label{sec:Rotational_Period}
We infer the stellar rotational period by combining the radius of the star with the projected stellar rotational velocity, $\vsini$, as well as fitting a periodic signal to the photometric monitoring data. The $\vsini$ for WASP-96 was determined by \cite{Hellier:2014}  using the Euler/CORALIE spectrograph. It was not well constrained but was found to be 1.5$\pm$1.3\,km\,s\textsuperscript{-1}. This provides a 3-$\sigma$ upper limit on $\vsini$ of 5.4\,km\,s\textsuperscript{-1}. Combining this with the stellar radius of 1.05$\pm$0.05\,R\textsubscript{$\odot$} \citep{Hellier:2014} we estimate a 3-$\sigma$ lower limit on rotation period of about 9.8 days. Thus, we scanned the photometric data for periodic peaks within a range of 9.8 to 300 days using Lomb-Scargle periodogram analysis \citep{Lomb1976}. With the binned combined TESS data the highest periodic peaks were 35.9, 37.7, and 31.2, and with the combined ASAS-SN data they were 28.3, 28.8, and 11.2. However, for all peaks the False Alarm Probability (FAP) were greater than 10\textsuperscript{-1}, thus no significant peaks were found with the periodogram analysis. This is likely because the photometric modulation is in general very small (see Sec.~\ref{subsec:Photo_Monitor} or Sec.~\ref{sec:PhotMod}). 

\begin{figure*}[htb]
    \centering
    \includegraphics[width=.8\textwidth]{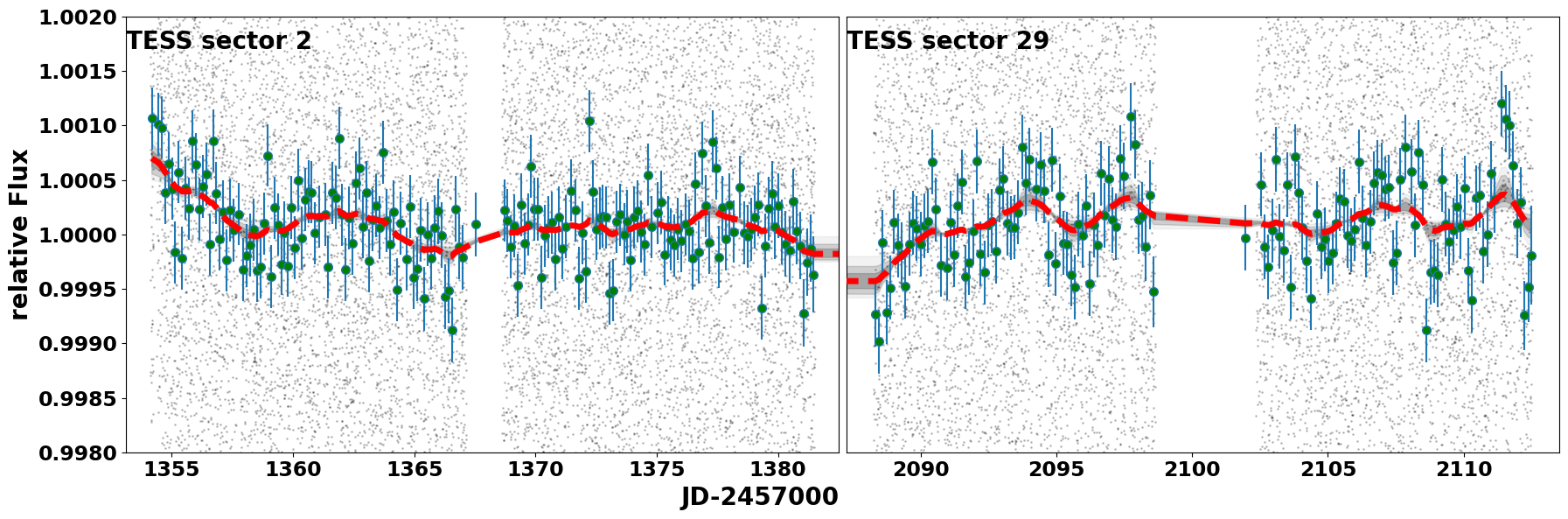}
    \includegraphics[width=.8\textwidth]{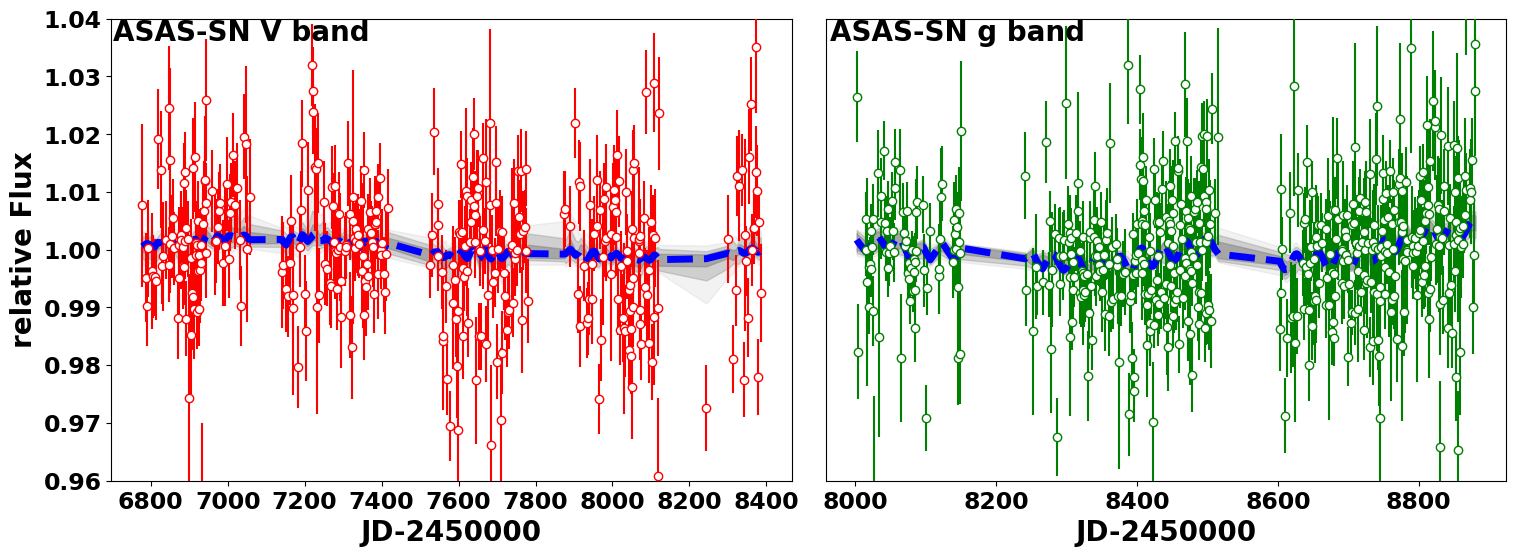}
    \caption{\textbf{Photometeric monitoring of WASP-96:} The top row is TESS monitoring. There the red dashed lines are the \texttt{Juliet} semi-periodic best fit. The grey dots are the original 30-minute cadence TESS observations and the green dots are the data at 100 ($\sim$3.34 hours) binning, with their associated error bars in blue. The right figure is sector 29 data and the left is from sector 2, where both sectors were combined to act as one dataset for the \texttt{Juliet} fit (the binned data was used for the fitting). The bottom row is ASAS-SN monitoring, where the blue dashed lines are the \texttt{Juliet} semi-periodic best fit. On the right, the green hollow circles with associated error bars correspond to the g band data. On the left, the red hollow circles with associated error bars correspond to the V band data. The V and g band data were used as separated datasets in the \texttt{Juliet} fit. For all \texttt{Juliet} fits (blue and red dotted lines), there is a gradient of shaded gray regions representing the 1,2,\& 3 sigma confidence intervals. For the TESS data, the confidence intervals are about the size of the dotted line.}
    \label{fig:PhotMon} 
\end{figure*}

We then jointly fit the ASAS-SN and binned TESS data using the \texttt{Juliet} package \citep{Espinoza2019_juliet} with a semi-periodic kernel. In this joint fit, only the period and timescale terms (see equation 9 of \cite{Espinoza2019_juliet}) are set common for all photometric campaigns. All other parameters were specific to the combined TESS (sector 2 and 29), the V band ASAS-SN, or the g band ASAS-SN data. When doing this we assume the periodicicity is due to stellar inhomogeneities \footnote{Even with a relatively quiet star like WASP-96, we are assuming that smaller spots/faculae can be used to measure modulation and rotational period.} coming in and out of view as the star rotates, which is often done \citep[e.g.,][]{Hirano_2012, Sing:2015, Newton:2018}. We use wide uniform priors on the period from 9.8 to 50 days, which is consistent with what the periodograms and $\vsini$ weakly suggest. The resulting period was found to be 31.3$^{+0.3}_{-3.4}$ days. Given the observed correlation of rotational period to activity levels \citep[e.g.][]{Pallavicini:1981,Reiners:2014}, this also implies that WASP-96 is a relatively quiet star. Figure~\ref{fig:PhotMon} shows the photometric monitoring campaigns along with the \texttt{Juliet} best fits.

\subsection{Ca\,\textsc{ii} H \& K lines} \label{sec:R_hk} 
\begin{figure*}[htb]
    \centering
    \includegraphics[width=.8\textwidth]{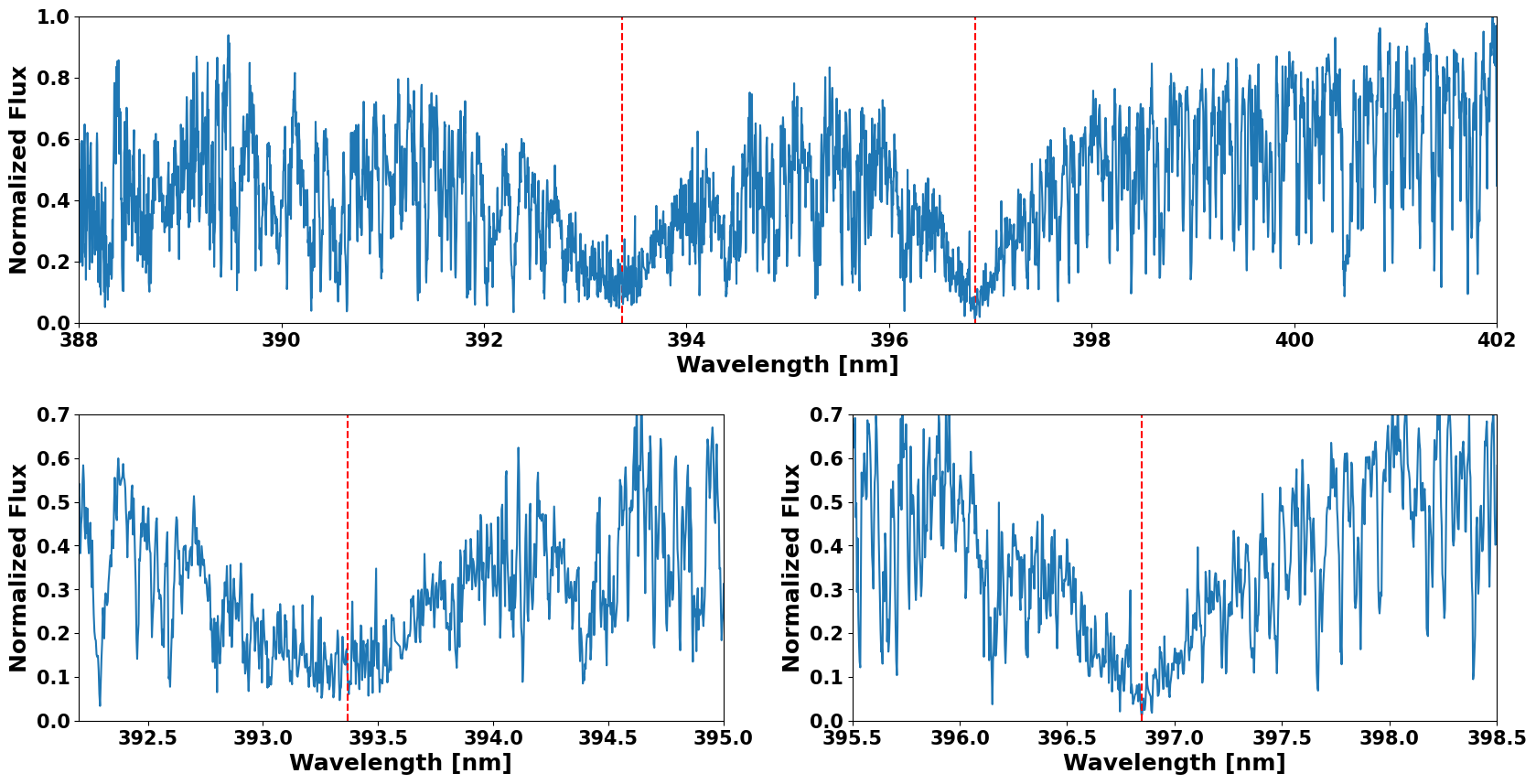}
    \caption{High-resolution FEROS spectrum of WASP-96: Both Ca\,\textsc{ii} H \& K lines (top), where the dashed red lines correspond to the K line core centered at 393.366\,nm and the H line core centred at 396.847\,nm. The left bottom panel is zoomed in on the K line and the bottom right is zoomed in on the H line, with the central cores highlighted with red dotted lines in both. There is no emission seen in either of the cores.}
    \label{fig:CaHnK} 
\end{figure*}

The Ca\,\textsc{ii} H \& K lines were measured using two R\,=\,48000 spectra collected with the MPG 2.2-m/FEROS spectrograph on 19 December 2016. Each spectrum has an average SNR of 27. The reduced data was acquired from \href{http://archive.eso.org/wdb/wdb/adp/phase3_main/form}{\texttt{ESO’s online archive}}. Figure \ref{fig:CaHnK} shows the Ca\,\textsc{ii} H \& K lines and we see no emission in the core of either of the lines, implying that WASP-96 is a relatively inactive star.

\subsection{Photometric Modulation}\label{sec:PhotMod}
The analysis in the above two subsections suggest that WASP-96 is a quiet G-type star. However, we use the amplitude of the photometric modulation of the star to quantify its level of activity in terms of spot covering fraction and spot temperature, which are the parameters needed for the retrievals (see Sec.~\ref{sec:Retrievals}). This is done by using equation 2 and Table 2 of \citet{Rackham:2019} and twice the TESS MAD of 0.0058 (see Sec. \ref{subsec:Photo_Monitor}), assuming that the MAD is approximately the amplitude of sinusoidal variation. We use the TESS data because the variability in the ASAS-SN data seems to be driven by more than just the star (likely low precision), which is apparent when comparing the MAD of the TESS and ASAS-SN monitoring campaigns of the same target. Following that procedure, we estimate a spot covering fraction of WASP-96 1.35$\pm$0.97\% assuming only spots are present and 1.40$\pm$1.09\% assuming spots and faculae are both present.

To capture these limitations in the retrieval analysis, we constrained the heterogeneity covering fraction, $f$, to have normal priors with a mean of 0.014 and standard deviation of 0.009. For a G8 star like WASP-96, we should expect the heterogeneity temperature contrast, $\Delta$\,T, to be roughly 1600\,K \citep[see ][]{Berdyugina2005,Rackham:2019}. We used wider uniform priors from -3000 to 3000\,K on $\Delta$\,T, see Appendix~\ref{Appx:RetrievalModelingPriors}. 

\section{Retrieval Analysis} \label{sec:Retrievals}
We combined each of the three optical spectra discussed in Section \ref{sec:Tran_Spec} (IMACS only, FORS2 only, and global spectra, see Fig. \ref{fig:OpticalSpec}) with the HST/WFC3 data. We ran each of these three optical to near-IR spectra against two retrievals: \href{https://platon.readthedocs.io/en/latest/}{\texttt{PLATON}} \citep[v3,][]{Zhang_2019PLATON} and \texttt{Exoretrievals} \citep[introduced in][]{Espinoza2019}. We used both \texttt{PLATON} and \texttt{Exoretrievals}, because their differing approaches of modeling exo-atmospheres provides different insights about the observed transmission spectra. The key differences between \texttt{PLATON} and \texttt{Exoretrievals} are: (1) \texttt{PLATON} includes collision induced absorption, where \texttt{Exoretrievals} does not, (2) \texttt{Exoretrievals} models the transmission spectrum using a semi-analytical formalism with an isothermal, isobaric, atmosphere and non-equilibrium chemistry, but \texttt{PLATON} uses an isothermal atmosphere and imposes equilibrium chemistry\footnote{Though equilibrium chemistry, might be an inaccurate assumption for observed transmission spectral features \citep[e.g.][]{Venot:2012, Komacek2019, Roudier:2021}, using it still provides useful insights that could not be obtained without this assumption.}, and (3) the bulk of the line list used by \texttt{PLATON} are from HITRAN \citep{Rothman2013}, whereas the majority of what \texttt{Exoretrievals} uses is from HITEMP \citep{Rothman2010} and ExoMol \citep{Yurchenko2013, Tennyson2016}. Additionally, testing a transmission spectrum against multiple retrievals provides a robustness against assumptions that are unique to each retrieval \citep[e.g.][]{McGruder:2020,Kirk:2021}. 

Both retrievals used nested sampling to explore their parameter space (\texttt{dynesty}, \citeauthor{Speagle_2020Dynesty} \citeyear{Speagle_2020Dynesty}, for \texttt{PLATON} and \texttt{PyMultiNest}, \citeauthor{2014BuchnerPyMultiNest} \citeyear{2014BuchnerPyMultiNest}, for \texttt{Exoretrievals}), as such we used the differences in log Bayesian evidences, $\Delta \ln Z$, to test which specific model was favored over another.  Following the same prescription as \citeauthor{McGruder:2020} \citeyear{McGruder:2020} \citep[from][]{2008Trotta, 2013Benneke}, we interpreted the $\Delta \ln Z$ values in a frequentist significance as: |$\Delta \ln Z$| of 0 to 2.5 is inconclusive with < 2.7$\sigma$ support for the higher evidence model, |$\Delta \ln Z$| of 2.5 to 5 corresponds to a moderately significant detection of 2.7$\sigma$ to 3.6$\sigma$, and |$\Delta \ln Z$| $\geq$ 5 corresponds to strong support for one model over the other.

With \texttt{Exoretrievals}, we tested the spectra against having either water, potassium, sodium, water and sodium, and all three species in the transmission spectra. Along with these molecular and atomic species, we tested if the spectra warranted high altitude scattering agents, stellar activity, and a combination of scatters and activity. Lastly, a model with no features (flat spectrum) and only activity features was tested. In total, we tested 22 different combinations of models. For all models, aside from the flat spectrum, a reference radius (parameterized with $f$) and reference pressure (P$_0$) were fit. These are the pressure and corresponding radius where the atmosphere is optically thick in all wavelengths. 


With \texttt{PLATON} we tested a model with scattering agents, stellar activity, models with both scattering agents and stellar activity, and a model without either (clear). All models fit for a reference radius (R$_0$) corresponding to the radius of the planet at an arbitrary reference pressure, which was set to 1\,bar. The models also fit for a pressure (P$_{cloud}$) where the atmosphere is optically thick.

The priors did not change for each spectrum we ran against the retrievals and we set the priors between \texttt{PLATON} and \texttt{Exoretrievals} as consistent to one another as possible. For all three spectra and both retrieval models we fit an offset between the optical and near-IR data\footnote{\texttt{PLATON} v3 could not fit an offset between different dataset, as such, we modified the code to do so.}, which is justified given that the detrending method used for the optical data does not preserve absolute depth (see Sec. \ref{sec:Synth}). 

\subsection{IMACS \& FORS2 Transmission Spectra} \label{sec:ImacsnFors2_TransSpec}
Tables of $\Delta \ln Z$ for both datasets against both retrievals are shown in Tables \ref{tab:IMACS_LnZs} \& \ref{tab:FORS2_LnZs}. In the tables the $\Delta \ln Z$ values for each model was determined by comparing the clear model, for \texttt{PLATON}, or the clear model and flat spectrum, for \texttt{Exoretrievals}, to the other models. One can also compare any given model from another by examining the difference of $\Delta \ln Z$'s between one another.

Using \texttt{Exoretrievals}, we find from both the IMACS and FORS2 spectra, independently, that the model with the highest $\Delta \ln Z$ is one with water and sodium (no potassium). We find that models with additional stellar activity or scatters do not make a significant difference in the $\Delta \ln Z$, aside for the IMACS dataset that has a significant decrease in $\Delta \ln Z$ when including scatters ($\Delta \ln Z$ decreases by 4, compared to a model with only water and sodium). The feature blue-ward of 500\,nm is likely what the retrievals with scatters and activity are attempting to fit. However, because the feature is not extreme, relative to the uncertainties (especially for the IMACS dataset), there is not sufficient support for the extra complexity of these models. As such we can only claim a tentative detection of a slight blue-ward slope either attributed from the star or a possible Rayleigh scattering slope. For the IMACS data the best fit model that included stellar activity found spot parameters of $\Delta T$=-150$^{+1300}_{-1700}$\,K and $f$=0.0114$^{+0.0085}_{-0.0072}$. For FORS2 those activity parameters were $\Delta T$=-1060$^{+1100}_{-1200}$\,K and $f$=0.0124$^{+0.0083}_{-0.0080}$. The best model which included a haze scattering slope for the IMACS data obtained a haze power law of $\gamma$=-9.3$^{+7.8}_{-3.4}$, and $\gamma$=-9.7$^{+5.1}_{-3.0}$ was obtained for the FORS2 best scatters model. With \texttt{Exoretrievals} $\gamma$ of -4 corresponds to a Rayleigh slope (see Appendix D of \citet{Espinoza2019}), implying that even though the retrieved slopes are unconstrained, they are consistent with Rayleigh scattering. The detections of $H_2O$ and Na were highly significant ($\Delta \ln Z >$ 11) for both datasets. The highest-evidence retrieval model parameters for this data subset and the others can be seen in Table \ref{tab:BestRetrievedPar}

\begin{deluxetable*}{|l|C|C|C|C|C|C|C|C|C|}[h!]
    \caption{$\Delta$ln~Z for various \texttt{Exoretrievals} (left) and \texttt{PLATON} (right) models relative to a clear (and flat for \texttt{Exoretrievals}'s case) spectrum with the subset of data that included only the Magellan/IMACS and HST/WFC3 data. The retrievals with water and sodium were heavily supported by \texttt{Exoretrievals}. Models that included scattering were the most supported with \texttt{PLATON}.}
    \label{tab:IMACS_LnZs}
    \tablehead{\multicolumn{6}{|c|}{\bfseries {\large Exoretrievals}} &&\multicolumn{2}{|c|}{\bfseries {\large PLATON}}}
    \startdata 
      \textbf{Model:}  &  \text{flat}  &  $H_2O$  &  $Na$  &  $K$  &  $H_2O +Na$ &  $H_2O + K +Na$  && \textbf{Model:} &\\ \hline
      clear &                0.0 &	6.99 &	-0.39 &	4.07 &	12.78 &	10.94 && \text{clear} & 0.0\\
      scatterers &          	--- &	5.5 &	-0.89 &	3.49 &	8.78 &	8.39 & &\text{scattering} & 4.54\\
      activity  &           0.35 &	5.97 &	-1.23 &	4.04 &	10.65 &	10.1 & &\text{activity } & 0.6\\
      Both  &	--- &	4.91 &	-1.92 &	3.34 &	9.63 &	-4.05 & &\text{Both} & 4.14 \\
    \enddata 
\end{deluxetable*}

Interestingly, with \texttt{PLATON} there is less consistency amongst the two datasets. All \texttt{PLATON} models with the FORS2 dataset were indistinguishable from one another, which is in agreement with what \texttt{Exoretrievals} found for the FORS2 dataset. That is the slope blue-ward of 500\,nm could be explained by either activity or a scattering slope, but neither is required for the data. For the FORS2 dataset the best fit model including activity found $\Delta T$=-1040$^{+1200}_{-1000}$ and $f$=0.0143$^{+0.0083}_{-0.0089}$; the best fit model including scatterers found a scattering slope, $\alpha$, of 5.7$^{+5.3}_{-6.0}$ ($\alpha$=4 is Rayleigh); and the highest evidence model (clear model) obtained a metallicity, $\log_{10}(Z/Z_{\odot})$, of 0.26$^{+0.75}_{-0.78}$\,dex, C/O of 1.11$^{+0.55}_{-0.39}$, $log_{10}(P_0)$ of 0.3$^{+1.7}_{-1.3}$\,bars, and T$_p$ of 987$^{+92}_{-52 }$. Contrarily to \texttt{PLATON}, the IMACS data obtained a significantly higher evidence ($\Delta \ln Z >$ 4) for the models that included scatterers. The retrieved values with IMACS differed from the fit with FORS2 likely because the Na feature was not as prominent in the IMACS data. Thus, the retrieval increases the metallicity ($log_{10}(Z/Z_{\odot})$=0.51$^{+0.53}_{-0.75}$\,dex) and decreases the temperature (T$_p$ of 752$^{+62}_{-94}$\,K) to mute the Na feature. In turn, the best retrieved spectrum is overall different from the FORS2 spectrum. 
Still the retrieved scattering slope of both datasets was consistent with a Rayleigh scattering slope, though not well constrained. 

\begin{deluxetable*}{|l|C|C|C|C|C|C|C|C|C|}[h!]
    \caption{Same as Table \ref{tab:IMACS_LnZs} but with the subset of data that included only the VLT/FORS2 and HST/WFC3 data. The retrievals with water and sodium were heavily supported by \texttt{Exoretrievals}. There was no model that had high enough evidence to favor it over another with \texttt{PLATON}.}
    \label{tab:FORS2_LnZs}
    \tablehead{\multicolumn{7}{|c|}{\bfseries {\large Exoretrievals}} &&\multicolumn{2}{|c|}{\bfseries {\large PLATON}}}
    \startdata 
      \textbf{Model:}  &  \text{flat}  &  $H_2O$  &  $Na$  &  $K$  &  $H_2O +Na$ &  $H_2O + K +Na$  && \textbf{Model:} &\\ \hline
      clear &                0.0 &	4.94 &	0.13 &	5.81 &	11.32 &	10.65 && \text{clear} & 0.0\\
      scatterers &          		--- &	3.54 &	-1.41 &	3.73 &	9.48 &	8.97 & &\text{scattering} & -1.18\\
      activity &           0.06 &	4.13 &	-1.43 &	4.18 &	11.38 &	11.11& &\text{activity} & -0.18\\
      Both & --- &	3.11 &	-2.12 &	3.11 &	9.13 &	-4.07 & &\text{Both} & -1.25 \\
    \enddata 
\end{deluxetable*}

\begin{deluxetable*}{|l|C|C|C|l|C|C|C|}[h!]
    \caption{Parameters obtained by the best-fit retrievals for each data subsets (IMACS with WFC3, FORS2 with WFC3, and combined optical with WFC3). The model which only included water and sodium was the highest-evidence model with \texttt{Exoretrievals} for each data subset (the evidence for the model with and without activity was nearly identical in the FORS2 subset). With \texttt{PLATON}, the model including scattering was favored with the combined dataset and the IMACS one, but for the FORS2 dataset a featureless model was marginally preferred. Here T\textsubscript{p}, P\textsubscript{0}, offset, $\alpha$, Z/Z$_{\odot}$, C/O, and $\log_{10}(H_2O)$, $\log_{10}(Na)$ correspond to planet terminator temperature [K], reference pressure at which the atmosphere is optically thick [bar], offset in transit depth [R\textsubscript{p}/R\textsubscript{s}], scattering slope ($\alpha$ of 4 is Rayleigh), metallicity of the star relative to solar, carbon-to-oxygen abundance ratio, and log mixing ratios of water and sodium, respectively.}
    \label{tab:BestRetrievedPar}
    \tablehead{\multicolumn{4}{|c|}{\bfseries {\large Exoretrievals}} &\multicolumn{4}{|c|}{\bfseries {\large PLATON}}}
    \startdata 
          &  \text{IMACS+WFC3}  &  \text{FORS2+WFC3}  &  \text{combined data} && \text{IMACS+WFC3}  &  \text{FORS2+WFC3}  &  \text{combined data}\\ \hline
     T\textsubscript{p} & 730$^{+180}_{-140}$ & 790$^{+220}_{-160}$ & 830$^{+160}_{-140}$  &  T\textsubscript{p}  & 752$^{+62}_{-94}$ &  987$^{+92}_{-52}$ & 877$\pm40$ \\ \hline
     $\log_{10}(P\textsubscript{0})$  & 0.64$^{+1.63}_{-1.93}$ &  0.4$^{+1.81}_{-2.17}$ & 0.29$^{+1.86}_{-2.02}$  & $\log_{10}(P\textsubscript{0})$   & 1.58$^{+0.84}_{-0.97}$ &  0.3$^{+1.7}_{-1.3}$ & 1.3$^{+1.0}_{-1.1}$ \\ \hline
      offset  & 0.01201$\pm$0.0004 &  0.00238$^{+0.00041}_{-0.00038}$ & 0.00547$^{+0.00031}_{-0.00032}$  &  offset &  0.00967$^{+0.00034}_{-0.00031}$ &  0.00199$^{+0.00044}_{-0.00043}$ & 0.00414$^{+0.00036}_{-0.00035}$ \\ \hline
      $\log_{10}(H_2O)$   & -3.9$^{+1.8}_{-2.0}$ &  -4.3$^{+1.9}_{-2.1}$ & -4.5$\pm2.0$  &  $\log_{10}(Z/Z_{\odot}$)  & 0.51$^{+0.53}_{-0.75}$ &  0.26$^{+0.75}_{-0.78}$ & -0.49$^{+1.00}_{-0.37}$ \\ \hline
      $\log_{10}(Na)$   & -5.5$^{+1.8}_{-1.9}$&  -5.0$^{+2.0}_{-2.1}$ & -5.4$^{+2.0}_{-1.9}$  &  C/O &  0.69$^{+0.72}_{-0.41}$ &  1.11$^{+0.55}_{-0.39}$ & 0.97$^{+0.65}_{-0.50}$ \\ \hline
      & &  &  & $\alpha$ &  3.7$^{+6.3}_{-5.0}$ &  --- & 10.4$^{+2.5}_{-4.5}$ \\
    \enddata 
\end{deluxetable*}
\subsection{Combined Transmission Spectrum} \label{sec:CombinedSpec}
The $\Delta \ln Z$ of all models run using the global data against both retrievals is shown in Table \ref{tab:Global_LnZs}. When running the retrievals against the combined data, we find a similar trend as that for the individual datasets (aside for IMACS with \texttt{PLATON}). That is, a major detection of water and sodium, and tentative signs of a blue-ward slope attributed to stellar activity or a scattering slope. 

\begin{figure*}[htb]
    \centering
    \includegraphics[width=.99\textwidth]{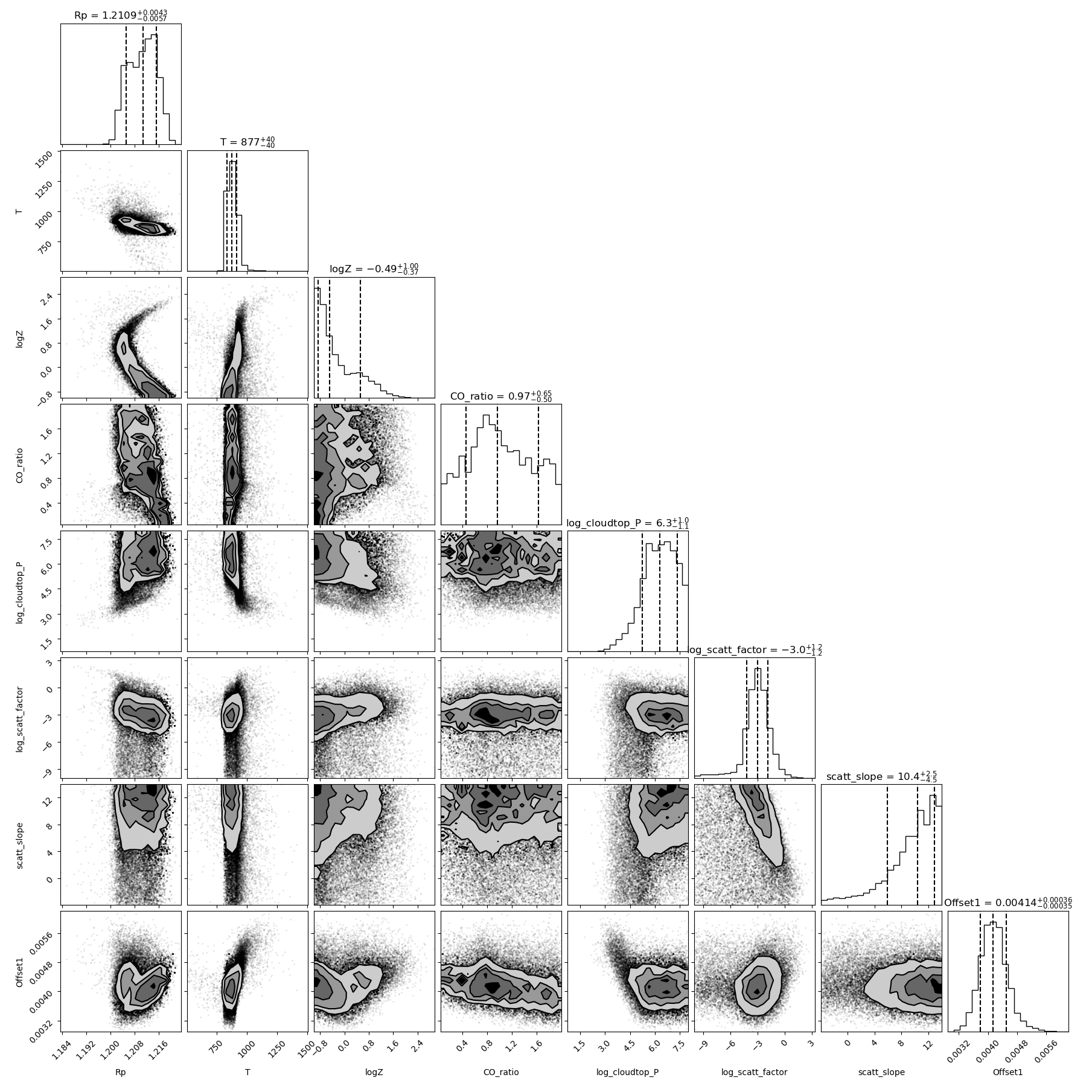}
    \caption{Corner plot of the \texttt{PLATON} best fit retrieval model with,  which is run against the combined Magellan/IMACS, VLT/FORS2, and HST/WFC3 data. Its corresponding transmission spectrum is shown in Figure \ref{fig:PlatonFits} (red model). Vertical dashed lines mark the 16\% and 84\% quantiles.}
    \label{fig:PlatonClearCornerPlot} 
\end{figure*}

\begin{figure*}[htb]
    \centering
    \includegraphics[width=.99\textwidth]{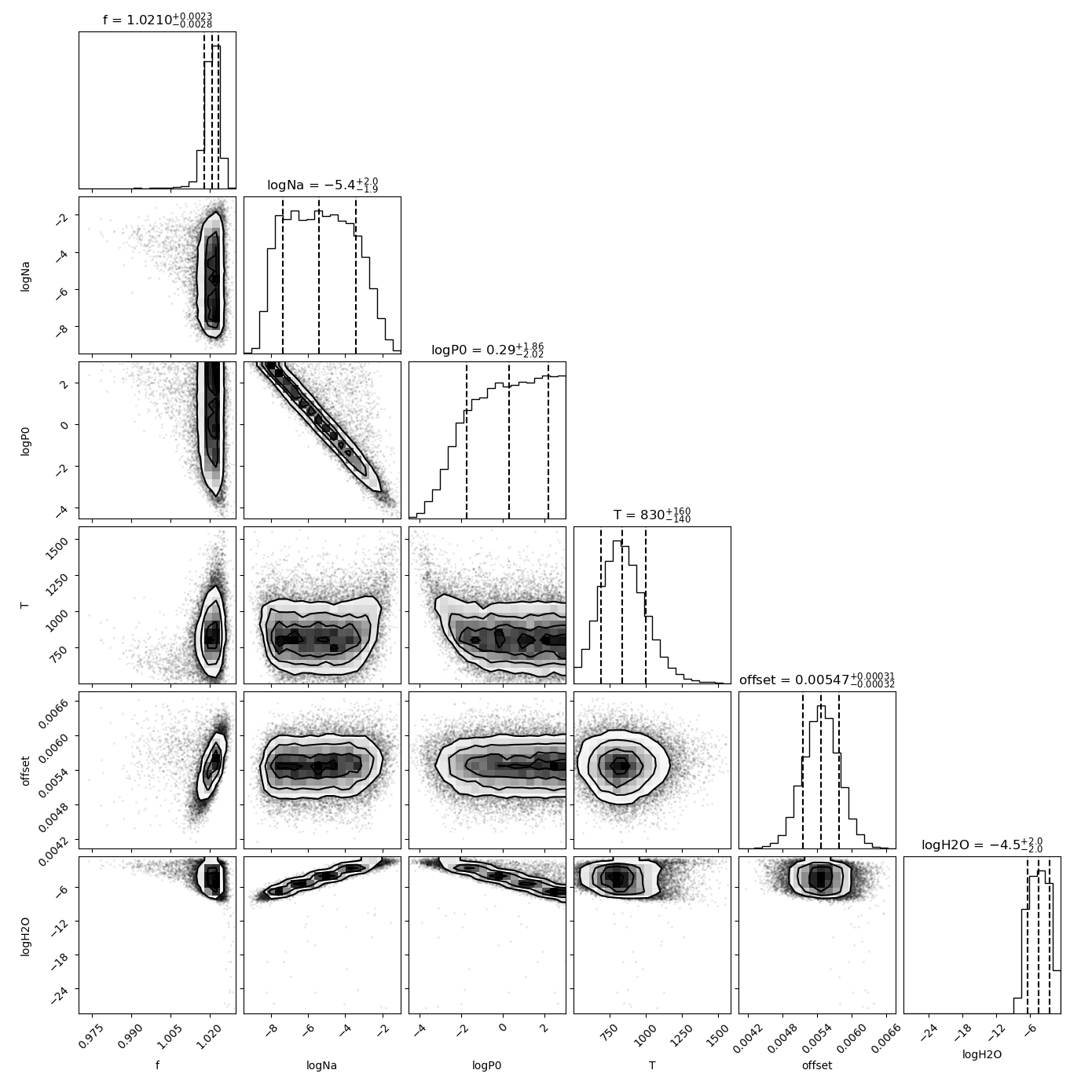}
    \caption{Same as Figure \ref{fig:PlatonClearCornerPlot}, but for \texttt{Exoretrievals}, where its corresponding transmission spectrum is shown in Figure \ref{fig:ExoretrievalFits} (blue colored model).}
    \label{fig:ExoretrievalClearCornerPlot} 
\end{figure*}

\begin{figure*}[htb]
    \centering
    \includegraphics[width=.99\textwidth]{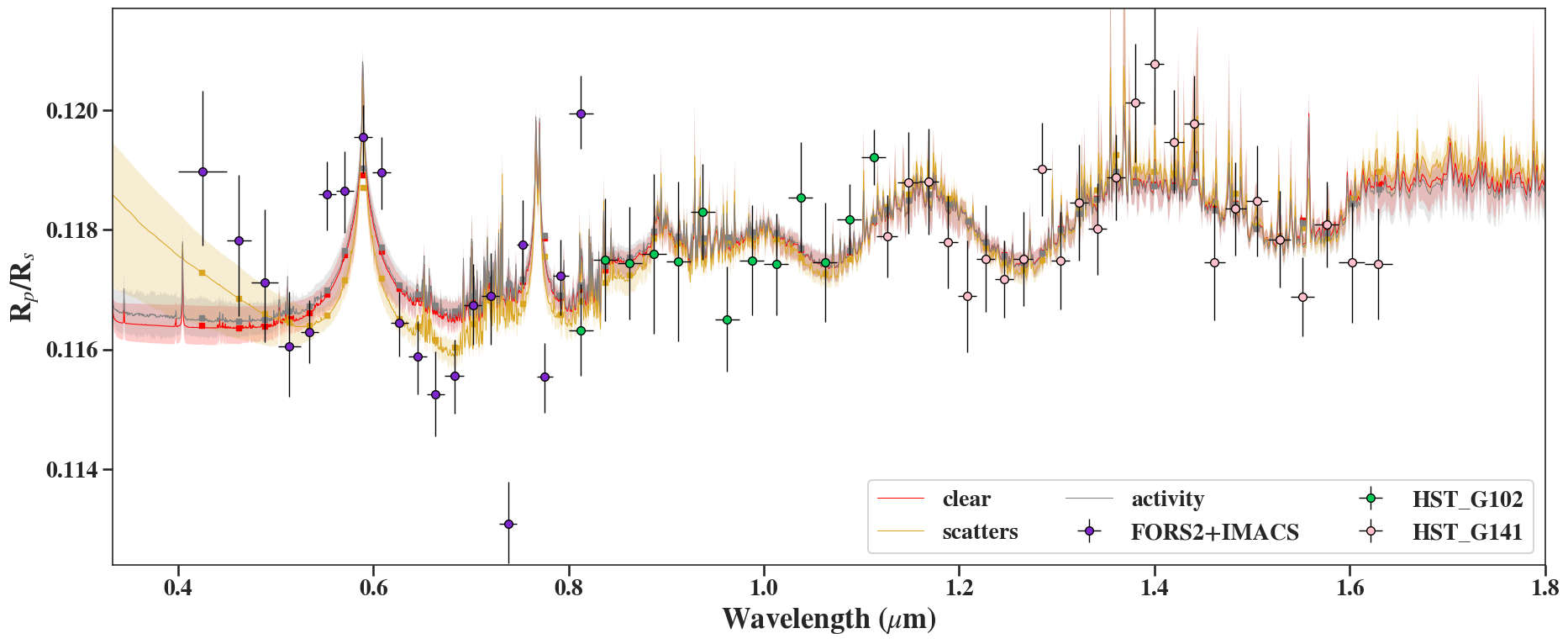}
    \caption{The final transmission spectra of WASP-96b, which include the combined FORS2 and IMACS data (purple) and the G102 (green) and G141 (pink) data both from \cite{Yip:2021}. Best fit \texttt{PLATON} retrieval models that include stellar activity (grey), scatters (gold), and a clear atmosphere (red) are also plotted, with there 1-$\sigma$ confidence interval shaded in the same colors. Because for each of the three models, the optical offsets were slightly different, what is shown here is the mean depth where the true offsets were 0.00476$^{+.00040}_{-.00039}$, 0.00414$^{+.00036}_{-.00035}$, and 0.00467$^{+.00039}_{-.00034}$ for the activity, scatters, and clear models, respectively. As can be seen in Table \ref{tab:Global_LnZs} the model with the highest evidence is one with scatters; however, no model has an evidence strong enough to favor it over another.}
    \label{fig:PlatonFits} 
\end{figure*}

\begin{figure*}[htb]
    \centering
    \includegraphics[width=.99\textwidth]{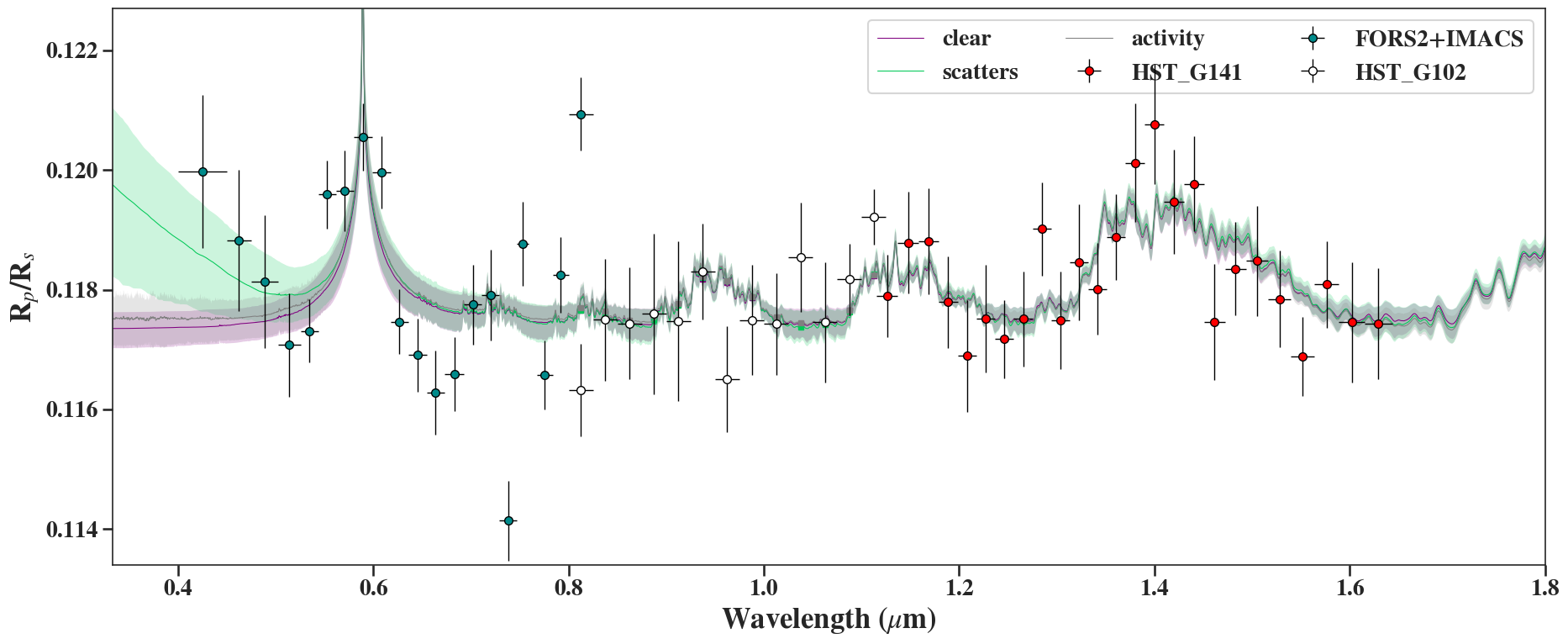}
    \caption{Same as Figure \ref{fig:PlatonFits}, but with \texttt{Exoretrievals}. In this figure the optical (combined FORS2 and IMACS), G102, and G141 observations are cyan, white, and red, respectively. Again, the evidence of each plotted model (activity - gray, scatters - green, and clear - purple) are indistinguishable from each other, but the clear atmosphere has the highest evidence. The offset with each fit were 0.00554$\pm$.00031, 0.00558$\pm$.00034, and 0.00547$^{+.00031}_{-.00032}$ for the activity, scatters, and clear models, respectively.}
    \label{fig:ExoretrievalFits} 
\end{figure*}

For \texttt{Exoretrievals} the evidence, $\Delta \ln Z$ = 19.45, is even stronger than the individual datasets, showing that combining the data does improve the overall detection of the species. The corner plot of the highest-evidence model retrieved by \texttt{PLATON} (one with scatters) and \texttt{Exoretrievals} (one without activity and scatters but including H$_2$O and Na) is shown in Figures \ref{fig:PlatonClearCornerPlot} and \ref{fig:ExoretrievalClearCornerPlot}. Figures \ref{fig:PlatonFits} and \ref{fig:ExoretrievalFits} show the global data with over-plotted models that either include stellar activity, a scattering slope, or none (flat and clear) for both retrievals. We elect to use the global transmission spectra which included all optical data to interpret the retrievals and atmosphere of WASP-96b, because the maximum relative retrieved evidence and transmission spectrum precision are higher when including both optical data.

\begin{deluxetable*}{|l|C|C|C|C|C|C|C|C|C|}[h!]
    \caption{Same as Table \ref{tab:IMACS_LnZs} but with the dataset that included the combined IMACS, FORS2, and HST data. The retrievals with water and sodium were heavily supported by \texttt{Exoretrievals} and no model was prefered with \texttt{PLATON}.}
    \label{tab:Global_LnZs}
    \tablehead{\multicolumn{7}{|c|}{\bfseries {\large Exoretrievals}} &&\multicolumn{2}{|c|}{\bfseries {\large PLATON}}}
    \startdata 
      \textbf{Model:}  &  \text{flat}  &  $H_2O$  &  $Na$  &  $K$  &  $H_2O+Na$ &  $H_2O+K+Na$  && \textbf{Model:} &\\ \hline
      clear &      0.0 &	5.73 &	-0.56 &	12.52 &	19.45 &	19.11 & & \text{clear} & 0.0\\
      scatterers &          		--- &	6.87 &	-0.53 &	11.63 &	18.35 &	17.46 &	& \text{scattering} & 0.8\\
      activity  &           	0.1 &	5.06 &	-1.45 &	12.36 &	18.33 &	18.14 &	& \text{activity } & -0.28\\
      Both &--- &	4.95 &	-1.92 &	10.79 &	17.62 &	-4.16 & &\text{Both } & 0.55 \\
    \enddata 
\end{deluxetable*}

\section{Retrieval Interpretation} \label{sec:Retrieval_Interp}
\subsection{Atomic and Molecular Features} \label{sec:Na+H2O}
The retrieved mixing ratios of sodium ($\log_{10}(Na)=-5.4^{+2.0}_{-1.9}$) and water ($\log_{10}(H_2O)=-4.5\pm2.0$) are in agreement both with what was obtained by \citeauthor{Yip:2021} \citeyear{Yip:2021} ($\log_{10}(Na)=-3.88^{+1.05}_{-0.82}$ and $\log_{10}(H_2O)=-3.65^{+0.90}_{-0.94}$) and what was obtained by \citeauthor{Nikolov:2018} \citeyear{Nikolov:2018} ($\log_{10}(Na)=-5.1^{+0.6}_{-0.4}$). The Na mixing ratios obtained here are consistent with solar abundance ($log_{10}(Na)=-5.78\pm0.03$; \citeauthor{Asplund:2020} \citeyear{Asplund:2020}) and WASP-96's stellar abundance \citep{Nikolov:2018}. The water abundance is also consistent with water abundances of Jupiter constrained by Juno ($\log_{10}(H_2O)=-2.6^{+0.27}_{-0.44}$; \citealt{2020:Li}). This implies that the formation process for WASP-96b might have been similar to our own Jovians, but that conclusion is limited by the uncertainties in the measured mixing ratios. 

Theoretical predictions for clear atmosphere planets predict absorption signatures from the optical alkali features of Na and K \citep{Seager:2000}. However, our observations of WASP-96b show no observable evidence of K absorption, even though strong ${\rm Na~I}$ and ${\rm H_2O}$ features imply WASP-96b has a clear transmission spectrum. There are multiple possibilities as to why potassium was not significantly detected. The most obvious hindrance in detecting K was the gap in the transmission spectrum (759.4--767.2\,nm) that was nearly aligned with the center of the most prominent K doublet feature (768.15\,nm). Essentially, this only allowed K to be constrained by its wings, which might not be enough even if K is present. The possibility of K being present in the atmosphere is hinted in Figure \ref{fig:PlatonFits}, where \texttt{PLATON}'s retrievals by default include K because of imposed equilibrium chemistry and relative abundances determined by metallicity. In those models we see that the transmission spectrum is somewhat consistent with the K doublets' features, where the blue-ward bin in the K wing is within the model's 1-$\sigma$ interval and the red-ward bin is not. The fact that the evidences of the \texttt{Exoretrievals} models including K are lower than those excluding it is likely only because the data could be explained with or without K, given the gap in the central doublet band. Another possibility is that the potassium is locked away in KCl, which has been suggested to have a highly efficient formation rate in this temperature regime \citep{Gao:2020}. Further space-based observations, or ground based high resolution observations will be needed to determine if abundant K absorption is truly present in the atmosphere of WASP-96b.

\subsection{Activity and Optical Slope} \label{sec:Slope+Activity}
For the models that included activity, retrieved $\Delta T$ temperatures from both retrievals and all datasets were consistent ($<$ 1-$\sigma$) with no active regions ($\Delta T$ = 0\,K), implying a non-detection of stellar activity. Therefore, stellar photometry (see sec. \ref{sec:Rotational_Period} \& \ref{sec:PhotMod}), stellar spectroscopy (see sec. \ref{sec:R_hk}), and the transmission spectrum retrieval analysis all support the idea that WASP-96 has very little activity.

H$_2$ Rayleigh scattering is expected blue-wards of $\sim$450--550\,nm in a hydrogen dominated atmosphere without high altitude clouds \citep[i.e.][]{Seager:2000,Etangs:2008}. The fact that we were unable to significantly identify one is likely due to limitations of the data, given that we only have two and a half broader bins in this wavelength range. The slope could be better constrained with HST/STIS \footnote{G430L has coverage approximately from 0.29--0.57\,$\mu$m \url{https://www.stsci.edu/hst/instrumentation/stis}} or HST/WFC3/UVIS \footnote{WFC3/UVIS G280 covers 0.2-0.8\,$\mu$m and has better throughput in the blue than STIS \citep{Wakeford2020}} observations. This would likely be sufficient to distinguish from stellar activity, because the star is relatively quiet. Therefore, when imposing these constraints, as was done in Section \ref{sec:PhotMod}, the only viable activity features produced are very shallow slopes. This is outlined in Figures \ref{fig:PlatonFits} and \ref{fig:ExoretrievalFits} where the models with activity fit are shown in gray for both figures. Furthermore, the activity of quiet stars, like this one, is dominated by faculae \citep{Reinhold2019, Rackham2022}, which produces the opposite signal of what is expected of a Rayleigh slope, when the activity regions are unocculted. Meaning that large cold unocculted spots, needed to mimic a Rayleigh slope, would be in direct contradiction to all other observations of WASP-96. Thus, it is clear that if a blue-ward scattering slope is found with additional observations, the most viable explanation would be attributing the feature to the planet. 

Though the data is not sufficient enough to significantly distinguish between a model with and without a scattering slope, all retrieved scattering slopes were found to be consistent ($<$ 2-$\sigma$) 
with a Rayleigh scattering slope. If this can be confirmed the Rayleigh like slope would best be attributed to H$_2$ scattering.

\subsection{Aerosols} \label{sec:Aerosol_free}
The higher reference pressures obtained by \texttt{PLATON} and \texttt{Exoretrievals} (a  few to tens of bars) both agree with one another and supports the idea that if WASP-96b hosts thick aerosol layers, they are confined to beneath the top of the atmosphere. The strong water features and broad ${\rm Na~I}$ absorption wings observed also support this conclusion. Additionally, when fitting for aerosols with \texttt{Exoretrievals} the cloud cross section, $\sigma_{cloud}$ (see \cite{Espinoza2019} Appendix D) was initially set to be very wide, with log-uniform priors from -80 to 80 for $\sigma_{cloud}$. In doing so, we found that across all 3 datasets (i.e. optical data from IMACS, FORS2, and combined) and all iterations of models that include scattering (i.e. no mater which molecules were included or if including activity) the means of  $\sigma_{cloud}$ were nearly the same at $\sim$ -55. This value is so low it effectively means that the retrievals find no evidence of clouds affecting the spectra, which is consistent with the strong water and sodium features, and the larger reference pressures. 

Aerosols are likely present in the atmosphere of all planets, but when aerosols are not warranted by the retrievals that implies that the aerosols are not thick enough at the high altitudes in which transmission spectroscopy probes to significantly mute the spectral features in the observed wavelengths, and/or they are not significantly present in the terminator. Also, though the retrievals, suggests that data can be explained without thick aerosols, color dependent scatters may still affect the transmission spectra. In particular, the possible Rayleigh scattering slope in the optical cannot be confidently refuted in the combined optical data.

\subsection{Temperature and C/O ratio} \label{sec:T/P+C/O}
The terminator temperatures found with \texttt{PLATON} (T$_p$=877$\pm$40\,K) and \texttt{Exoretrievals} (T$_p$=830$^{+160}_{-140}$\,K) are within agreement of one another and with what \cite{Yip:2021} retrieved (T$_p$=954$^{+198}_{-195}$\,K), given their uncertainties. However, these values are vastly different from the temperature retrieved by \citealt{Nikolov:2018} (T$_p$=1710$^{+150}_{-200}$\,K), who used the ATMO models \citep{Amundsen2014, Goyal2018} to perform retrieval analysis on their data. Our retrieved temperatures are also far from the 0 albedo equilibrium temperature of 1285$\pm40$\,K \citep{Hellier:2014}. A lower retrieved terminator temperature is often found in the literature and is expected when applying a 1D model to probe a 3D atmosphere. 1D models are useful for fitting features found in transmission spectra substantially more quickly than 3D models. Unfortunately, they can artificially shift the retrieved temperature up to hundreds of Kelvin cooler than the true average temperature \citep{McDonald2020, Pluriel2020}. Other physical assumptions prescribed in the retrieval frameworks could also contribute to a discrepancy in retrieved temperatures \citep{Welbanks2021}.

The only feature in the observed spectrum that could directly constrain C or O was H$_2$O, thus, the quoted C/O ratios should be taken lightly, given that it is only using the water features and chemical equilibrium constraints to obtain the ratio. Including the two Spitzer/IRAC points centered at 3.6 and 4.5\,$\mu$m (program ID 14255), as was done by \cite{Yip:2021}, would include data near carbon bearing features (i.e. CO, CO$_2$, and CH$4$). However, we elected to exclude these points because the two points have larger uncertainties (in wavelength and depth), which would not significantly constrain any of these carbon species, as can be seen in \cite{Yip:2021}.  
Given the lack of a constrained C/O ratio, it is hard to ascertain the formation region of WASP-96b, but JWST observations would be able to refine the C/O ratio with higher precision and higher spectral resolution observations in the near- to mid-IR wavelengths. The observed lack of aerosols obscuring features in the optical to near-IR spectrum of WASP-96b makes it an ideal target for such observations, which has the potential to provide key insight of where hot Jupiters formed and when/if they migrated. 

\section{WASP-96b in context} \label{sec:Contxt}
\subsection{Mass-Metallicity Trend} \label{Mass-Met}
\begin{figure*}[htb]
    \centering
    \includegraphics[width=.8\textwidth]{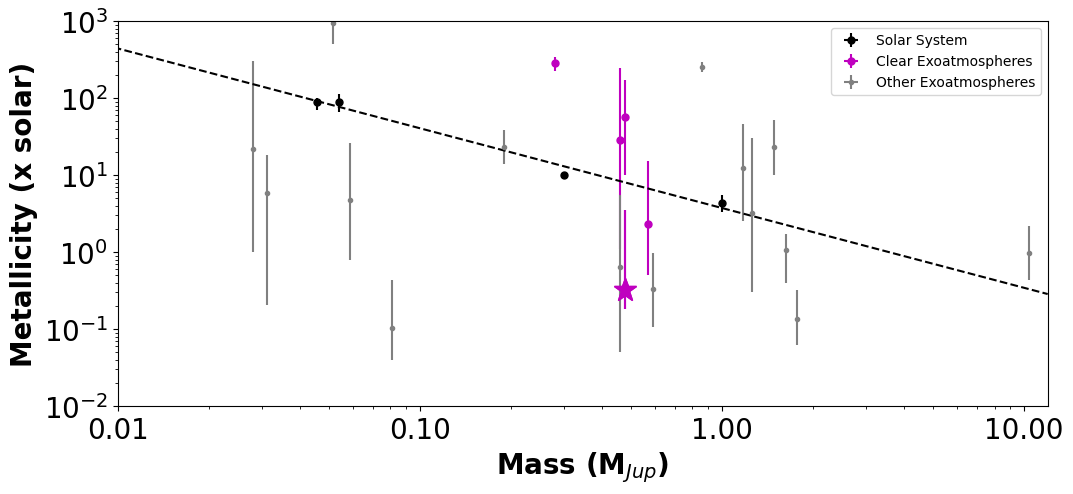}
    \caption{Observed mass–metallicity trend for transiting exoplanets. The bulk of the data was acquired from \cite{Wakeford:2020} \footnote{\url{https://stellarplanet.org/science/mass-metallicity/}}, but we added HAT-P-32b \citep{Alam2020}, WASP-17b (\texttt{PLATON} run with data from \cite{Sing:2016}), WASP-62b (using Na abundance from \cite{Alam:2021} as a proxy), WASP-94b \citep{Ahrer:2022}, and WASP-96 (this work, magenta star). The dashed black line corresponds to a linear fit in log–log space to the solar system points (black dots). We highlight the clear planets with magenta and all other planets are gray.}
    \label{fig:Mass_met} 
\end{figure*}

Our metallicity with Platon is consistent with the constraints from \citealt{Nikolov:2018} ($\log_{10}(Z/Z_{\odot})=0.4^{+0.7}_{-0.5}$). Furthermore, it is within 3\,$\sigma$ agrement to the solar system mass-metallicity trend \citep[explored for exoplanets by e.g.,][]{Wakeford:2018,Alam2020,Wakeford:2020}, shown in Figure \ref{fig:Mass_met}. In this figure we plot the metallicites and masses of the solar system giants and a linear fit of that data in log-log space. We also show planets with metallicities derived from molecular and atomic abundance of one to a few different species. 

Looking at all exoplanets in the figure, the trend found for the solar system does not persist among the exoplanet sample. However, given that high-altitude aerosols make it more difficult to accurately determine molecular abundances, we also highlighted the five planets which are thought to have little high-altitude aerosols obscuring their spectra. When only including these planets, four of the five planets (WASP-17b, WASP-62b, WASP-94b, and WASP-96b) are less than 3\,$\sigma$ from the predicted values. Though most of the clear atmospheres prove to be consistent with the solar system mass-metallicity trend, it is still hard to claim that we can interpret this as most exoplanets undergo the same formation mechanisms as our solar system for multiple reasons. For one, all three of the consistent metallicities are in similar mass regimes, therefore, giving perspective only on a small fraction of exoplanets. Additionally, most of these metallicities were obtained assuming the abundances of one to a few species can be directly translated to the bulk metallicity of the atmosphere. How far this assumption is skewed from reality is unknown, since interactions among transport, chemistry, and phase changes may mean the elements are not well-mixed in the atmosphere \citep{Zhang:2020}. Furthermore, most of the planets in Figure \ref{fig:Mass_met} are hot jupiters/neptunes, which are outliers that are not representative to exoplanets in general. For these reasons, and others, there is much headway required in order to obtain a more complete grasp of exoplanet formation mechanisms and trends. Observing more planets like WASP-96b, with relatively clear atmospheres, in many wavelengths and improving our analysis processes around these observations is the best path forward.

Interestingly, though the work from this paper and others \citep[e.g.][]{Alam2020,Wakeford:2020} find no clear trend amongst atmospheric composition and mass, there has been a found mass-metallicity trend in terms of planet bulk composition \citep[][, inferred from structure models]{Thorngren2016}. Still, much work is needed in order to determine if there truly is or is not an atmospheric metallicity-mass trend before the root of this potential discrepancy can be explored further.

\subsection{Correlations to aerosol formation rates} \label{ExoplanetForcast}
Understanding why WASP-96b is one of the few planets that has an observed transmission spectrum unobscured by aerosols is vital for advances in exoplanetology. This would allow astronomers to a priori determine what key characteristics make an exoplanet transmission spectrum clear and allow for more targeted exo-atmosphere surveys. 

\begin{figure*}[htb]
    \centering
    \includegraphics[width=.8\textwidth]{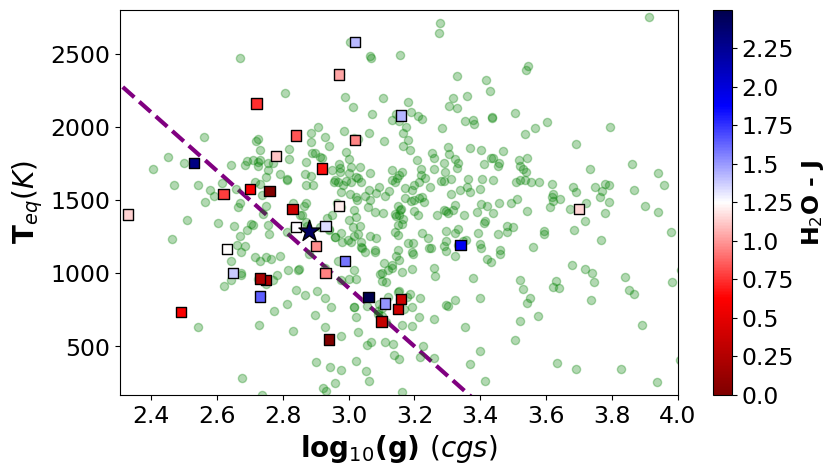}
    \caption{The H$2$O-J indices (color coded squares) defined by \cite{Stevenson:2016} relative to the planetary surface gravities and equilibrium temperatures. Other planets with surface gravities and effective temperatures obtained from \href{astro.keele.ac.uk/jkt/tepcat}{\texttt{TEPCat}} are plotted as green circles for context. We also plot the dividing line in T$_{equ}$-log$_{10}$(g) phase space that \cite{Stevenson:2016} proposed delineates between cloudy and clear planets. Where the authors say H$2$O-J indices greater than 2 is likely completely clear, and indices less than 1 are greatly obscured by clouds. The only planets with indices greater than 2 are HAT-P-11b (2.5$\pm$0.5), WASP-96 (2.4$\pm$0.7), and WASP-17b (2.3$\pm$0.9). WASP-96 (star) is shown to be one of the clearest observed planets, both through this index and in the optical, yet it sits very close to the dividing line. The data in this table was produced in \cite{Alam2020}.}
    \label{fig:Cloudy_vsT&logG} 
\end{figure*}

Figure \ref{fig:Cloudy_vsT&logG} puts WASP-96b in a broader context, it was initially used by \cite{Stevenson:2016}, who proposed a temperature-gravity trend in the formation of high-altitude clouds. They used the H$_{2}$O-J index as a proxy for the cloud levels, which inherently assumes a similar absolute water abundance of all hot Jupiters. HAT-P-11b is a prime example of the limitations of using this index as a proxy for cloud formation, where its water feature is extremely strong (2.499$\pm$0.505), but its optical spectra revealed it to have a cloudy atmosphere \citep{Chachan:2019}. Nonetheless, the H$_{2}$O-J index provides an easy index to parameterize cloud levels. Multiple other planets have been added to the original figure \citep[e.g.,][etc.]{Alam2020,Weaver:2021}, and make it clear that the trend does not strictly exist. As seen in our Figure \ref{fig:Cloudy_vsT&logG}, WASP-96b (marked as a star) highlights the lack of an obvious trend even further, where it is near the fitted temperature-log(g) boundary line but is in fact one of the most clear atmospheres observed to-date. Thus, there is still extended work needed to isolate what causes high-altitude aerosol formation.

\section{Summary \& Conclusion} \label{sec:Conclusion}
We observed two transmission spectra of WASP-96b with Magellan/IMACS as part of the ACCESS survey. In the process of analyzing the data, we tested the precisions and accuracies of two commonly used spectroscopic light curve detrending techniques: A) Common mode correction followed by a polynomial fitting (CMC+Poly) and B) a Gaussian processes (GP) routine. Both routines were tested against simulated data, where we find that for data without substantial chromatic systemics the CMC+Poly procedure produces more accurate depths with higher precision. Additionally, we find that neither method (worse for CMC+Poly) was able to consistently reproduce absolute depths. This provided justification of fitting for offsets amongst transmission spectra from different nights and instruments.  

The transmission spectrum from IMACS was then added with reanalyzed transmission spectra from FORS2, both of which were reduced using the CMC+Poly routine. The optical data was combined with a previously published HST/WFC3 (G102 and G141) transmission spectrum \citep{Yip:2021}, collectively producing a nearly continuous coverage transmission spectrum from 400--1237\,nm. This spectrum was run against two retrievals: \texttt{PLATON} and \texttt{Exoretrievals}. Both retrievals found that the terminator of WASP-96b was shrouded by little or no aerosols, as such it is still one of the most clear exo-atmospheres observed. More specifically \texttt{PLATON} found a metallicity, C/O ratio, reference pressure, and terminator temperature of $\log_{10}(Z/Z_{\odot}) = -0.49^{1.0}_{-0.37}$\,dex, C/O = 0.97$^{+0.65}_{-0.50}$, $\log_{10}(P_0)$ = 1.3$^{+1.0}_{-1.1}$\,bars, and T$_p$ = 877$\pm$40\,K, respectively. \texttt{Exoretrievals} found a terminator temperature, reference pressure, and water and sodium mixing ratios of T$_p$ = 830$^{+160}_{-140}$\,K, $\log_{10}(P_0)$ = 0.29$^{+1.86}_{-2.02}$\,bars, $\log_{10}(H_2O)$ = -4.5$\pm$2.0, and $\log_{10}(Na)$ = -5.4$^{+2.0}_{-1.9}$, respectively. With the constraints on stellar activity imposed from photometric monitoring, neither retrieval had strong support for stellar activity explaining the data. Though there is a hint of a slight slope blue-ward of 550\,nm, there was not substantial evidence supporting the need for a scattering slope. However, bluer, space-borne observations of the planet could discern if the hint of a blue-ward slope is indeed a true feature.

Finally, in an attempt to put WASP-96b in a broader context, we found that it along with three other clear planets (WASP-62b, WASP-94b, and WASP-17b) are consistent with the mass-metallicity trend observed in the solar system. However this sample is not complete enough to make overarching claims about the extrasolar giants' formation mechanisms compared to giant planet formation pathways in the solar system. We also find no clear correlation to what causes the aerosol formation rate in exo-atmospheres, and recommend more in-depth analysis of a higher sample of planets in order to find such a correlation.

\acknowledgments 
The results reported herein benefited from support, collaborations and information exchange within NASA's Nexus for Exoplanet System Science (NExSS), a research coordination network sponsored by NASA's Science Mission Directorate. 
This material is partly based upon work supported by the National Aeronautics and Space Administration under Agreement No. 80NSSC21K0593 for the program “Alien Earths”. 
This paper includes data gathered with the 6.5 meter Magellan Telescopes located at Las Campanas Observatory, Chile. We thank the staff at the Magellan Telescopes and Las Campanas Observatory for their ongoing input and support to make the ACCESS observations presented in this work possible. 
This work uses observations collected at the European Organization for Astronomical Research in the Southern Hemisphere under European Southern Observatory programme 199.C-0467(H).
We also appreciate the support from the NSF Graduate Research Fellowship (GRFP), grant number DGE1745303. 
The computations in this paper were conducted on the Smithsonian High Performance Cluster (SI/HPC), Smithsonian Institution. \url{https://doi.org/10.25572/SIHPC}. 

B.V.R. thanks the Heising-Simons Foundation for support.
C.M. thanks Neale Gibson for conversations on marginalization techniques.
A.J. acknowledges support from ANID -- Millennium  Science  Initiative -- ICN12\_009 and from FONDECYT project 1210718.

\software{Astropy \citep{astropy2013}, corner \citep{corner2016}, Matplotlib \citep{matplotlib2007}, NumPy \citep{numpy2006}, Multinest \citep{multinest2009}, PyMultiNest \citep{2014BuchnerPyMultiNest}, SciPy \citep{scipy2001}, batman \citep{Kreidberg2015_batman}, george \citep{Mackey2014_george}} dynesty \citep{Speagle_2020Dynesty}, PLATON \citep{Zhang_2019PLATON}, Juliet \citep{Espinoza2019_juliet}

\facilities{Magellan:Baade, Smithsonian Institution High Performance Cluster
(SI/HPC)}

\bibliographystyle{yahapj}
\bibliography{References}

\appendix 
\section{Second Order Contamination} \label{Appx:2ndOrderCorr}
As mentioned in Section~\ref{subsec:Setup}, we did not use a blocking filter during the UT170804 transit, but we added the GG495 blocking filter in the UT171108 observations after we learned from other ACCESS observations that second order light introduces contamination in some cases.

\begin{figure*}[htb]
    \centering
    \includegraphics[width=1\textwidth]{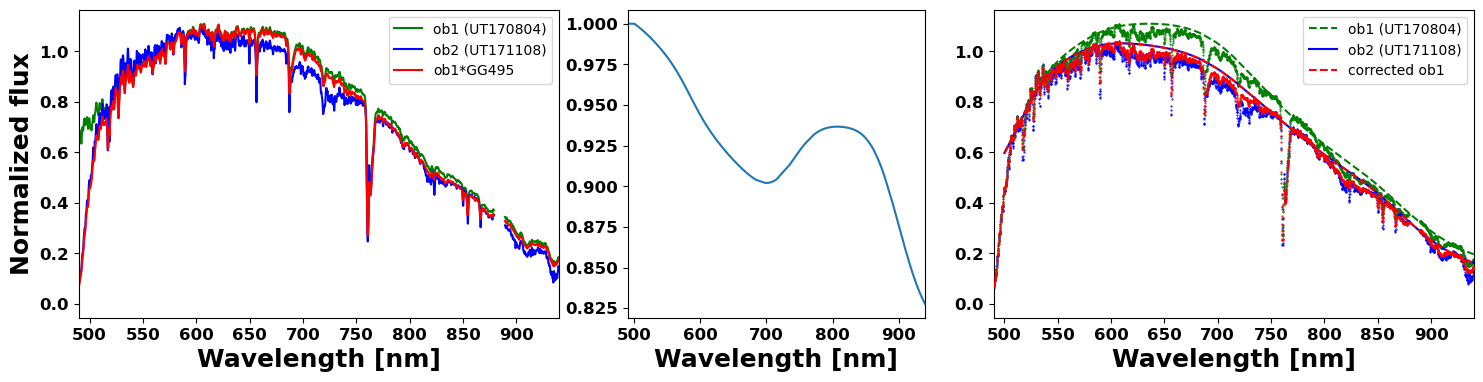}
    \caption{The process used to correct for second order light. The left panel shows the normalized IMACS observations taken on UT170804, which had no filter (red); UT170804 convolved with the GG495 transmission profile (green); and UT171108, which had the GG495 filter (blue). The middle panel shows the normalized second order light correction, determined by dividing the filter convolved first night (green) by the second night (blue).  The right panel shows normalized UT170804 (red), UT170804 convolved with the GG495 profile and corrected for second order light (green), and UT171108 (blue). Dotted lines of each spectra’s modeled continua is overlayed in the same respective colors as their spectra. These steps are also applied to each comparison so every spectrum is corrected from second order light.}
    \label{fig:2nd_OrderCor} 
\end{figure*}

To check for possible  contamination in the UT170804 observation, we modeled its effect by convolving the unfiltered stellar spectra, which inherently is a convolution of the instrument's throughput and WASP-96’s stellar spectra, with the G495 filter’s throughput. This gave us the spectral profile of the UT170804 observation, if the G495 filter had been added, but with any second order light still present. Then we modeled the normalized continuum of this light curve against the normalized continuum of the UT171108 observations light curve. Because we assumed that the only difference between both continuum structures should be from the 2nd order light, we used the ratio of the two continua as a correction term for the second order light. This process is illustrated in Figure~\ref{fig:2nd_OrderCor}. We applied this correction method to each comparison star and WASP-96 spectrum observed on UT170804 (the exact correction term is unique per spectrum) to determine the effect of the second order light. We then used both the corrected and uncorrected spectra to produce a white-light curve and binned light curves, which are constructed by integrating either all wavelengths of light (white-light) or specific band passes (binned) in a given spectrum, and plotting each integrated spectral counts relative to time. Next, we compared the final white-light and binned transits with and without the correction, and found an insignificant difference between the two. This is likely, because the relative effect of the second order light is negated when dividing by the comparison stars (or using a PCA correction with the comparison stars). Given that the correction had minimal effects, in our final analysis we used the uncorrected spectra. 

\section{Detrending routines} \label{Appx:Detrending_Routines}
\subsection{Principle Component Analysis (PCA) and Gaussian Processes (GP)} \label{appx:PCA}
The PCA+GP routine we used has been implemented often in recent years \citep{Weaver:2020,Yan2020,McGruder:2020,Weaver:2021}, but we outline it here. We first model out commmon systematics found in all the comparison stars by performing singular value decomposition on a matrix composed of comparison light curves ($L_k(t)$) in the form:
\begin{align}
    L_k(t)=\sum_{i=1}^{K}A_{k,i}s_i(t),
\end{align}
where $s_i(t)$ is a set of signals representing the systematics affecting a given light curve and $A_{k,i}$ is the weight for each of those signals. 
This allows us to identify the principal components (i.e. PCA) of the light curve, which is driven by systematics. This is used rather than just dividing the target by the sum or mean light curve of the comparison stars, because when doing that you combine systematics that are unique to a specific star, instead of isolating which systematics persist amongst all stars. Therefore, this method is less likely to incorrectly divide out systematics in the comparison star that were not present in the target star. 

To model out systemtics unique to the target's light curve we used a Gaussian process (GP) regression with a joint probability distribution of the form $\mathcal{N}[0,\Sigma]$, where the covariance function ($\Sigma$) is defined as
$K_{SE}(x_i,x_j) +\sigma^2_w\delta_{i,j}$. Here, $\sigma^2_w$ and $\delta_{i,j}$ are a jitter term and the Kroenecker delta function, respectively. $K_{SE}(x_i,x_j)$ is a multidimensional squared-exponential kernel of the form:
\begin{align}
 K_{SE}(x_i, x_j)=\sigma^2_{GP} \exp\Bigg(-\sum_{d=1}^{D} \alpha_d(x_{d,i}-x_{d,j})^2\Bigg),
\end{align}
where $\sigma^2_{GP}$ is the amplitude of the GP and $\alpha_d$ are the inverse (squared) length-scales of each components of the GP. The priors on the jitter term and $\sigma^2_{GP}$ were both log-uniform, where the jitter term ranged from 0.01 to 100 ppm and $\sigma^2_{GP}$ from 0.01 to 100 mmag. The prior on each $\alpha_d$ was an exponential function, similar to what was done by \cite{Gibson2017WASP31} and \cite{Evans:2018}. 

In the above equations index i denotes each time-stamp and d corresponds to a set of different time-dependent external (auxiliary) parameters used. For the Magellan/IMACS transits these auxiliary parameters were variation of air mass, full-width at half-maximum (FWHM) of the spectra, mean sky flux, position of the central pixel trace (perpendicular to the dispersion axis), and drift of the wavelength solution. For the VLT/FORS2 transits they were dispersion and cross-dispersion drift, variation of the FWHM of the spectra, air mass, mean sky flux, and change in rotator angle.

The PCA and GP components were combined using the following equation: 
\begin{align}
 M_k(t)=c_k + \sum_{i=1}^{N_k} A_{k,i} s_i(t) - 2.51*log_{10}T(t|\phi) + \epsilon,
\label{equ:PCA+GP+transit}
\end{align}
where M$_k(t)$ is the (mean-subtracted) magnitude of the target star in the k$^{th}$ model, $c_k$ is a magnitude offset, $N_k$ is the number of PCA signals $s_i(t)$, $A_{i,k}$ is the weight for each signal in each models, $T(t|\phi)$ is the transit model with parameters $\phi$\footnote{We used the Python package \href{https://lweb.cfa.harvard.edu/~lkreidberg/batman/}{\texttt{batman}} \citep{Kreidberg2015_batman} to produce the analytic transit model.}, and $\epsilon$ our GP component \footnote{We used \href{https://github.com/dfm/george}{\texttt{george}} \citep{Mackey2014_george} to evaluate the GP marginalized likelihoods}.

We used \cite{2014BuchnerPyMultiNest}'s Nested sampling routine, \href{https://github.com/JohannesBuchner/PyMultiNest}{\texttt{PyMultiNest}} (a python wrapper of \texttt{MultiNest}, \citealt{multinest2009}), to explore the posteriors of our PCA+GP models. Because Nested sampling provides Bayesian evidences of models, we used those evidences as weights to combine posterior distributions of each M$_k(t)$ model in a technique called Bayesian Model Averaging \citep[BMA,][]{Gibson2014_modelAvg}.

In the limiting case where there is only one comparison star, the PCA portion of the routine cannot be performed. In that case equation \ref{equ:PCA+GP+transit} becomes 
 
\begin{align}
 M(t)=c_0 + A*m_c(t) - 2.51*log_{10}T(t|\phi) + \epsilon,
\label{equ:GP+transit}
\end{align}

where M(t), $T(t|\phi)$, $c_0$, and $\epsilon$ are the same as M$_k(t)$, $T(t|\phi)$, $c_k$, and $\epsilon$ from equation \ref{equ:PCA+GP+transit}, but without summing over different $N_k$ PCA signals. A is now a weight for the mean-subtracted comparison star magnitude, m$_c$. This equation is what was used by \cite{Yan2020}, because they only had one comparison star.

When this detrending routine can be used to its fullest (i.e. with multiple comparison stars) it is referred as ``PCA+GP.'' In the case when there is only one comparison star, like in the FORS2 data, it is referred to as a ``GP'' routine. As discussed in Section \ref{sec:Synth} and Appendix \ref{Appx:Synethic_Spectra}, the synthetic data only had one comparison star, so we in fact are comparing the GP and CMC+Poly routines in that analysis. 

This general procedure can be used to detrend both the white-light curves and the binned-light curves. When producing the binned-light curves solely with PCA+GP we first used this method on the white-light data, then fixed all parameters based on the PCA+GP white-light fit (aside from $q_1$, $q_2$, and $R_p/R_s$), then ran the PCA+GP routine against the binned-light curves. When producing the binned-light curves with the CMC+Poly routine we again used the PCA+GP method on the white-light data, used that white-light fit to obtain the CMC term, fixed all binned-light curve parameters based on the PCA+GP white-light fit (aside from $q_1$, $q_2$, and $R_p/R_s$), and ran the CMC+Poly routine (see sec. \ref{Appx:CMC}).

\subsection{Common Mode Correction (CMC)} \label{Appx:CMC}
The general process in CMC is to first produce a partially detrended white-light curve, where the larger systematics common to the target and the comparison(s) are removed. Two common ways to do this is either by dividing the light curve of the target by the light curve of the star(s) or with PCA. Next, a best fit transit model for the light curve is found using a routine that fits for both a transit and additional systematics (e.g. GPs). This transit model is divided by the uncorrected white-light curve to produce a constant systematic correction term (common mode). The correction term is then used for each individual bin, because the systematics are expected to be relatively consistent throughout each spectral band pass. After the common systematics are removed, each bin has an additional detrending routine (e.g. GPs or polynomial correction) to correct for any residual chromatic systematics. Doing this technique has proven to produce relatively high precision (though accuracy was not tested) for retrieved transit depths of each band pass \citep{Gibson:2013,Nikolov2016,Gibson2017WASP31,Nikolov:2018,Carter2020}.

\begin{figure*}[htb]
    \centering
    \includegraphics[width=.8\textwidth]{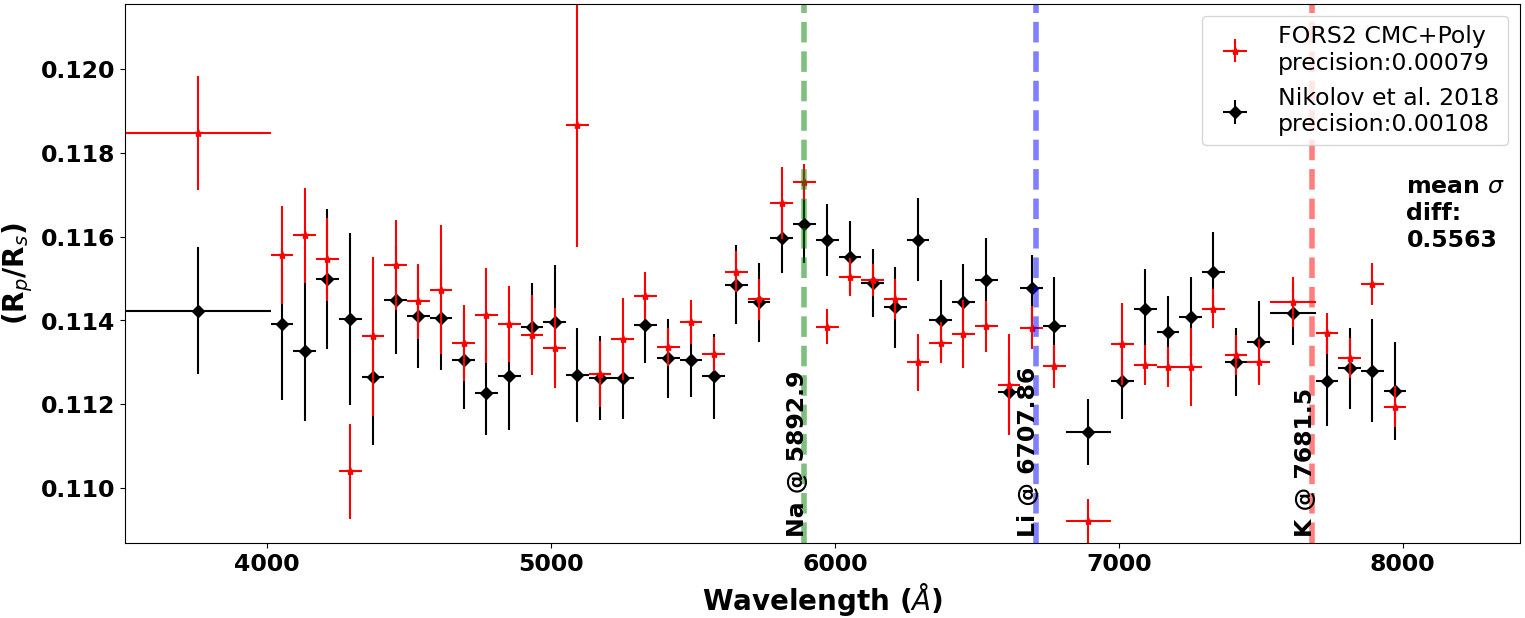}
    \caption{The original transmission spectrum of WASP-96b analyzed by \cite{Nikolov:2018} (black), and a re-reduction of the same data (VLT/FORS2) with the same binning scheme using our CMC+Poly analysis discussed in section \ref{Appx:CMC}. The two analyses produce similar mean R$_p$/R$_s$ precisions (0.00108 for the original and 0.0079 our reanalysis) and have an average difference of depth of only 0.56-$\sigma$, suggesting strong agreement of the two data. An offset is applied so the means of both datasets are the same.}
    \label{fig:Nik_vs_Redo} 
\end{figure*}

Our best fit white-light curve transit model (used for the common mode term) is obtained with the same steps of Appendix~\ref{appx:PCA}. After the common mode correction is applied we fit the remaining systematics using 1st and 2nd order polynomial fits of all external parameters. We individually explore the posterior space of all possible combinations of systematic corrections. This means for each bin, including a fit without external parameters (just a single constant coefficient), 729 and 243 models were fit for the IMACS and FORS2 data, respectively. The posterior space was explored in three steps: first, we fit for just the polynomial systematic coefficients on the out-of-transit data using \href{https://docs.scipy.org/doc/scipy/reference/generated/scipy.optimize.minimize.html}{\texttt{scipy.optimize.minimize}} \citep{scipy2001} with the 'Powell' method, coefficient bounds of -5 to 5, and initial points of 0. Then we use the found coefficients as the initial start points of another \texttt{scipy.optimize.minimize} run which includes the in transit data and a transit model fit. Lastly, we use \texttt{PyMultiNest} as our final posterior exploration where the transit parameters found with \texttt{scipy} were used as priors on the mean value, while maintaining the prior bounds (see Sec. \ref{sec:LC}), and the found polynomial coefficients were used as the mean values for a normal prior distributions with a standard deviations of 0.05. Again, we used BMA to best combine posteriors from all explored posterior spaces. For the fits we held $t_0$ fixed to what the white-light curve fit found, thus, the only transit parameters fit were the LD parameters and $R_p/R_s$. There are differences in how this technique is implemented compared to \cite{Nikolov:2018}'s method; however, it produces a nearly identical transmission spectrum, as shown in Figure~\ref{fig:Nik_vs_Redo}.
\section{Detrending Plots} \label{Appx:DetrendPlots}
\begin{figure*}[htb]
    \centering
    \includegraphics[width=.8\textwidth]{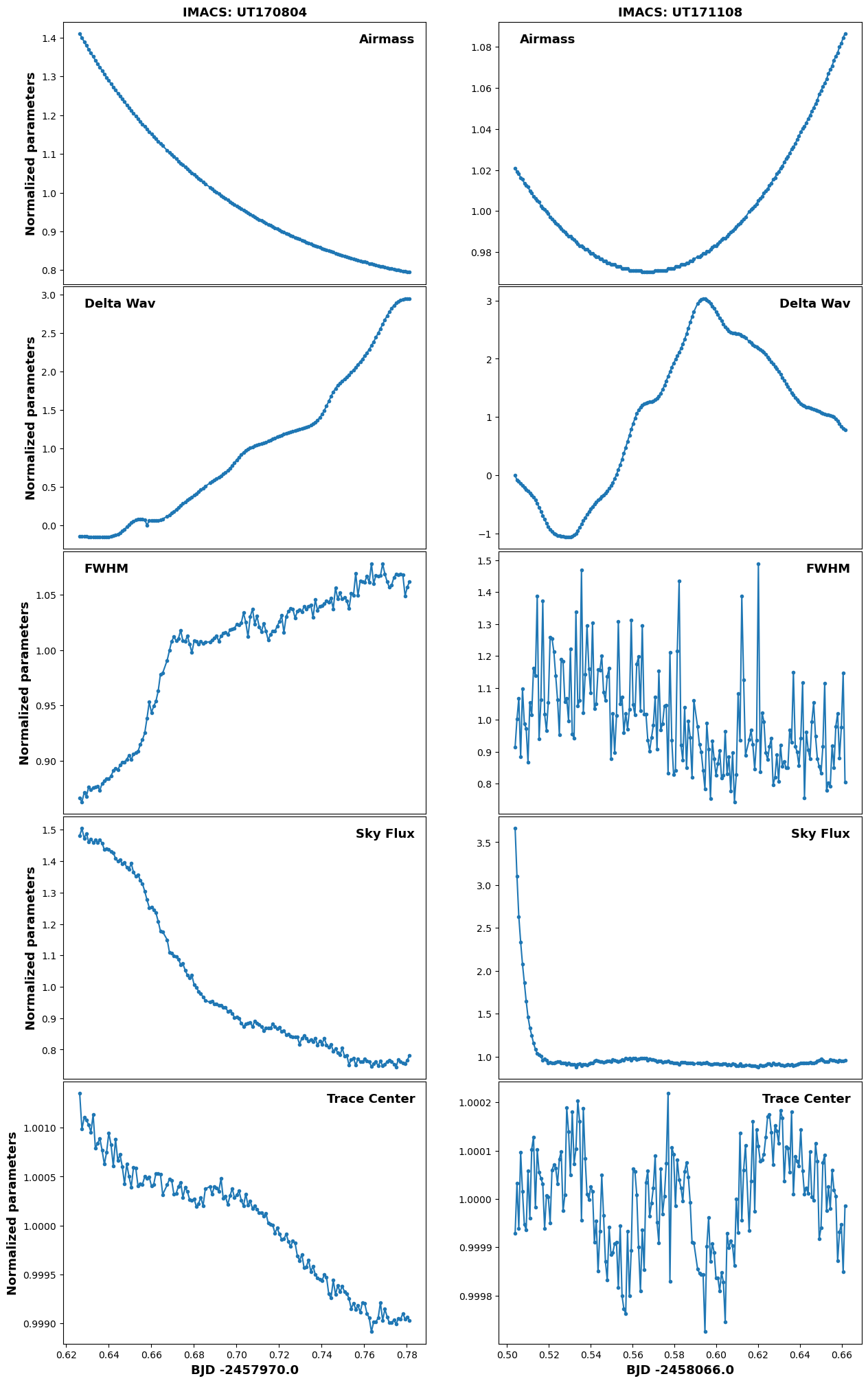}
    \caption{The mean normalized external parameters of the IMACS observations, used to correct for systematics in both the detrending methods (PCA+GP and CMC+Poly). The left column are the external parameters for UT170804 and the right is for UT171108. The rows correspond to the change in air mass, drift of the wavelength solution, FWHM of the target spectra, mean sky flux, and position of the central pixel trace (perpendicular to the dispersion axis), respectively. Plots of the FORS2 external parameters can be seen in \cite{Nikolov:2018} Extended Data Fig. 4}
    \label{fig:ePars} 
\end{figure*}

\begin{figure*}[htb]
    \centering
    \includegraphics[width=.8\textwidth]{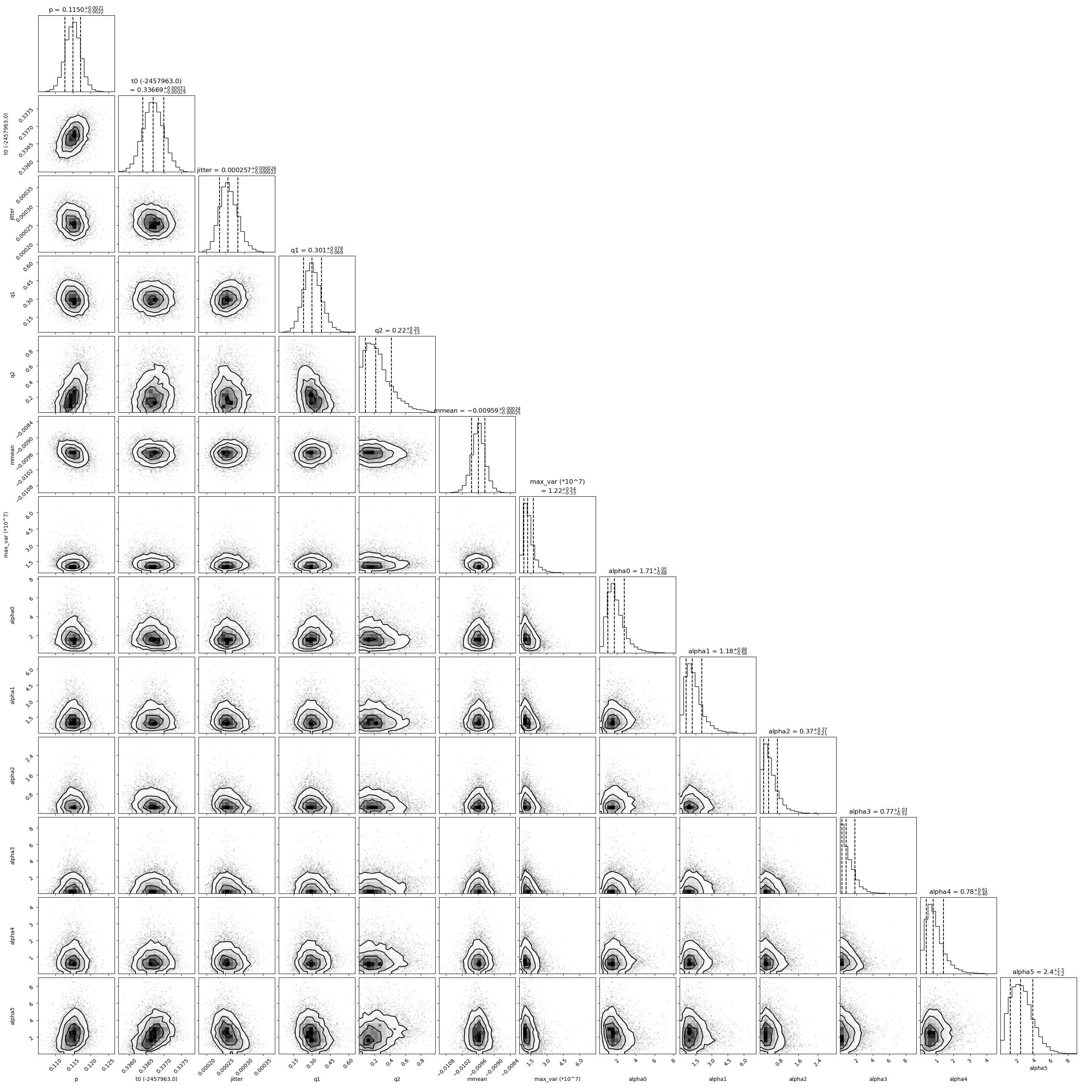}
    \caption{Corner plot of the GP fit on the FORS2 600B blue grism (UT170729) white-light light curve. The white light curve fit is used to create the final CMC term and this posterior corresponds to the fit shown in the first column of Figure \ref{fig:WLCs}. This would also be the same detrending routine used to detrend the binned data, if the GP (PCA+GP for the IMACS data) routine was used rather than the CMC routine. The alphas are the GP inverse squared length-scales on the external parameter regressors. Alphas 0 to 5 correspond to the target dispersion drift, cross-dispersion drift, variation of the FWHM, air mass, mean sky flux, rotator angle.}
    \label{fig:cornerB} 
\end{figure*}

\begin{figure*}[htb]
    \centering
    \includegraphics[width=.8\textwidth]{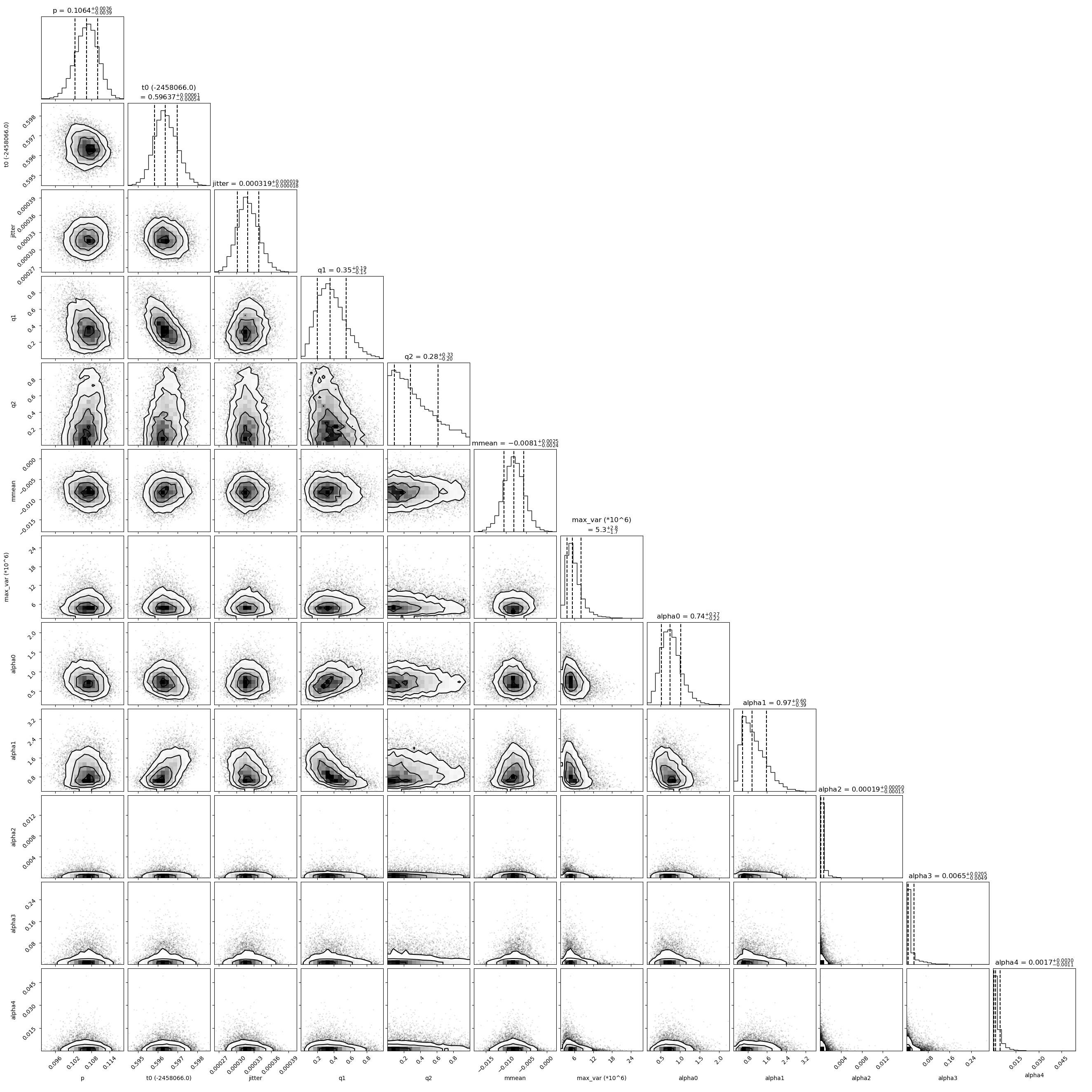}
    \caption{ Same as Figure \ref{fig:cornerB} but for the second IMACS transit (UT171108). Here Alphas 0 to 4 correspond to air mass, drift of the wavelength solution, FWHM, mean sky flux, and central trace.}
    \label{fig:cornerOb2} 
\end{figure*}
\section{Light Curves} \label{Appx:Lgt_curves}
For each bin we estimated the theoretical photon noise precision $\sigma_w$. This was calculated by taking the mean counts (per star) over each exposure, summing all the counts (N) in a given bin, and using $\sqrt{N}$ to estimate the shot noise for each star. The error of each star was propagated where the final (uncorrected) transit light curve was produced by dividing WASP-96's flux by the sum of the comparison stars' flux. 

We also estimate the noise of the corrected binned light curves. This was done with two approaches. First we calculate the root mean squared (r.m.s.) of the residuals. To also include an estimate that is not dependent on the best fit transit model, we take the standard deviation of the out-of-transit data. Both methods produced similar precisions and we use the higher of the two for our estimates of the the corrected light curve precision.  

We quantify the red noise by taking the ratio of the detrended light curve's noise level and the theoretical photon noise. We call this ratio $\beta$, as it is similar to the $\beta$ term used by \cite{Winn:2008} (though derived differently). The average $\beta$ is 1.122, 1.671, 1.089, and 1.528 for each night in chronological order. Their corresponding $R_p/R_s$ uncertainties are 0.001262, 0.002284, 0.000935, and 0.001485, correlated to the total light curve precision (red and white noise) but overall determined by how well the posterior space can be constrained, given the data. 
\begin{figure*}[h!]
    \centering
    \includegraphics[width=1\textwidth]{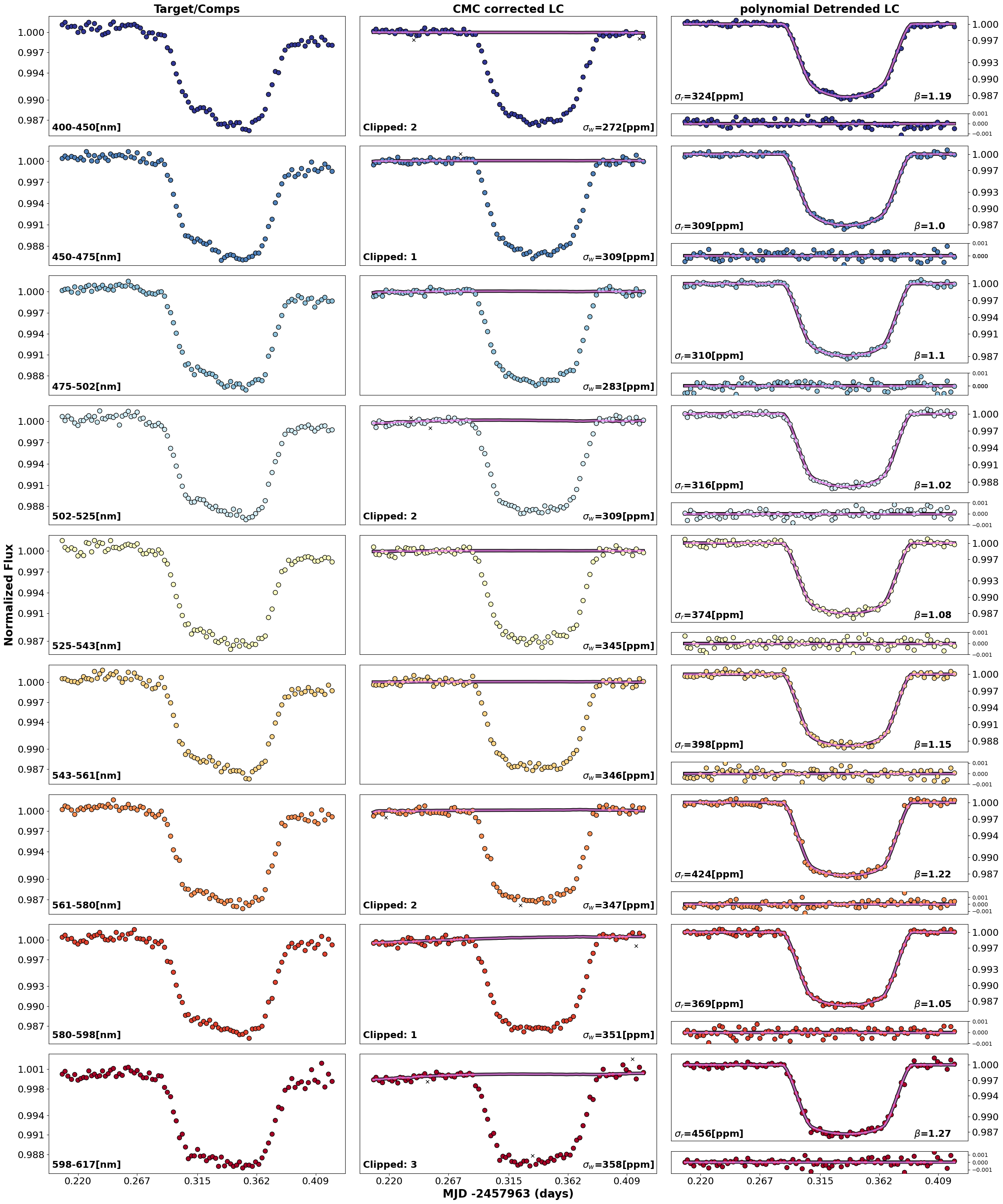}
    \caption{\textbf{Left:} Raw binned light curves (LC) of VLT/FORS2 transit UT170729 produced by dividing the comparison star LC (or the sum of the comparison star LC, in the case of the IMACS data) by the target's. The wavelength range of each bin is printed in the bottom left corner. \textbf{Middle:} LC produced by dividing the raw LC by the Common Mode Correction (CMC) term shown in Figure \ref{fig:WLCs} (final row). The number of sigma clipped points is printed in the bottom left and those points are marked with a black x, the theoretical photon noise precision is also printed as $\sigma_w$. The additional polynomial systematic correction specifically fit for each bin is shown as a magenta line. \textbf{Right:} The final CMC and polynomial corrected light curve, with the best fit transit model in magenta and the residuals of the detrended data from the best-fit transit model on the bottom. The standard deviation ($\sigma_r$) of the residuals in ppm is printed in the bottom left corner and the ratio of $\sigma_r$ and $\sigma_w$ ($\beta$) is printed.}
    \label{fig:LC_ut170729} 
\end{figure*}
\begin{figure*}[h!]
    \centering
    \includegraphics[width=1\textwidth]{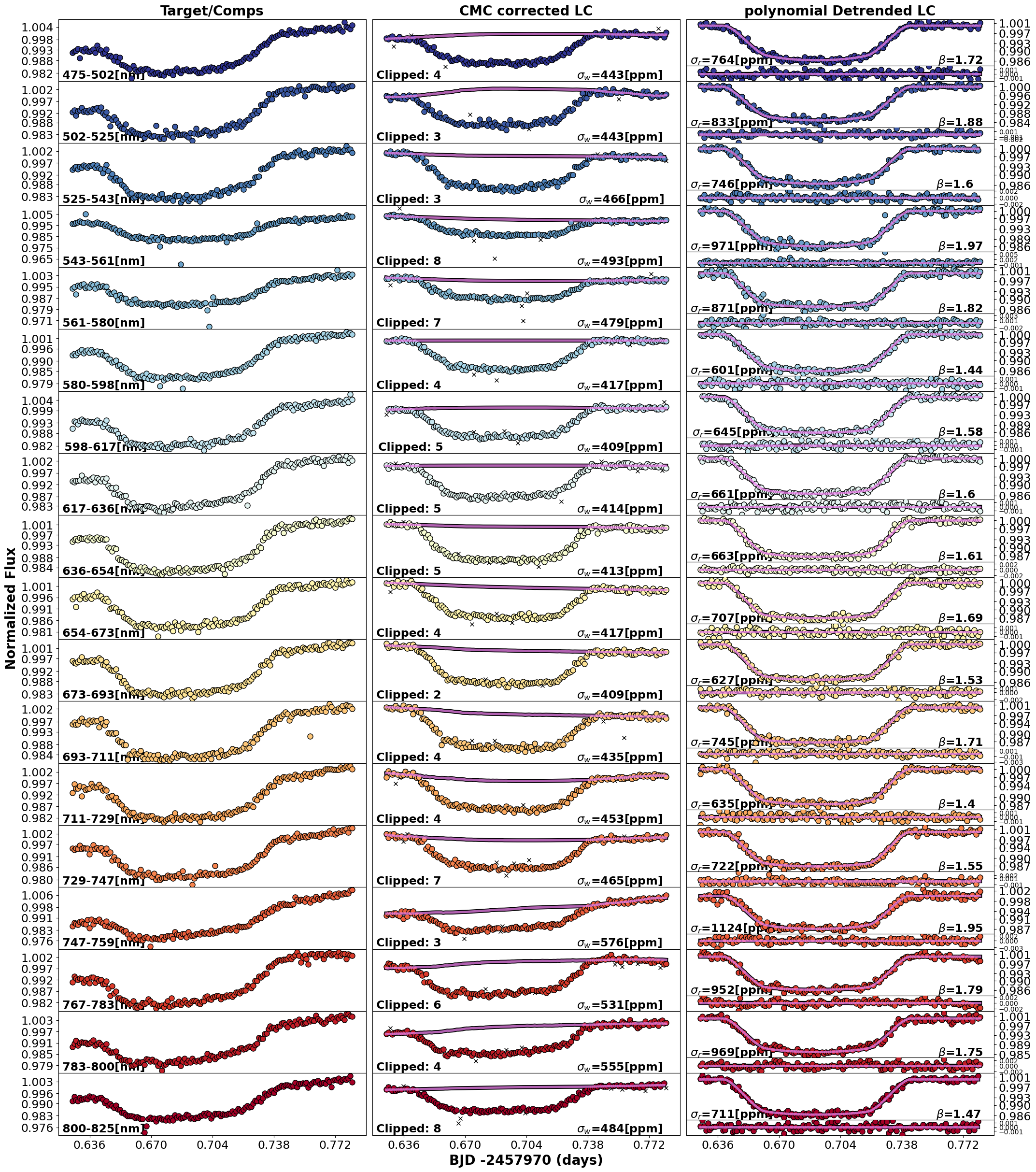}
    \caption{Same as Figure \ref{fig:LC_ut170729}, but for Magellan/IMACS transit UT170804.}
    \label{fig:LC_ut170804} 
\end{figure*}
\begin{figure*}[h!]
    \centering
    \includegraphics[width=1\textwidth]{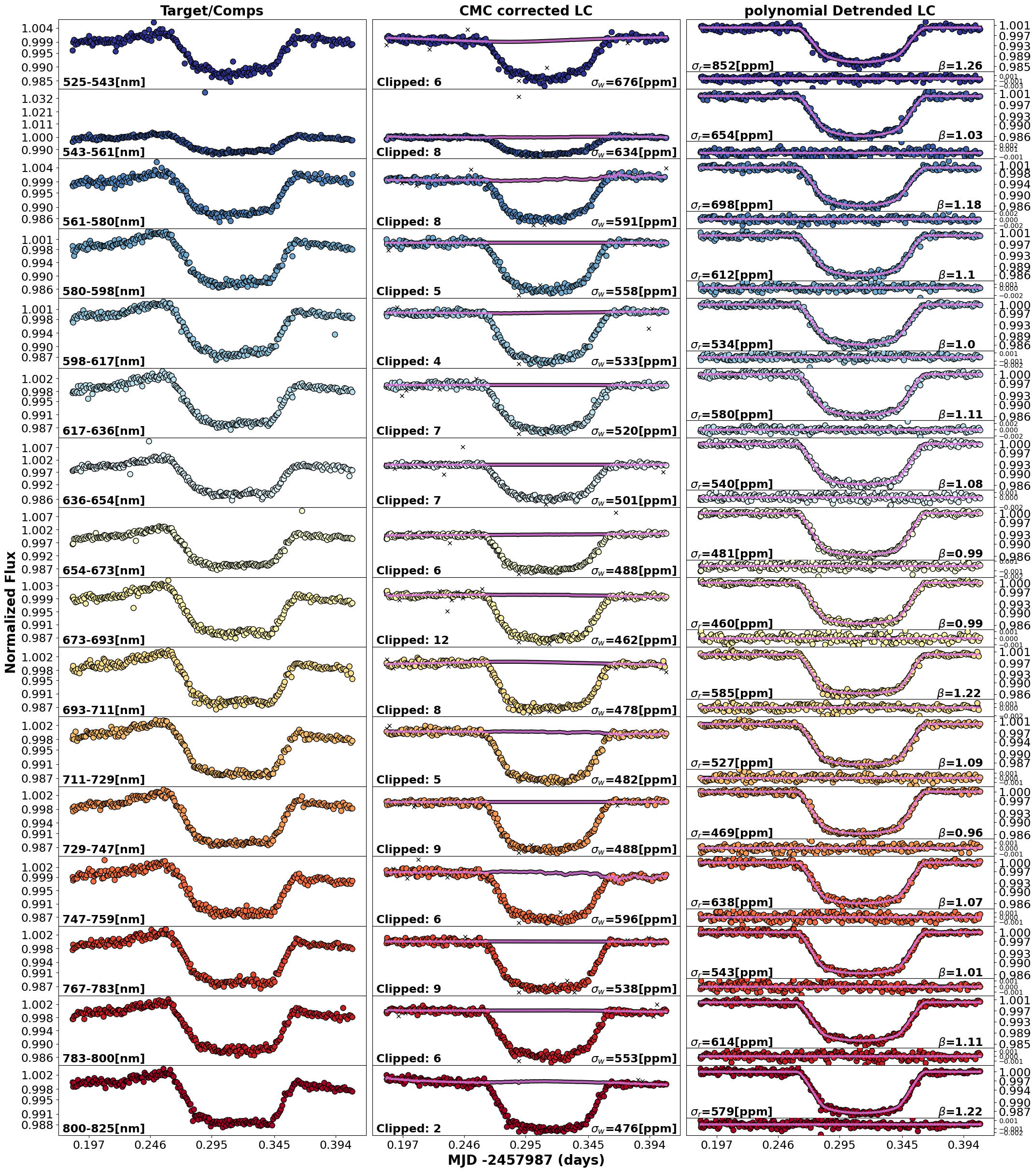}
    \caption{Same as Figure \ref{fig:LC_ut170804}, but for VLT/FORS2 transit UT170822. Though there are a few cases (8th, 9th, and 12th bins) where $\beta$ is slightly less than one. We do not interpret this as the light curve was over corrected, but given that they are nearly one, we see this as we achieved photon noise precision for those bins.}
    \label{fig:LC_ut170822} 
\end{figure*}
\begin{figure*}[h!]
    \centering
    \includegraphics[width=1\textwidth]{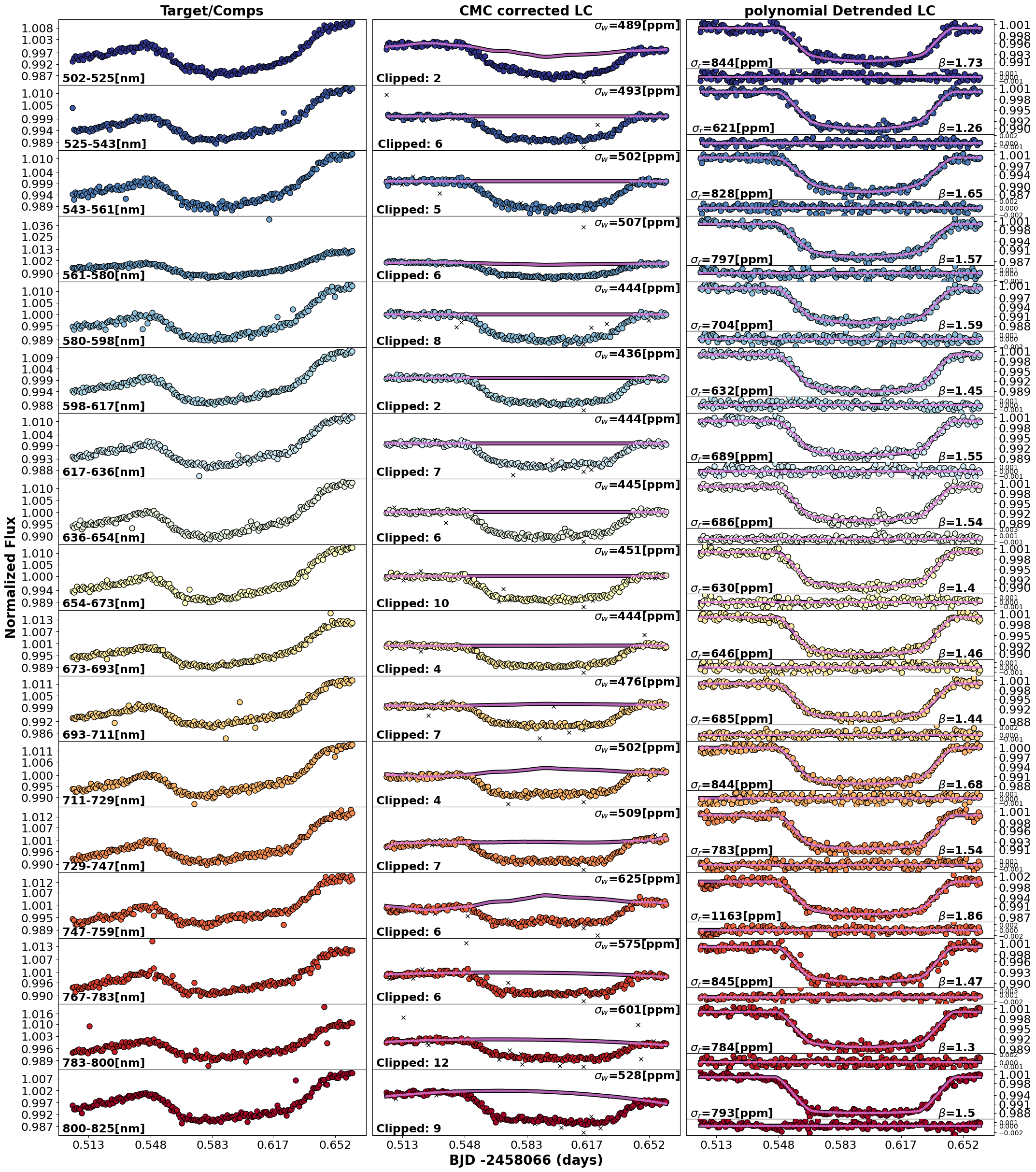}
    \caption{Same as Figure \ref{fig:LC_ut170822}, but for Magellan/IMACS transit UT171108.}
    \label{fig:LC_ut171108} 
\end{figure*}

\begin{deluxetable*}{|C|C|C|C|}[htb]
    \caption{Combined Magellan/IMACS (UT170804 and UT171108) optical transmission spectrum (\textit{R\textsubscript{p}/R\textsubscript{s}}), combined VLT/FORS2 (UT170729 and UT170822) optical transmission spectrum, and global combined optical spectrum (all four transits). The data were produced implementing the reduction and detrending processes discussed in Section~\ref{sec:Tran_Spec}. These depths are offset so the mean depth is 0.1158.}
    \label{tab:Optical_Spec}
    \tablehead{\textbf{Wavelength~(\si{\angstrom})} & \textbf{Magellan/IMACS} & \textbf{VLT/FORS2} & \textbf{Global}}
    \startdata
4000.0-4500.0 &  ---- & 0.1173^{+0.0014}_{-0.0013} & 0.1177^{+0.0014}_{-0.0013} \\ \hline
4500.0-4750.0 &  ---- & 0.1161^{+0.0011}_{-0.0013} & 0.1165^{+0.0011}_{-0.0013} \\ \hline
4750.0-5025.0 & 0.1165^{+0.0036}_{-0.0043} & 0.1153^{+0.0011}_{-0.0013} & 0.1158^{+0.0013}_{-0.001} \\ \hline
5025.0-5255.0 & 0.1143^{+0.0014}_{-0.0015} & 0.1149^{+0.0011}_{-0.0013} & 0.1147\pm0.0009 \\ \hline
5255.0-5435.0 & 0.1138^{+0.0011}_{-0.001} & 0.1153\pm0.0006 & 0.1149^{+0.0006}_{-0.0005} \\ \hline
5435.0-5615.0 & 0.1207^{+0.0011}_{-0.0012} & 0.1158\pm0.0007 & 0.1173\pm0.0006 \\ \hline
5615.0-5800.5 & 0.1185^{+0.0023}_{-0.002} & 0.1169^{+0.0007}_{-0.0008} & 0.1174\pm0.0007 \\ \hline
5800.5-5985.5 & 0.1184\pm0.0009 & 0.1185^{+0.0007}_{-0.0009} & 0.1183^{+0.0005}_{-0.0006} \\ \hline
5985.5-6175.0 & 0.1186^{+0.0011}_{-0.001} & 0.1173\pm0.0008 & 0.1177\pm0.0006 \\ \hline
6175.0-6360.0 & 0.1139\pm0.0009 & 0.1162^{+0.0008}_{-0.0007} & 0.1151^{+0.0005}_{-0.0006} \\ \hline
6360.0-6545.0 & 0.1139\pm0.0009 & 0.1156^{+0.001}_{-0.0008} & 0.1145^{+0.0006}_{-0.0007} \\ \hline
6545.0-6730.0 & 0.1135^{+0.001}_{-0.0009} & 0.1151^{+0.0012}_{-0.0011} & 0.1139\pm0.0007 \\ \hline
6730.0-6930.0 & 0.1141\pm0.0009 & 0.1147^{+0.001}_{-0.0009} & 0.1142^{+0.0006}_{-0.0007} \\ \hline
6930.0-7110.0 & 0.1155^{+0.0015}_{-0.0018} & 0.115^{+0.0007}_{-0.0008} & 0.1154\pm0.0007 \\ \hline
7110.0-7290.0 & 0.1167^{+0.0014}_{-0.0015} & 0.115^{+0.001}_{-0.0009} & 0.1156^{+0.0007}_{-0.0009} \\ \hline
7290.0-7470.0 & 0.1073^{+0.0012}_{-0.001} & 0.1154^{+0.001}_{-0.0009} & 0.1116\pm0.0007 \\ \hline
7470.0-7594.0 & 0.1168^{+0.0012}_{-0.0011} & 0.1161^{+0.0009}_{-0.001} & 0.1165^{+0.0008}_{-0.0007} \\ \hline
7672.0-7836.0 & 0.1134^{+0.0008}_{-0.0009} & 0.1152^{+0.0009}_{-0.0008} & 0.1142\pm0.0006 \\ \hline
7836.0-8000.0 & 0.116\pm0.0009 & 0.1163^{+0.001}_{-0.0009} & 0.1159^{+0.0006}_{-0.0007} \\ \hline
8000.0-8250.0 & 0.1225\pm0.0008 & 0.114^{+0.001}_{-0.0011} & 0.1187^{+0.0007}_{-0.0006} \\ \hline
    \enddata 
\end{deluxetable*}

\begin{deluxetable*}{|C|C|c|C|C|}[htb]
    \caption{Near-IR transmission spectrum obtained from \cite{Yip:2021}. The G102 grism is from 8000--11250 \,{\AA} and the G141 grism is from 11372--12180\,{\AA}.}

    \label{tab:Trans_Spec_Combin}
    \tablehead{\textbf{Wavelength~(\si{\angstrom})} & \textbf{R\textsubscript{p}/R\textsubscript{s}} & &\textbf{Wavelength~(\si{\angstrom})} & \textbf{R\textsubscript{p}/R\textsubscript{s}}}
    \startdata 
8000.00 - 8250.00  & 0.116327 \pm 0.000772 & & 12370.00 - 12559.00 & 0.117175 \pm 0.000650 \\ \hline 
8250.00 - 8500.00 & 0.117499 \pm 0.001021 & & 12559.00 - 12751.00 & 0.117512 \pm 0.000790 \\ \hline 	
8500.00 - 8750.00 & 0.117439 \pm 0.000940 & & 12751.00 - 12944.00 & 0.119013 \pm 0.000780 \\ \hline 
8750.00 - 9000.00 & 0.117597 \pm 0.001338 & & 12944.00 - 13132.00 & 0.117490 \pm 0.000816 \\ \hline 
9000.00 - 9250.00 & 0.117473 \pm 0.001331 & & 13132.00 - 13320.00 & 0.118453 \pm 0.000971 \\ \hline 	
9250.00 - 9500.00 & 0.118305 \pm 0.000801 & & 13320.00 - 13509.00 & 0.118013 \pm 0.000765 \\ \hline 
9500.00 - 9750.00 & 0.116508 \pm 0.000883 & & 13509.00 - 13701.00 & 0.118878 \pm 0.000717 \\ \hline 
9750.00 - 10000.00 & 0.117490 \pm 0.000919 & & 13701.00 - 13900.00 & 0.120121 \pm 0.000995 \\ \hline 
10000.00 - 10250.00 & 0.117427 \pm 0.000850 & & 13900.00 - 14100.00 & 0.120768 \pm 0.001002 \\ \hline 
10250.00 - 10500.00 & 0.118545 \pm 0.000915 & & 14100.00 - 14303.00 & 0.119470 \pm 0.000865 \\ \hline 
10500.00 - 10750.00 & 0.117456 \pm 0.001000 & & 14303.00 - 14509.00 & 0.119771 \pm 0.000800 \\ \hline 
10750.00 - 11000.00 & 0.118174 \pm 0.000594 & & 14509.00 - 14721.00 & 0.117461 \pm 0.000970 \\ \hline 
11000.00 - 11250.00 & 0.119214 \pm 0.000466 & & 14721.00 - 14941.00 & 0.118351 \pm 0.000776 \\ \hline 
11153.00 - 11372.00 & 0.117894 \pm 0.000689 & & 14941.00 - 15165.00 & 0.118482 \pm 0.000924 \\ \hline 
11372.00 - 11583.00 & 0.118786 \pm 0.000845 & & 15165.00 - 15395.00 & 0.117843 \pm 0.000809 \\ \hline 
11583.00 - 11789.00 & 0.118811 \pm 0.000887 & & 15395.00 - 15636.00 & 0.116880 \pm 0.000661 \\ \hline 
11789.00 - 11987.00 & 0.117792 \pm 0.000771 & & 15636.00 - 15889.00 & 0.118089 \pm 0.000722 \\ \hline 
11987.00 - 12180.00 & 0.116893 \pm 0.000932 & & 15889.00 - 16153.00 & 0.117456 \pm 0.001005 \\ \hline 
12180.00 - 12370.00 & 0.117520 \pm 0.000897 & & 16153.00 - 16436.00 & 0.117435 \pm 0.000928 \\ \hline 
    \enddata 
\end{deluxetable*}

\newpage
\clearpage

\section{Synthetic Spectra} \label{Appx:Synethic_Spectra}
\begin{figure*}[htb]
    \centering
    \includegraphics[width=.8\textwidth]{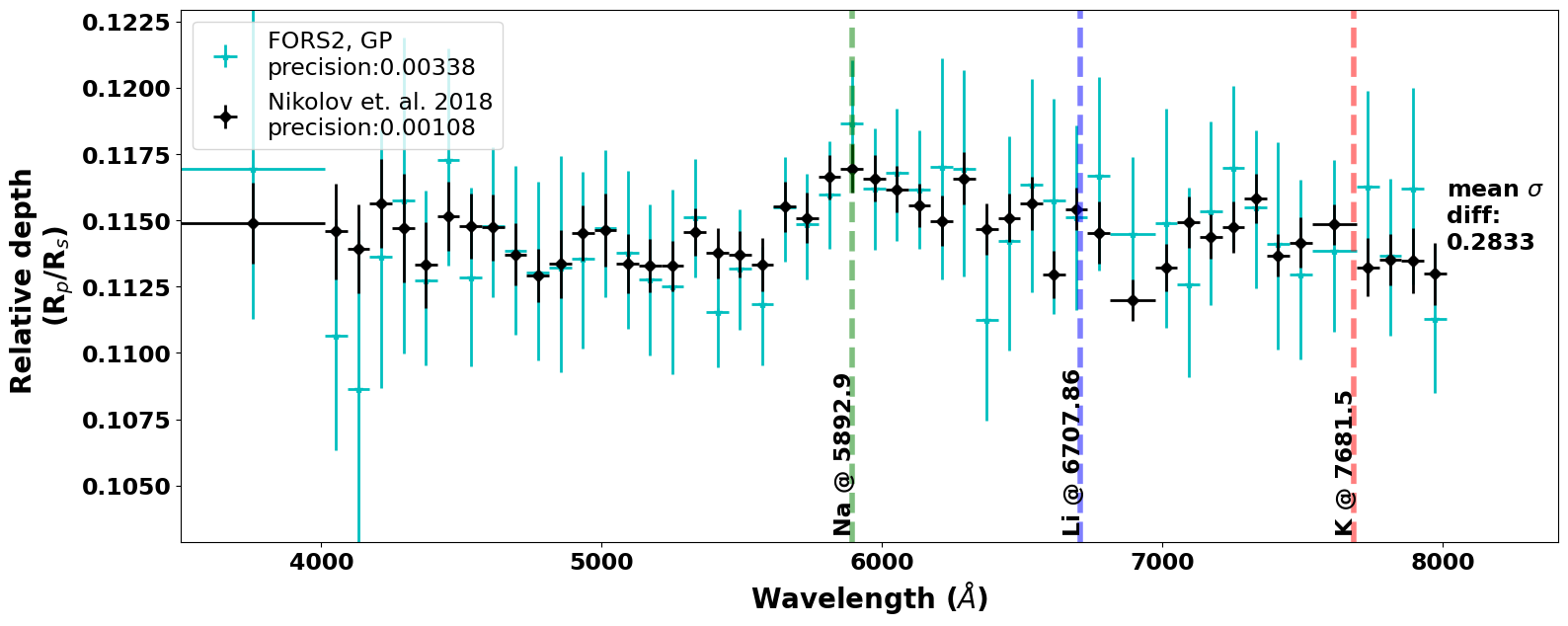}
    \caption{The transmission spectra of VLT/FORS2 data. The black diamonds with uncertainties is the analysis done by \cite{Nikolov:2018}, and the cyan dots with uncertainties is our initial analysis of the VLT/FORS2 data with the same wavelength bins and using the GP detrending method. An offset is applied so their mean depths are equal to one another. Each spectrum is a weighted average of observations taken with the B filter (350--617.3\,nm) on the night of UT170729 and the R filter (529.3--801.3\,nm) on the night of UT170822. The average deviation of each point relative to their uncertainties is 0.28$\sigma$, suggesting that the spectral structures are the same.}
    \label{fig:PCAvsCMC} 
\end{figure*}

In our initial analysis, we reduced the VLT/FORS2 transmission spectrum using the PCA+GP routine (just `GP', given that there is only one comparison),  discussed in appendix~\ref{appx:PCA}, to reduce the white light curves and the binned light curves. When comparing the VLT/FORS2 transmission spectrum produced with this method to the original spectrum \citep{Nikolov:2018}, shown in Figure~\ref{fig:PCAvsCMC}. We found that the uncertainties from the GP method were over 3 times larger than when \cite{Nikolov:2018} used their CMC+Poly method on the same data. This was independent of if the squared exponential kernel (see Appendix~\ref{appx:PCA}) or the \texttt{george} Matern32Kernel was used. Naturally, this is concerning, because though both methods produce similar overall structures, Figure~\ref{fig:PCAvsCMC} implies that either CMC+Poly underestimates the uncertainties or GP overestimates it. As such, we test the accuracy and precision of both methods using synthetic data. 

The first step to produce our synthetic data was to generate a transit light curve with parameters similar to what produces the transit light curve of WASP-96b. For generating all light curves we used \href{https://lweb.cfa.harvard.edu/~lkreidberg/batman/}{\texttt{batman}} \citep{Kreidberg2015_batman} with the same time stamps as FORS2 transit UT170729 (i.e. 90 points covering $\sim$5 hours). The exact light curve parameters used are shown in Table \ref{tab:synth_initial_pars}. 

\begin{figure}[h!]
    \centering
     \includegraphics[width=.65\textwidth]{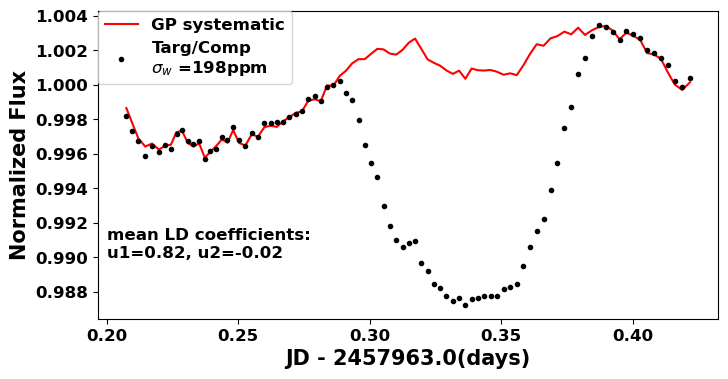}
    \caption{An example synthetic white light curve (black dots) constructed by adding all 10 binned light curves in Figure~\ref{fig:Synth_Binned}. Given that all of the individual chromatic systematics are smaller relative to the white-light systematics and the transit, we do not see their effect in this light curve. This is exactly what we would see in a true light curve (e.g. see Fig. \ref{fig:WLCs}. The red line is the systematics produced by a random draw from a GP distribution created using airmass, full-width at half-maximum (FWHM), and rotational angle as inputs. $\sigma_w$ is the residuals of the out-of-transit data from the GP systematic model, which would be 0 if there was no white noise added.}
    \label{fig:Synth_WL} 
\end{figure}

Next, we generated a realization of a Gaussian Process, GP(t, instrumental\_systematics), for the common systematics that affect all spectrophotometric bands. Here the 'instrumental\_systematics' were the FORS2 transit UT170729 auxiliary parameters of airmass, full-width at half-maximum (FWHM), and rotational angle. Using \href{https://george.readthedocs.io/en/latest/}{\texttt{george}} \citep{Mackey2014_george}, we combined three squared exponential kernels each with an inverse natural-log length scale of -2 and a constant term of -5, -12, and -12 for the GP inputs of airmass, FWHM, and rotational angle, respectively. 
The values of the inverse length scale and constant terms were empirically deduced in order to produce systematics with amplitudes and structure similar to what was seen in the FORS2 transit UT170729 light curves. An example of the white light curve and its GP systematics is shown in Figure \ref{fig:Synth_WL}.

\begin{figure*}[h!]
    \centering
    \includegraphics[width=1\textwidth]{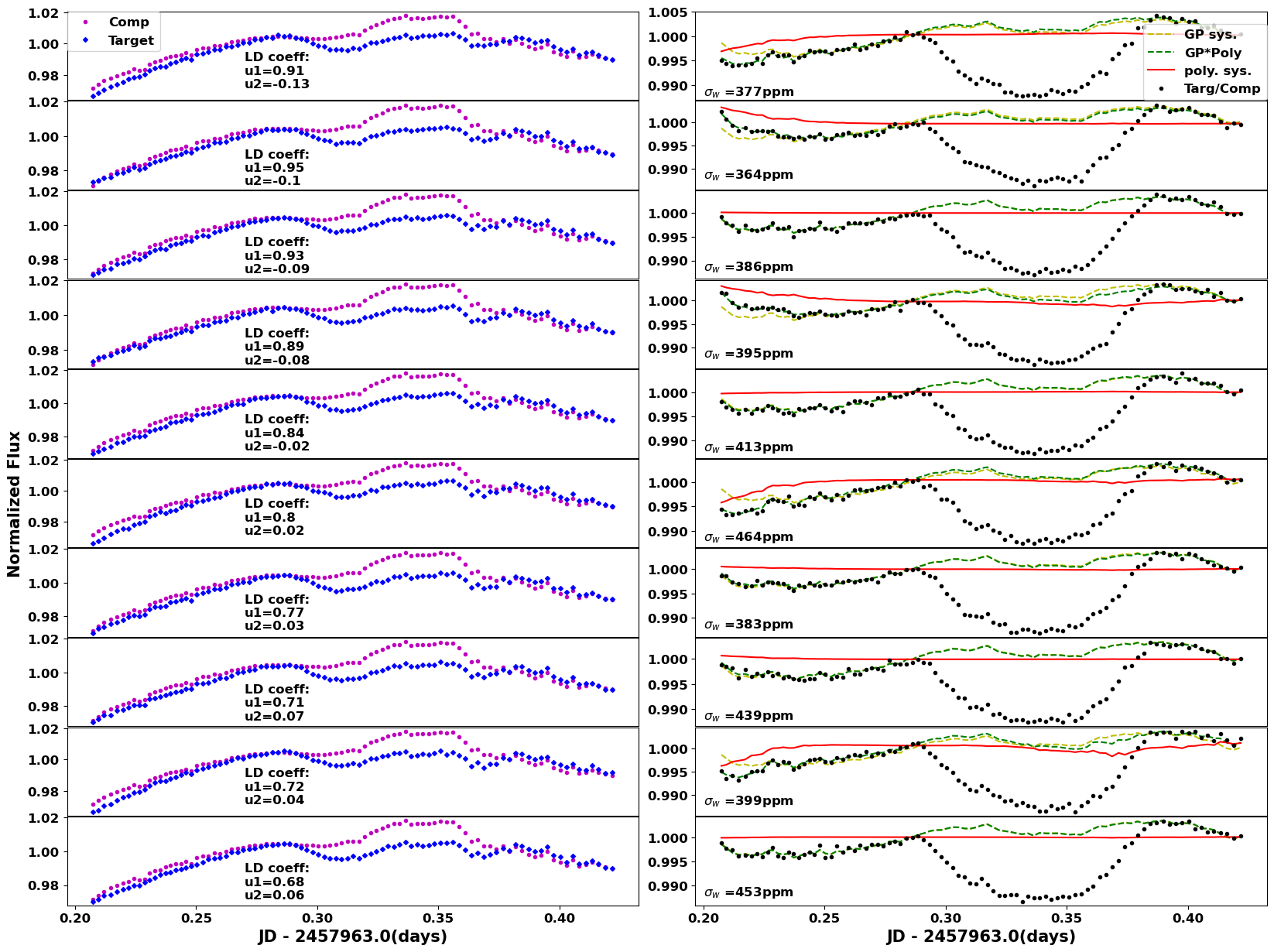}
    \caption{The specific binned light curve used to construct the WLC in Figure~\ref{fig:Synth_WL}. The left column shows the binned target light curves (blue) and the binned comparison light curves (purple). The target light curve is composed of the transit, systematics common to all bins and the comparions, systematics common to all bins and just the target, and systematics unique to each bin. The LD coefficients used for each bin are printed below the light curves, and were determined with \texttt{PyLDTk}. The right column shows the 'raw' (target/comparison) light curve in black dots. The larger amplitude GP systematics (constant for each bin) is shown as a yellow dashed line, the smaller amplitude polynomial systematics (unique per bin) is shown in red and the combined GP and polynomial systematics is shown as green dashes. The residuals of the out-of-transit data from the green dashed line is printed in the bottom left corner ($\sigma_w$).}
    \label{fig:Synth_Binned} 
\end{figure*}

Our third step, was to produce the binned light curves. This was done by generating 10 light curves each containing the same GP systematic realization as above, but with quadratic limb darkening (LD) coefficients determined using \href{https://github.com/hpparvi/ldtk}{\texttt{PyLDTk}} \citep{Parviainen2015} with a stellar effective temperature of 5500~K, stellar log$_{10}$(g) of 4.42, metallicity (log$_{10}$(Z)/log$_{10}$(Z$_{\odot}$)) of 0.25~dex, and assuming each bin was 150{\AA} wide starting from 3800 to 5300{\AA}. Each binned light curve had a unique realization of shot noise with a 400~ppm standard deviation. In addition, each binned light curve had an unique systematic added by generating a first order polynomial on dispersion drift, a second order polynomial on cross-dispersion drift, and a first order polynomial on rotator angle. The coefficients of each polynomial on each bin were randomly drawn from a uniform distribution from -0.1 to 0.1. Each polynomial function was standardized \citep[a common practice done by e.g.][etc.]{Evans2017, Espinoza2019, Kirk:2021} and their product was used as the chromatic systematics. An example of binned light curves and their polynomial systematics can be seen in Figure \ref{fig:Synth_Binned}.

The final step in producing the synthetic data was to decompose the light curves into two components, the target light curves and the comparison star light curves. When making the comparison star light curve we take a constant light curve with normalized value of 1 and add a separate draw from the above GP systematics, i.e. using the same auxiliary parameters and kernels, but a different random draw from the GP distribution. Because it is assumed that the systematics affecting the comparison star's light curve should also affect the targets, we multiple this GP draw by the original target light curve. Thus, when the target is divided by the comparison, the effect of the comparison's systematics are completely divided out. This might not happen in real datasets, because the comparision could add additional systematics, but for our synthetic data, we assume any such systematics is included in the initial construction of the light curve (i.e. the first GP draw). The first column of Figure \ref{fig:Synth_Binned} shows examples of synthesized binned target light curves and comparison light curves. The white light curve can then be constructed by combining all 10 bins, and assuming each bin had the same number of counts. When producing all synthetic data we followed these steps 50 times, where each white light curve was produced with a different draw from the GP distribution, unique realizations of shot noise, and unique coefficients for the polynomial systematics on each bin.




\begin{deluxetable*}{Cccc}[htb]
    \caption{The initial transit parameters to construct the synthetic data (column 2), the white-light priors used when fitting the data (column 3), and the binned priors (column 4). The period (P) of 3.42526\,days and mid transit time (t0) of 2457963.336499\,days were held fixed for all fits. Here 'b' is the impact parameter, which can be transformed to inclination (i) via equation \ref{equ:inclination} using $a/R_s$, eccentricity (fixed at 0), and $\omega$ (fixed at 90$^{\circ}$). The limb-darkening parameters were determined using \texttt{PyLDTk} and were different for each bin, what is shown in the table below is the average.}
    \label{tab:synth_initial_pars}
    \tablehead{\colhead{parameter} & \colhead{value} & \colhead{white-light} & \colhead{binned}}
    \startdata 
    $R_p/R_s$         & 0.1157    & uniform: 0.1--0.14       & normal: m=(WL fit), $\sigma_n$= 0.05   \\
    <q_1>                    & 0.8208                & uniform: 0--1          & uniform: 0--1   \\
    <q_2>               & -0.0201                & uniform: 0--1          & uniform: 0--1 \\ 
    b               & 0.7456                & uniform: 0.5--1          & Fix to WL fit \\ 
    $a/R_s$               & 9.0                & uniform: 8--10          & Fix to WL fit \\ 
    \enddata 
\end{deluxetable*}

\section{Retrieval Modeling Priors} \label{Appx:RetrievalModelingPriors}
\begin{deluxetable*}{|C|C|C|C|C|C|}[htb]
    \caption{The priors for \texttt{Exoretrievals} and \texttt{PLATON}. These priors were set to allow for a wide parameter space to be surveyed, but contained within physical regimes. Not all parameters were included in each model fit (see Tab.~\ref{tab:Global_LnZs}). We used 5000 live points for all runs. For further description of the parameters of \texttt{Exoretrievals}, please refer to the Appendix D of \citet{Espinoza2019}. The log cloud absorbing ($\sigma_{cloud}$) parameter was fixed because across all datasets and all models which included aerosols, this retrieved parameter was near -55. As such, we decided to reduce the dimensionality of the explored posterior space by fixing it to this value, effectively turning off clouds.}
    \label{tab:Priors}
    \tablehead{\multicolumn{3}{|c|}{\bfseries {\large Exoretrievals}} &\multicolumn{3}{|c|}{\bfseries {\large PLATON}}}
    \startdata 
    \textbf{parameter}                          & \textbf{function} & \textbf{bounds} & \textbf{parameter}           & \textbf{function} & \textbf{bounds} \\ \hline
    \text{reference pressure (P\textsubscript{0}, bars)}                            & \text{log-uniform}      & \text{-8 to 3}        &  \text{reference pressure (P\textsubscript{clouds}, Pa)}           & \text{log-uniform} & \text{-3.99 to 7.99} \\ \hline
    \text{planetary atmospheric}                & \text{uniform}          & \text{500 to 1600K}   & \text{planetary atmospheric} & \text{uniform} & \text{500 to 1600K} \\
    \text{temperature (T\textsubscript{p})}                      &                         &                    & \text{temperature (T\textsubscript{p})}        &                & \\ \hline
    \text{stellar temperature}& \text{uniform} & \text{5300 to 5780K}& \text{stellar temperature }& \text{gaussian}         & $\mu$\text{=5540K, }$\sigma$\text{=150K} \\ 
    \text{(T\textsubscript{occ})}   &      &                & \text{(T\textsubscript{star})}   &  & \\ \hline
    \text{stellar heterogeneities}              & \text{uniform}          & \text{2540 to 8540K}& \text{stellar heterogeneities} & \text{uniform} & \text{2540 to 8540K} \\
    \text{temperature (T\textsubscript{het})}   &                         &  & \text{temperature (T\textsubscript{spot})} &    &   \\\hline
\text{heterogeneity covering}                        & \text{gaussian}          & $\mu$\text{=0.014, }$\sigma$\text{=0.009}          & \text{heterogeneity covering}                                      & \text{gaussian} &$\mu$\text{=0.014, }$\sigma$\text{=0.009}  \\ 
    \text{fraction (f\textsubscript{het})}      &                         &               & \text{fraction (f\textsubscript{spot}) }                                      &                   &                   \\ \hline
    \text{offset (depth)}                       & \text{gaussian}         & $\mu$\text{=0, }$\sigma$\text{=2000ppm}& \text{offset (depth)}                   & \text{uniform} & \text{-6000 to 6000ppm} \\ \hline
    \text{haze amplitude ($a$)}                 & \text{log-uniform}         & \text{-1 to 10}           & \text{scattering factor}                                           & \text{log-uniform} & \text{-10 to 10}\\ \hline
    \text{haze power law (}$\gamma$\text{)} $^*$
    & \text{uniform }         & \text{-14 to 4}            & \text{scattering slope (}$\alpha$\text{)} $^{\ddagger}$ 
    & \text{uniform} & \text{-4 to 14}\\ \hline
    \text{log cloud absorbing} & \text{Fixed} & \text{-55}        & \text{metallicity (Z/Z}$_{\odot}$)     & \text{log-uniform} & \text{-1 to 3}\\
    \text{cross-section (}$\sigma$\text{\textsubscript{cloud})} &                   &      &                                            &  & \\ \hline
    \text{trace molecules'}       & \text{log-uniform}      & \text{-30 to 0 }             & \text{C/O}                                                         & \text{uniform} & \text{0.05 to 2}\\
    \text{mixing ratios}       &     &            &                                                        &  & \\ \hline
    \text{reference radius factor} ($f$) $^{\diamond}$  & \text{uniform }         & \text{0.8 to 1.2}            & \text{1\,bar, reference radius (R\textsubscript{0})}                       & \text{uniform} & \text{1 to 1.4R\textsubscript{j}}\\ \hline
    \enddata
\tablenotetext{*}{This is the exponent of the scattering slope power law, where $-4$ is a Rayleigh scattering slope.}
\tablenotetext{\ddagger}{This is the wavelength dependence of scattering, with 4 being Rayleigh.}
\tablenotetext{\diamond}{This is a factor multiplied by the inputted planetary radius to produce the reference radius, i.e. R$_0=f$R$_p$}
\end{deluxetable*}

\end{document}